\documentclass[a4paper, traditabstract, longauth]{aa} 

\pdfoutput=1

\usepackage[english]{babel} 
\usepackage[utf8]{inputenc} 
\usepackage[T1]{fontenc}
\usepackage[mathscr]{eucal}
\usepackage{graphicx,amsmath}
\usepackage{epstopdf}
\usepackage{epsf,color}

\usepackage[breaklinks, colorlinks, citecolor=blue]{hyperref}
\usepackage[nonamebreak]{natbib}
\usepackage{longtable,lscape}
\usepackage{txfonts}
\usepackage{amsfonts}
\usepackage{amssymb}
\usepackage{newlfont}
\usepackage{textcomp}
\usepackage{times}
\usepackage{mathrsfs}
\usepackage{mathbbol}
\bibpunct{(}{)}{;}{a}{}{,} 

\def\setsymbol#1#2{\expandafter\def\csname #1\endcsname{#2}}
\def\getsymbol#1{\csname #1\endcsname}

\def\Planck{\textit{Planck}}

\def\all2013resultspapers{\nocite{planck2013-p01, planck2013-p02, planck2013-p02a, planck2013-p02d, planck2013-p02b, planck2013-p03, planck2013-p03c, planck2013-p03f, planck2013-p03d, planck2013-p03e, planck2013-p01a, planck2013-p06, planck2013-p03a, planck2013-pip88, planck2013-p08, planck2013-p11, planck2013-p12, planck2013-p13, planck2013-p14, planck2013-p15, planck2013-p05b, planck2013-p17, planck2013-p09, planck2013-p09a, planck2013-p20, planck2013-p19, planck2013-pipaberration, planck2013-p05, planck2013-p05a, planck2013-pip56, planck2013-p06b}}
\def\respaps{\nocite{planck2013-p01, planck2013-p02, planck2013-p02a, planck2013-p02d, planck2013-p02b, planck2013-p03, planck2013-p03c, planck2013-p03f, planck2013-p03d, planck2013-p03e, planck2013-p06b, planck2013-p06, planck2013-p03a, planck2013-pip88, planck2013-p08, planck2013-p11, planck2013-p12, planck2013-p13, planck2013-p14, planck2013-p15, planck2013-p05b, planck2013-p17, planck2013-p09, planck2013-p09a, planck2013-p20, planck2013-p19, planck2013-pipaberration, planck2013-p05, planck2013-p05a, planck2013-pip56, planck2013-p01a}}

\newbox\tablebox    \newdimen\tablewidth
\def\leaderfil{\leaders\hbox to 5pt{\hss.\hss}\hfil}
\def\tablenote#1 #2\par{\begingroup \parindent=0.8em
    \abovedisplayshortskip=0pt\belowdisplayshortskip=0pt
    \noindent
    $$\hss\vbox{\hsize\tablewidth \hangindent=\parindent \hangafter=1 \noindent
    \hbox to \parindent{$^#1$\hss}\strut#2\strut\par}\hss$$
    \endgroup}

\def\L2{\ifmmode L_2\else $L_2$\fi}

\def\DeltaT{\ifmmode \Delta T\else $\Delta T$\fi}
\def\deltat{\ifmmode \Delta t\else $\Delta t$\fi}
\def\fknee{\ifmmode f_{\rm knee}\else $f_{\rm knee}$\fi}
\def\Fmax{\ifmmode F_{\rm max}\else $F_{\rm max}$\fi}
\def\solar{\ifmmode{\rm M}_{\mathord\odot}\else${\rm M}_{\mathord\odot}$\fi}
\def\Msolar{\ifmmode{\rm M}_{\mathord\odot}\else${\rm M}_{\mathord\odot}$\fi}
\def\Lsolar{\ifmmode{\rm L}_{\mathord\odot}\else${\rm L}_{\mathord\odot}$\fi}

\def\inv{\ifmmode^{-1}\else$^{-1}$\fi}
\def\mo{\ifmmode^{-1}\else$^{-1}$\fi}
\def\sup#1{\ifmmode ^{\rm #1}\else $^{\rm #1}$\fi}
\def\expo#1{\ifmmode \times 10^{#1}\else $\times 10^{#1}$\fi}
\def\,{\thinspace}
\def\lsim{\mathrel{\raise .4ex\hbox{\rlap{$<$}\lower 1.2ex\hbox{$\sim$}}}}
\def\gsim{\mathrel{\raise .4ex\hbox{\rlap{$>$}\lower 1.2ex\hbox{$\sim$}}}}

\def\simprop{\mathrel{\raise .4ex\hbox{\rlap{$\propto$}\lower 1.2ex\hbox{$\sim$}}}}
\def\deg{\ifmmode^\circ\else$^\circ$\fi}
\def\pdeg{\ifmmode $\setbox0=\hbox{$^{\circ}$}\rlap{\hskip.11\wd0 .}$^{\circ}
          \else \setbox0=\hbox{$^{\circ}$}\rlap{\hskip.11\wd0 .}$^{\circ}$\fi}
\def\arcs{\ifmmode {^{\scriptstyle\prime\prime}}
          \else $^{\scriptstyle\prime\prime}$\fi}
\def\arcm{\ifmmode {^{\scriptstyle\prime}}
          \else $^{\scriptstyle\prime}$\fi}
\newdimen\sa  \newdimen\sb
\def\parcs{\sa=.07em \sb=.03em
     \ifmmode \hbox{\rlap{.}}^{\scriptstyle\prime\kern -\sb\prime}\hbox{\kern -\sa}
     \else \rlap{.}$^{\scriptstyle\prime\kern -\sb\prime}$\kern -\sa\fi}
\def\parcm{\sa=.08em \sb=.03em
     \ifmmode \hbox{\rlap{.}\kern\sa}^{\scriptstyle\prime}\hbox{\kern-\sb}
     \else \rlap{.}\kern\sa$^{\scriptstyle\prime}$\kern-\sb\fi}
\def\ra[#1 #2 #3.#4]{#1\sup{h}#2\sup{m}#3\sup{s}\llap.#4}
\def\dec[#1 #2 #3.#4]{#1\deg#2\arcm#3\arcs\llap.#4}
\def\deco[#1 #2 #3]{#1\deg#2\arcm#3\arcs}
\def\rra[#1 #2]{#1\sup{h}#2\sup{m}}

\def\dots{\relax\ifmmode \ldots\else $\ldots$\fi}
\def\WHzsr{\ifmmode $W\,Hz\mo\,sr\mo$\else W\,Hz\mo\,sr\mo\fi}
\def\mHz{\ifmmode $\,mHz$\else \,mHz\fi}
\def\GHz{\ifmmode $\,GHz$\else \,GHz\fi}
\def\mKs{\ifmmode $\,mK\,s$^{1/2}\else \,mK\,s$^{1/2}$\fi}
\def\muKs{\ifmmode \,\mu$K\,s$^{1/2}\else \,$\mu$K\,s$^{1/2}$\fi}
\def\muKRJs{\ifmmode \,\mu$K$_{\rm RJ}$\,s$^{1/2}\else \,$\mu$K$_{\rm RJ}$\,s$^{1/2}$\fi}
\def\muKHz{\ifmmode \,\mu$K\,Hz$^{-1/2}\else \,$\mu$K\,Hz$^{-1/2}$\fi}
\def\MJysr{\ifmmode \,$MJy\,sr\mo$\else \,MJy\,sr\mo\fi}
\def\MJysrmK{\ifmmode \,$MJy\,sr\mo$\,mK$_{\rm CMB}\mo\else \,MJy\,sr\mo\,mK$_{\rm CMB}\mo$\fi}
\def\microns{\ifmmode \,\mu$m$\else \,$\mu$m\fi}

\def\muK{\ifmmode \,\mu$K$\else \,$\mu$\hbox{K}\fi}
\def\microK{\ifmmode \,\mu$K$\else \,$\mu$\hbox{K}\fi}
\def\muW{\ifmmode \,\mu$W$\else \,$\mu$\hbox{W}\fi}
\def\kms{\ifmmode $\,km\,s$^{-1}\else \,km\,s$^{-1}$\fi}
\def\kmsMpc{\ifmmode $\,\kms\,Mpc\mo$\else \,\kms\,Mpc\mo\fi}
\setsymbol{LFI:center:frequency:70GHz:units}{70.3\,GHz}
\setsymbol{LFI:center:frequency:44GHz:units}{44.1\,GHz}
\setsymbol{LFI:center:frequency:30GHz:units}{28.5\,GHz}

\setsymbol{LFI:center:frequency:70GHz}{70.3}
\setsymbol{LFI:center:frequency:44GHz}{44.1}
\setsymbol{LFI:center:frequency:30GHz}{28.5}

\setsymbol{LFI:center:frequency:LFI18:Rad:M:units}{71.7\GHz}
\setsymbol{LFI:center:frequency:LFI19:Rad:M:units}{67.5\GHz}
\setsymbol{LFI:center:frequency:LFI20:Rad:M:units}{69.2\GHz}
\setsymbol{LFI:center:frequency:LFI21:Rad:M:units}{70.4\GHz}
\setsymbol{LFI:center:frequency:LFI22:Rad:M:units}{71.5\GHz}
\setsymbol{LFI:center:frequency:LFI23:Rad:M:units}{70.8\GHz}
\setsymbol{LFI:center:frequency:LFI24:Rad:M:units}{44.4\GHz}
\setsymbol{LFI:center:frequency:LFI25:Rad:M:units}{44.0\GHz}
\setsymbol{LFI:center:frequency:LFI26:Rad:M:units}{43.9\GHz}
\setsymbol{LFI:center:frequency:LFI27:Rad:M:units}{28.3\GHz}
\setsymbol{LFI:center:frequency:LFI28:Rad:M:units}{28.8\GHz}
\setsymbol{LFI:center:frequency:LFI18:Rad:S:units}{70.1\GHz}
\setsymbol{LFI:center:frequency:LFI19:Rad:S:units}{69.6\GHz}
\setsymbol{LFI:center:frequency:LFI20:Rad:S:units}{69.5\GHz}
\setsymbol{LFI:center:frequency:LFI21:Rad:S:units}{69.5\GHz}
\setsymbol{LFI:center:frequency:LFI22:Rad:S:units}{72.8\GHz}
\setsymbol{LFI:center:frequency:LFI23:Rad:S:units}{71.3\GHz}
\setsymbol{LFI:center:frequency:LFI24:Rad:S:units}{44.1\GHz}
\setsymbol{LFI:center:frequency:LFI25:Rad:S:units}{44.1\GHz}
\setsymbol{LFI:center:frequency:LFI26:Rad:S:units}{44.1\GHz}
\setsymbol{LFI:center:frequency:LFI27:Rad:S:units}{28.5\GHz}
\setsymbol{LFI:center:frequency:LFI28:Rad:S:units}{28.2\GHz}

\setsymbol{LFI:center:frequency:LFI18:Rad:M}{71.7}
\setsymbol{LFI:center:frequency:LFI19:Rad:M}{67.5}
\setsymbol{LFI:center:frequency:LFI20:Rad:M}{69.2}
\setsymbol{LFI:center:frequency:LFI21:Rad:M}{70.4}
\setsymbol{LFI:center:frequency:LFI22:Rad:M}{71.5}
\setsymbol{LFI:center:frequency:LFI23:Rad:M}{70.8}
\setsymbol{LFI:center:frequency:LFI24:Rad:M}{44.4}
\setsymbol{LFI:center:frequency:LFI25:Rad:M}{44.0}
\setsymbol{LFI:center:frequency:LFI26:Rad:M}{43.9}
\setsymbol{LFI:center:frequency:LFI27:Rad:M}{28.3}
\setsymbol{LFI:center:frequency:LFI28:Rad:M}{28.8}
\setsymbol{LFI:center:frequency:LFI18:Rad:S}{70.1}
\setsymbol{LFI:center:frequency:LFI19:Rad:S}{69.6}
\setsymbol{LFI:center:frequency:LFI20:Rad:S}{69.5}
\setsymbol{LFI:center:frequency:LFI21:Rad:S}{69.5}
\setsymbol{LFI:center:frequency:LFI22:Rad:S}{72.8}
\setsymbol{LFI:center:frequency:LFI23:Rad:S}{71.3}
\setsymbol{LFI:center:frequency:LFI24:Rad:S}{44.1}
\setsymbol{LFI:center:frequency:LFI25:Rad:S}{44.1}
\setsymbol{LFI:center:frequency:LFI26:Rad:S}{44.1}
\setsymbol{LFI:center:frequency:LFI27:Rad:S}{28.5}
\setsymbol{LFI:center:frequency:LFI28:Rad:S}{28.2}

\setsymbol{LFI:white:noise:sensitivity:70GHz:units}{134.7\muKs}
\setsymbol{LFI:white:noise:sensitivity:44GHz:units}{164.7\muKs}
\setsymbol{LFI:white:noise:sensitivity:30GHz:units}{143.4\muKs}

\setsymbol{LFI:white:noise:sensitivity:70GHz}{134.7}
\setsymbol{LFI:white:noise:sensitivity:44GHz}{164.7}
\setsymbol{LFI:white:noise:sensitivity:30GHz}{143.4}

\setsymbol{LFI:white:noise:sensitivity:LFI18:Rad:M:units}{512.0\muKs}
\setsymbol{LFI:white:noise:sensitivity:LFI19:Rad:M:units}{581.4\muKs}
\setsymbol{LFI:white:noise:sensitivity:LFI20:Rad:M:units}{590.8\muKs}
\setsymbol{LFI:white:noise:sensitivity:LFI21:Rad:M:units}{455.2\muKs}
\setsymbol{LFI:white:noise:sensitivity:LFI22:Rad:M:units}{492.0\muKs}
\setsymbol{LFI:white:noise:sensitivity:LFI23:Rad:M:units}{507.7\muKs}
\setsymbol{LFI:white:noise:sensitivity:LFI24:Rad:M:units}{462.2\muKs}
\setsymbol{LFI:white:noise:sensitivity:LFI25:Rad:M:units}{413.6\muKs}
\setsymbol{LFI:white:noise:sensitivity:LFI26:Rad:M:units}{478.6\muKs}
\setsymbol{LFI:white:noise:sensitivity:LFI27:Rad:M:units}{277.7\muKs}
\setsymbol{LFI:white:noise:sensitivity:LFI28:Rad:M:units}{312.3\muKs}
\setsymbol{LFI:white:noise:sensitivity:LFI18:Rad:S:units}{465.7\muKs}
\setsymbol{LFI:white:noise:sensitivity:LFI19:Rad:S:units}{555.6\muKs}
\setsymbol{LFI:white:noise:sensitivity:LFI20:Rad:S:units}{623.2\muKs}
\setsymbol{LFI:white:noise:sensitivity:LFI21:Rad:S:units}{564.1\muKs}
\setsymbol{LFI:white:noise:sensitivity:LFI22:Rad:S:units}{534.4\muKs}
\setsymbol{LFI:white:noise:sensitivity:LFI23:Rad:S:units}{542.4\muKs}
\setsymbol{LFI:white:noise:sensitivity:LFI24:Rad:S:units}{399.2\muKs}
\setsymbol{LFI:white:noise:sensitivity:LFI25:Rad:S:units}{392.6\muKs}
\setsymbol{LFI:white:noise:sensitivity:LFI26:Rad:S:units}{418.6\muKs}
\setsymbol{LFI:white:noise:sensitivity:LFI27:Rad:S:units}{302.9\muKs}
\setsymbol{LFI:white:noise:sensitivity:LFI28:Rad:S:units}{285.3\muKs}

\setsymbol{LFI:white:noise:sensitivity:LFI18:Rad:M}{512.0}
\setsymbol{LFI:white:noise:sensitivity:LFI19:Rad:M}{581.4}
\setsymbol{LFI:white:noise:sensitivity:LFI20:Rad:M}{590.8}
\setsymbol{LFI:white:noise:sensitivity:LFI21:Rad:M}{455.2}
\setsymbol{LFI:white:noise:sensitivity:LFI22:Rad:M}{492.0}
\setsymbol{LFI:white:noise:sensitivity:LFI23:Rad:M}{507.7}
\setsymbol{LFI:white:noise:sensitivity:LFI24:Rad:M}{462.2}
\setsymbol{LFI:white:noise:sensitivity:LFI25:Rad:M}{413.6}
\setsymbol{LFI:white:noise:sensitivity:LFI26:Rad:M}{478.6}
\setsymbol{LFI:white:noise:sensitivity:LFI27:Rad:M}{277.7}
\setsymbol{LFI:white:noise:sensitivity:LFI28:Rad:M}{312.3}
\setsymbol{LFI:white:noise:sensitivity:LFI18:Rad:S}{465.7}
\setsymbol{LFI:white:noise:sensitivity:LFI19:Rad:S}{555.6}
\setsymbol{LFI:white:noise:sensitivity:LFI20:Rad:S}{623.2}
\setsymbol{LFI:white:noise:sensitivity:LFI21:Rad:S}{564.1}
\setsymbol{LFI:white:noise:sensitivity:LFI22:Rad:S}{534.4}
\setsymbol{LFI:white:noise:sensitivity:LFI23:Rad:S}{542.4}
\setsymbol{LFI:white:noise:sensitivity:LFI24:Rad:S}{399.2}
\setsymbol{LFI:white:noise:sensitivity:LFI25:Rad:S}{392.6}
\setsymbol{LFI:white:noise:sensitivity:LFI26:Rad:S}{418.6}
\setsymbol{LFI:white:noise:sensitivity:LFI27:Rad:S}{302.9}
\setsymbol{LFI:white:noise:sensitivity:LFI28:Rad:S}{285.3}

\setsymbol{LFI:knee:frequency:70GHz:units}{29.5\mHz}
\setsymbol{LFI:knee:frequency:44GHz:units}{56.2\mHz}
\setsymbol{LFI:knee:frequency:30GHz:units}{113.7\mHz}

\setsymbol{LFI:knee:frequency:70GHz}{29.5}
\setsymbol{LFI:knee:frequency:44GHz}{56.2}
\setsymbol{LFI:knee:frequency:30GHz}{113.7}

\setsymbol{LFI:knee:frequency:LFI18:Rad:M:units}{16.3\mHz}
\setsymbol{LFI:knee:frequency:LFI19:Rad:M:units}{15.1\mHz}
\setsymbol{LFI:knee:frequency:LFI20:Rad:M:units}{18.7\mHz}
\setsymbol{LFI:knee:frequency:LFI21:Rad:M:units}{37.2\mHz}
\setsymbol{LFI:knee:frequency:LFI22:Rad:M:units}{12.7\mHz}
\setsymbol{LFI:knee:frequency:LFI23:Rad:M:units}{34.6\mHz}
\setsymbol{LFI:knee:frequency:LFI24:Rad:M:units}{46.2\mHz}
\setsymbol{LFI:knee:frequency:LFI25:Rad:M:units}{24.9\mHz}
\setsymbol{LFI:knee:frequency:LFI26:Rad:M:units}{67.6\mHz}
\setsymbol{LFI:knee:frequency:LFI27:Rad:M:units}{187.4\mHz}
\setsymbol{LFI:knee:frequency:LFI28:Rad:M:units}{122.2\mHz}
\setsymbol{LFI:knee:frequency:LFI18:Rad:S:units}{17.7\mHz}
\setsymbol{LFI:knee:frequency:LFI19:Rad:S:units}{22.0\mHz}
\setsymbol{LFI:knee:frequency:LFI20:Rad:S:units}{8.7\mHz}
\setsymbol{LFI:knee:frequency:LFI21:Rad:S:units}{25.9\mHz}
\setsymbol{LFI:knee:frequency:LFI22:Rad:S:units}{15.8\mHz}
\setsymbol{LFI:knee:frequency:LFI23:Rad:S:units}{129.8\mHz}
\setsymbol{LFI:knee:frequency:LFI24:Rad:S:units}{100.9\mHz}
\setsymbol{LFI:knee:frequency:LFI25:Rad:S:units}{38.9\mHz}
\setsymbol{LFI:knee:frequency:LFI26:Rad:S:units}{58.9\mHz}
\setsymbol{LFI:knee:frequency:LFI27:Rad:S:units}{104.4\mHz}
\setsymbol{LFI:knee:frequency:LFI28:Rad:S:units}{40.7\mHz}

\setsymbol{LFI:knee:frequency:LFI18:Rad:M}{16.3}
\setsymbol{LFI:knee:frequency:LFI19:Rad:M}{15.1}
\setsymbol{LFI:knee:frequency:LFI20:Rad:M}{18.7}
\setsymbol{LFI:knee:frequency:LFI21:Rad:M}{37.2}
\setsymbol{LFI:knee:frequency:LFI22:Rad:M}{12.7}
\setsymbol{LFI:knee:frequency:LFI23:Rad:M}{34.6}
\setsymbol{LFI:knee:frequency:LFI24:Rad:M}{46.2}
\setsymbol{LFI:knee:frequency:LFI25:Rad:M}{24.9}
\setsymbol{LFI:knee:frequency:LFI26:Rad:M}{67.6}
\setsymbol{LFI:knee:frequency:LFI27:Rad:M}{187.4}
\setsymbol{LFI:knee:frequency:LFI28:Rad:M}{122.2}
\setsymbol{LFI:knee:frequency:LFI18:Rad:S}{17.7}
\setsymbol{LFI:knee:frequency:LFI19:Rad:S}{22.0}
\setsymbol{LFI:knee:frequency:LFI20:Rad:S}{8.7}
\setsymbol{LFI:knee:frequency:LFI21:Rad:S}{25.9}
\setsymbol{LFI:knee:frequency:LFI22:Rad:S}{15.8}
\setsymbol{LFI:knee:frequency:LFI23:Rad:S}{129.8}
\setsymbol{LFI:knee:frequency:LFI24:Rad:S}{100.9}
\setsymbol{LFI:knee:frequency:LFI25:Rad:S}{38.9}
\setsymbol{LFI:knee:frequency:LFI26:Rad:S}{58.9}
\setsymbol{LFI:knee:frequency:LFI27:Rad:S}{104.4}
\setsymbol{LFI:knee:frequency:LFI28:Rad:S}{40.7}

\setsymbol{LFI:slope:70GHz:units}{$-1.03$\mHz}
\setsymbol{LFI:slope:44GHz:units}{$-0.89$\mHz}
\setsymbol{LFI:slope:30GHz:units}{$-0.87$\mHz}

\setsymbol{LFI:slope:70GHz}{$-1.03$}
\setsymbol{LFI:slope:44GHz}{$-0.89$}
\setsymbol{LFI:slope:30GHz}{$-0.87$}

\setsymbol{LFI:slope:LFI18:Rad:M:units}{$-1.04$\mHz}
\setsymbol{LFI:slope:LFI19:Rad:M:units}{$-1.09$\mHz}
\setsymbol{LFI:slope:LFI20:Rad:M:units}{$-0.69$\mHz}
\setsymbol{LFI:slope:LFI21:Rad:M:units}{$-1.56$\mHz}
\setsymbol{LFI:slope:LFI22:Rad:M:units}{$-1.01$\mHz}
\setsymbol{LFI:slope:LFI23:Rad:M:units}{$-0.96$\mHz}
\setsymbol{LFI:slope:LFI24:Rad:M:units}{$-0.83$\mHz}
\setsymbol{LFI:slope:LFI25:Rad:M:units}{$-0.91$\mHz}
\setsymbol{LFI:slope:LFI26:Rad:M:units}{$-0.95$\mHz}
\setsymbol{LFI:slope:LFI27:Rad:M:units}{$-0.87$\mHz}
\setsymbol{LFI:slope:LFI28:Rad:M:units}{$-0.88$\mHz}
\setsymbol{LFI:slope:LFI18:Rad:S:units}{$-1.15$\mHz}
\setsymbol{LFI:slope:LFI19:Rad:S:units}{$-1.00$\mHz}
\setsymbol{LFI:slope:LFI20:Rad:S:units}{$-0.95$\mHz}
\setsymbol{LFI:slope:LFI21:Rad:S:units}{$-0.92$\mHz}
\setsymbol{LFI:slope:LFI22:Rad:S:units}{$-1.01$\mHz}
\setsymbol{LFI:slope:LFI23:Rad:S:units}{$-0.95$\mHz}
\setsymbol{LFI:slope:LFI24:Rad:S:units}{$-0.73$\mHz}
\setsymbol{LFI:slope:LFI25:Rad:S:units}{$-1.16$\mHz}
\setsymbol{LFI:slope:LFI26:Rad:S:units}{$-0.79$\mHz}
\setsymbol{LFI:slope:LFI27:Rad:S:units}{$-0.82$\mHz}
\setsymbol{LFI:slope:LFI28:Rad:S:units}{$-0.91$\mHz}

\setsymbol{LFI:slope:LFI18:Rad:M}{$-1.04$}
\setsymbol{LFI:slope:LFI19:Rad:M}{$-1.09$}
\setsymbol{LFI:slope:LFI20:Rad:M}{$-0.69$}
\setsymbol{LFI:slope:LFI21:Rad:M}{$-1.56$}
\setsymbol{LFI:slope:LFI22:Rad:M}{$-1.01$}
\setsymbol{LFI:slope:LFI23:Rad:M}{$-0.96$}
\setsymbol{LFI:slope:LFI24:Rad:M}{$-0.83$}
\setsymbol{LFI:slope:LFI25:Rad:M}{$-0.91$}
\setsymbol{LFI:slope:LFI26:Rad:M}{$-0.95$}
\setsymbol{LFI:slope:LFI27:Rad:M}{$-0.87$}
\setsymbol{LFI:slope:LFI28:Rad:M}{$-0.88$}
\setsymbol{LFI:slope:LFI18:Rad:S}{$-1.15$}
\setsymbol{LFI:slope:LFI19:Rad:S}{$-1.00$}
\setsymbol{LFI:slope:LFI20:Rad:S}{$-0.95$}
\setsymbol{LFI:slope:LFI21:Rad:S}{$-0.92$}
\setsymbol{LFI:slope:LFI22:Rad:S}{$-1.01$}
\setsymbol{LFI:slope:LFI23:Rad:S}{$-0.95$}
\setsymbol{LFI:slope:LFI24:Rad:S}{$-0.73$}
\setsymbol{LFI:slope:LFI25:Rad:S}{$-1.16$}
\setsymbol{LFI:slope:LFI26:Rad:S}{$-0.79$}
\setsymbol{LFI:slope:LFI27:Rad:S}{$-0.82$}
\setsymbol{LFI:slope:LFI28:Rad:S}{$-0.91$}

\setsymbol{LFI:FWHM:70GHz:units}{13\parcm01}
\setsymbol{LFI:FWHM:44GHz:units}{27\parcm92}
\setsymbol{LFI:FWHM:30GHz:units}{32\parcm65}

\setsymbol{LFI:FWHM:70GHz}{13.01}
\setsymbol{LFI:FWHM:44GHz}{27.92}
\setsymbol{LFI:FWHM:30GHz}{32.65}

\setsymbol{LFI:FWHM:LFI18:units}{13\parcm39}
\setsymbol{LFI:FWHM:LFI19:units}{13\parcm01}
\setsymbol{LFI:FWHM:LFI20:units}{12\parcm75}
\setsymbol{LFI:FWHM:LFI21:units}{12\parcm74}
\setsymbol{LFI:FWHM:LFI22:units}{12\parcm87}
\setsymbol{LFI:FWHM:LFI23:units}{13\parcm27}
\setsymbol{LFI:FWHM:LFI24:units}{22\parcm98}
\setsymbol{LFI:FWHM:LFI25:units}{30\parcm46}
\setsymbol{LFI:FWHM:LFI26:units}{30\parcm31}
\setsymbol{LFI:FWHM:LFI27:units}{32\parcm65}
\setsymbol{LFI:FWHM:LFI28:units}{32\parcm66}

\setsymbol{LFI:FWHM:LFI18}{13.39}
\setsymbol{LFI:FWHM:LFI19}{13.01}
\setsymbol{LFI:FWHM:LFI20}{12.75}
\setsymbol{LFI:FWHM:LFI21}{12.74}
\setsymbol{LFI:FWHM:LFI22}{12.87}
\setsymbol{LFI:FWHM:LFI23}{13.27}
\setsymbol{LFI:FWHM:LFI24}{22.98}
\setsymbol{LFI:FWHM:LFI25}{30.46}
\setsymbol{LFI:FWHM:LFI26}{30.31}
\setsymbol{LFI:FWHM:LFI27}{32.65}
\setsymbol{LFI:FWHM:LFI28}{32.66}

\setsymbol{LFI:FWHM:uncertainty:LFI18:units}{0.170\arcm}
\setsymbol{LFI:FWHM:uncertainty:LFI19:units}{0.174\arcm}
\setsymbol{LFI:FWHM:uncertainty:LFI20:units}{0.170\arcm}
\setsymbol{LFI:FWHM:uncertainty:LFI21:units}{0.156\arcm}
\setsymbol{LFI:FWHM:uncertainty:LFI22:units}{0.164\arcm}
\setsymbol{LFI:FWHM:uncertainty:LFI23:units}{0.171\arcm}
\setsymbol{LFI:FWHM:uncertainty:LFI24:units}{0.652\arcm}
\setsymbol{LFI:FWHM:uncertainty:LFI25:units}{1.075\arcm}
\setsymbol{LFI:FWHM:uncertainty:LFI26:units}{1.131\arcm}
\setsymbol{LFI:FWHM:uncertainty:LFI27:units}{1.266\arcm}
\setsymbol{LFI:FWHM:uncertainty:LFI28:units}{1.287\arcm}

\setsymbol{LFI:FWHM:uncertainty:LFI18}{0.170}
\setsymbol{LFI:FWHM:uncertainty:LFI19}{0.174}
\setsymbol{LFI:FWHM:uncertainty:LFI20}{0.170}
\setsymbol{LFI:FWHM:uncertainty:LFI21}{0.156}
\setsymbol{LFI:FWHM:uncertainty:LFI22}{0.164}
\setsymbol{LFI:FWHM:uncertainty:LFI23}{0.171}
\setsymbol{LFI:FWHM:uncertainty:LFI24}{0.652}
\setsymbol{LFI:FWHM:uncertainty:LFI25}{1.075}
\setsymbol{LFI:FWHM:uncertainty:LFI26}{1.131}
\setsymbol{LFI:FWHM:uncertainty:LFI27}{1.266}
\setsymbol{LFI:FWHM:uncertainty:LFI28}{1.287}

\setsymbol{HFI:center:frequency:100GHz:units}{100\,GHz}
\setsymbol{HFI:center:frequency:143GHz:units}{143\,GHz}
\setsymbol{HFI:center:frequency:217GHz:units}{217\,GHz}
\setsymbol{HFI:center:frequency:353GHz:units}{353\,GHz}
\setsymbol{HFI:center:frequency:545GHz:units}{545\,GHz}
\setsymbol{HFI:center:frequency:857GHz:units}{857\,GHz}

\setsymbol{HFI:center:frequency:100GHz}{100}
\setsymbol{HFI:center:frequency:143GHz}{143}
\setsymbol{HFI:center:frequency:217GHz}{217}
\setsymbol{HFI:center:frequency:353GHz}{353}
\setsymbol{HFI:center:frequency:545GHz}{545}
\setsymbol{HFI:center:frequency:857GHz}{857}

\setsymbol{HFI:Ndetectors:100GHz}{8}
\setsymbol{HFI:Ndetectors:143GHz}{11}
\setsymbol{HFI:Ndetectors:217GHz}{12}
\setsymbol{HFI:Ndetectors:353GHz}{12}
\setsymbol{HFI:Ndetectors:545GHz}{3}
\setsymbol{HFI:Ndetectors:857GHz}{4}

\setsymbol{HFI:FWHM:Maps:100GHz:units}{9\parcm88}
\setsymbol{HFI:FWHM:Maps:143GHz:units}{7\parcm18}
\setsymbol{HFI:FWHM:Maps:217GHz:units}{4\parcm87}
\setsymbol{HFI:FWHM:Maps:353GHz:units}{4\parcm65}
\setsymbol{HFI:FWHM:Maps:545GHz:units}{4\parcm72}
\setsymbol{HFI:FWHM:Maps:857GHz:units}{4\parcm39}
\setsymbol{HFI:FWHM:Maps:100GHz}{9.88}
\setsymbol{HFI:FWHM:Maps:143GHz}{7.18}
\setsymbol{HFI:FWHM:Maps:217GHz}{4.87}
\setsymbol{HFI:FWHM:Maps:353GHz}{4.65}
\setsymbol{HFI:FWHM:Maps:545GHz}{4.72}
\setsymbol{HFI:FWHM:Maps:857GHz}{4.39}

\setsymbol{HFI:beam:ellipticity:Maps:100GHz}{1.15}
\setsymbol{HFI:beam:ellipticity:Maps:143GHz}{1.01}
\setsymbol{HFI:beam:ellipticity:Maps:217GHz}{1.06}
\setsymbol{HFI:beam:ellipticity:Maps:353GHz}{1.05}
\setsymbol{HFI:beam:ellipticity:Maps:545GHz}{1.14}
\setsymbol{HFI:beam:ellipticity:Maps:857GHz}{1.19}

\setsymbol{HFI:FWHM:Mars:100GHz:units}{9\parcm37}
\setsymbol{HFI:FWHM:Mars:143GHz:units}{7\parcm04}
\setsymbol{HFI:FWHM:Mars:217GHz:units}{4\parcm68}
\setsymbol{HFI:FWHM:Mars:353GHz:units}{4\parcm43}
\setsymbol{HFI:FWHM:Mars:545GHz:units}{3\parcm80}
\setsymbol{HFI:FWHM:Mars:857GHz:units}{3\parcm67}

\setsymbol{HFI:FWHM:Mars:100GHz}{9.37}
\setsymbol{HFI:FWHM:Mars:143GHz}{7.04}
\setsymbol{HFI:FWHM:Mars:217GHz}{4.68}
\setsymbol{HFI:FWHM:Mars:353GHz}{4.43}
\setsymbol{HFI:FWHM:Mars:545GHz}{3.80}
\setsymbol{HFI:FWHM:Mars:857GHz}{3.67}

\setsymbol{HFI:beam:ellipticity:Mars:100GHz}{1.18}
\setsymbol{HFI:beam:ellipticity:Mars:143GHz}{1.03}
\setsymbol{HFI:beam:ellipticity:Mars:217GHz}{1.14}
\setsymbol{HFI:beam:ellipticity:Mars:353GHz}{1.09}
\setsymbol{HFI:beam:ellipticity:Mars:545GHz}{1.25}
\setsymbol{HFI:beam:ellipticity:Mars:857GHz}{1.03}

\setsymbol{HFI:CMB:relative:calibration:100GHz}{$\lsim 1\%$}
\setsymbol{HFI:CMB:relative:calibration:143GHz}{$\lsim 1\%$}
\setsymbol{HFI:CMB:relative:calibration:217GHz}{$\lsim 1\%$}
\setsymbol{HFI:CMB:relative:calibration:353GHz}{$\lsim 1\%$}
\setsymbol{HFI:CMB:relative:calibration:545GHz}{}
\setsymbol{HFI:CMB:relative:calibration:857GHz}{}

\setsymbol{HFI:CMB:absolute:calibration:100GHz}{$\lsim 2\%$}
\setsymbol{HFI:CMB:absolute:calibration:143GHz}{$\lsim 2\%$}
\setsymbol{HFI:CMB:absolute:calibration:217GHz}{$\lsim 2\%$}
\setsymbol{HFI:CMB:absolute:calibration:353GHz}{$\lsim 2\%$}
\setsymbol{HFI:CMB:absolute:calibration:545GHz}{}
\setsymbol{HFI:CMB:absolute:calibration:857GHz}{}

\setsymbol{HFI:FIRAS:gain:calibration:accuracy:statistical:100GHz}{}
\setsymbol{HFI:FIRAS:gain:calibration:accuracy:statistical:143GHz}{}
\setsymbol{HFI:FIRAS:gain:calibration:accuracy:statistical:217GHz}{}
\setsymbol{HFI:FIRAS:gain:calibration:accuracy:statistical:353GHz}{2.5\%}
\setsymbol{HFI:FIRAS:gain:calibration:accuracy:statistical:545GHz}{1\%}
\setsymbol{HFI:FIRAS:gain:calibration:accuracy:statistical:857GHz}{0.5\%}

\setsymbol{HFI:FIRAS:gain:calibration:accuracy:systematic:100GHz}{}
\setsymbol{HFI:FIRAS:gain:calibration:accuracy:systematic:143GHz}{}
\setsymbol{HFI:FIRAS:gain:calibration:accuracy:systematic:217GHz}{}
\setsymbol{HFI:FIRAS:gain:calibration:accuracy:systematic:353GHz}{}
\setsymbol{HFI:FIRAS:gain:calibration:accuracy:systematic:545GHz}{7\%}
\setsymbol{HFI:FIRAS:gain:calibration:accuracy:systematic:857GHz}{7\%}

\setsymbol{HFI:FIRAS:zero:point:accuracy:100GHz:units}{0.8\MJysr}
\setsymbol{HFI:FIRAS:zero:point:accuracy:143GHz:units}{}
\setsymbol{HFI:FIRAS:zero:point:accuracy:217GHz:units}{}
\setsymbol{HFI:FIRAS:zero:point:accuracy:353GHz:units}{1.4\MJysr}
\setsymbol{HFI:FIRAS:zero:point:accuracy:545GHz:units}{2.2\MJysr}
\setsymbol{HFI:FIRAS:zero:point:accuracy:857GHz:units}{1.7\MJysr}

\setsymbol{HFI:FIRAS:zero:point:accuracy:100GHz}{0.8}
\setsymbol{HFI:FIRAS:zero:point:accuracy:143GHz}{}
\setsymbol{HFI:FIRAS:zero:point:accuracy:217GHz}{}
\setsymbol{HFI:FIRAS:zero:point:accuracy:353GHz}{1.4}
\setsymbol{HFI:FIRAS:zero:point:accuracy:545GHz}{2.2}
\setsymbol{HFI:FIRAS:zero:point:accuracy:857GHz}{1.7}

\setsymbol{HFI:unit:conversion:100GHz:units}{0.2415\MJysrmK}
\setsymbol{HFI:unit:conversion:143GHz:units}{0.3694\MJysrmK}
\setsymbol{HFI:unit:conversion:217GHz:units}{0.4811\MJysrmK}
\setsymbol{HFI:unit:conversion:353GHz:units}{0.2883\MJysrmK}
\setsymbol{HFI:unit:conversion:545GHz:units}{0.05826\MJysrmK}
\setsymbol{HFI:unit:conversion:857GHz:units}{0.002238\MJysrmK}

\setsymbol{HFI:unit:conversion:100GHz}{0.2415}
\setsymbol{HFI:unit:conversion:143GHz}{0.3694}
\setsymbol{HFI:unit:conversion:217GHz}{0.4811}
\setsymbol{HFI:unit:conversion:353GHz}{0.2883}
\setsymbol{HFI:unit:conversion:545GHz}{0.05826}
\setsymbol{HFI:unit:conversion:857GHz}{0.002238}

\setsymbol{HFI:colour:correction:alpha=-2:V1.01:100GHz}{0.9893}
\setsymbol{HFI:colour:correction:alpha=-2:V1.01:143GHz}{0.9759}
\setsymbol{HFI:colour:correction:alpha=-2:V1.01:217GHz}{1.0007}
\setsymbol{HFI:colour:correction:alpha=-2:V1.01:353GHz}{1.0028}
\setsymbol{HFI:colour:correction:alpha=-2:V1.01:545GHz}{1.0019}
\setsymbol{HFI:colour:correction:alpha=-2:V1.01:857GHz}{0.9889}

\setsymbol{HFI:colour:correction:alpha=0:V1.01:100GHz}{1.0008}
\setsymbol{HFI:colour:correction:alpha=0:V1.01:143GHz}{1.0148}
\setsymbol{HFI:colour:correction:alpha=0:V1.01:217GHz}{0.9909}
\setsymbol{HFI:colour:correction:alpha=0:V1.01:353GHz}{0.9888}
\setsymbol{HFI:colour:correction:alpha=0:V1.01:545GHz}{0.9878}
\setsymbol{HFI:colour:correction:alpha=0:V1.01:857GHz}{1.0014}

\providecommand{\sorthelp}[1]{}

\graphicspath{{./Figures/}{../Figures/}}

\def\reff@jnl#1{{#1\/}}
\def\apj{\reff@jnl{ApJ}}       
\def\apjs{\reff@jnl{ApJS}}     
\def\aaps{\reff@jnl{A\&AS}}    
\def\mnras{\reff@jnl{MNRAS}}   
\def\prd{\reff@jnl{Phys.\ Rev.\ D}}    

\newcommand{\Nside}{\ensuremath{N_{\mathrm{side}}}} 

\newcommand{\hi}{{\sc Hi}}
{\begin{enumerate}\setlength{\itemsep}{0mm}}%
{\end{enumerate}}
{\begin{enumerate}\setlength{\itemsep}{0mm}}%
{\end{enumerate}}

\newcommand{\bc}{\begin{center}}
\newcommand{\ec}{\end{center}}
\newcommand{\bi}{\begin{itemize}}
\newcommand{\ei}{\end{itemize}}
\newcommand{\ben}{\begin{enumerate}}
\newcommand{\een}{\end{enumerate}}

\newfont{\gwpfont}{cmssq8 scaled 1000}

\def\GHz{\ifmmode $GHz$\else \,GHz\fi}
\def\MHz{\ifmmode $MHz$\else \,MHz\fi}
\def\Hz{\ifmmode $Hz$\else \,Hz\fi}

\def\microm{\ifmmode \,\mu$m$\else \,$\mu$\hbox{m}\fi}
\newcommand{\healpix}{\texttt{HEALPix}}
\newcommand{\polkapix}{\texttt{polkapix}}
\newcommand{\bogopix}{\texttt{bogopix}}

\newcommand{\zodi}{\texttt{zodi}}
\newcommand{\WMAP}{\textit{WMAP\/}}
\newcommand{\COBE}{\textit{COBE\/}}

\begin{document}
\title{\Planck\ 2013 results. VIII. HFI photometric calibration and mapmaking}
 
\author{
\author{\small
Planck Collaboration:
P.~A.~R.~Ade\inst{84}
\and
N.~Aghanim\inst{58}
\and
C.~Armitage-Caplan\inst{89}
\and
M.~Arnaud\inst{71}
\and
M.~Ashdown\inst{67, 6}
\and
F.~Atrio-Barandela\inst{18}
\and
J.~Aumont\inst{58}
\and
C.~Baccigalupi\inst{83}
\and
A.~J.~Banday\inst{92, 10}
\and
R.~B.~Barreiro\inst{64}
% ...
}
\institute{\small
APC, AstroParticule et Cosmologie, Universit\'{e} Paris Diderot, CNRS/IN2P3, CEA/lrfu, Observatoire de Paris, Sorbonne Paris Cit\'{e}, 10, rue Alice Domon et L\'{e}onie Duquet, 75205 Paris Cedex 13, France\\
\and
Aalto University Mets\"{a}hovi Radio Observatory and Dept of Radio Science and Engineering, P.O. Box 13000, FI-00076 AALTO, Finland\\
\and
African Institute for Mathematical Sciences, 6-8 Melrose Road, Muizenberg, Cape Town, South Africa\\
\and
Agenzia Spaziale Italiana Science Data Center, Via del Politecnico snc, 00133, Roma, Italy\\
\and
Agenzia Spaziale Italiana, Viale Liegi 26, Roma, Italy\\
\and
Astrophysics Group, Cavendish Laboratory, University of Cambridge, J J Thomson Avenue, Cambridge CB3 0HE, U.K.\\
\and
Astrophysics \& Cosmology Research Unit, School of Mathematics, Statistics \& Computer Science, University of KwaZulu-Natal, Westville Campus, Private Bag X54001, Durban 4000, South Africa\\
\and
Atacama Large Millimeter/submillimeter Array, ALMA Santiago Central Offices, Alonso de Cordova 3107, Vitacura, Casilla 763 0355, Santiago, Chile\\
\and
CITA, University of Toronto, 60 St. George St., Toronto, ON M5S 3H8, Canada\\
\and
CNRS, IRAP, 9 Av. colonel Roche, BP 44346, F-31028 Toulouse cedex 4, France\\
\and
California Institute of Technology, Pasadena, California, U.S.A.\\
\and
Centre for Theoretical Cosmology, DAMTP, University of Cambridge, Wilberforce Road, Cambridge CB3 0WA, U.K.\\
\and
Centro de Estudios de F\'{i}sica del Cosmos de Arag\'{o}n (CEFCA), Plaza San Juan, 1, planta 2, E-44001, Teruel, Spain\\
\and
Computational Cosmology Center, Lawrence Berkeley National Laboratory, Berkeley, California, U.S.A.\\
\and
DSM/Irfu/SPP, CEA-Saclay, F-91191 Gif-sur-Yvette Cedex, France\\
\and
DTU Space, National Space Institute, Technical University of Denmark, Elektrovej 327, DK-2800 Kgs. Lyngby, Denmark\\
\and
D\'{e}partement de Physique Th\'{e}orique, Universit\'{e} de Gen\`{e}ve, 24, Quai E. Ansermet,1211 Gen\`{e}ve 4, Switzerland\\
\and
Departamento de F\'{\i}sica Fundamental, Facultad de Ciencias, Universidad de Salamanca, 37008 Salamanca, Spain\\
\and
Department of Astronomy and Astrophysics, University of Toronto, 50 Saint George Street, Toronto, Ontario, Canada\\
\and
Department of Astrophysics/IMAPP, Radboud University Nijmegen, P.O. Box 9010, 6500 GL Nijmegen, The Netherlands\\
\and
Department of Electrical Engineering and Computer Sciences, University of California, Berkeley, California, U.S.A.\\
\and
Department of Physics \& Astronomy, University of British Columbia, 6224 Agricultural Road, Vancouver, British Columbia, Canada\\
\and
Department of Physics and Astronomy, Dana and David Dornsife College of Letter, Arts and Sciences, University of Southern California, Los Angeles, CA 90089, U.S.A.\\
\and
Department of Physics and Astronomy, University College London, London WC1E 6BT, U.K.\\
\and
Department of Physics, Florida State University, Keen Physics Building, 77 Chieftan Way, Tallahassee, Florida, U.S.A.\\
\and
Department of Physics, Gustaf H\"{a}llstr\"{o}min katu 2a, University of Helsinki, Helsinki, Finland\\
\and
Department of Physics, Princeton University, Princeton, New Jersey, U.S.A.\\
\and
Department of Physics, University of California, One Shields Avenue, Davis, California, U.S.A.\\
\and
Department of Physics, University of California, Santa Barbara, California, U.S.A.\\
\and
Department of Physics, University of Illinois at Urbana-Champaign, 1110 West Green Street, Urbana, Illinois, U.S.A.\\
\and
Dipartimento di Fisica e Astronomia G. Galilei, Universit\`{a} degli Studi di Padova, via Marzolo 8, 35131 Padova, Italy\\
\and
Dipartimento di Fisica e Scienze della Terra, Universit\`{a} di Ferrara, Via Saragat 1, 44122 Ferrara, Italy\\
\and
Dipartimento di Fisica, Universit\`{a} La Sapienza, P. le A. Moro 2, Roma, Italy\\
\and
Dipartimento di Fisica, Universit\`{a} degli Studi di Milano, Via Celoria, 16, Milano, Italy\\
\and
Dipartimento di Fisica, Universit\`{a} degli Studi di Trieste, via A. Valerio 2, Trieste, Italy\\
\and
Dipartimento di Fisica, Universit\`{a} di Roma Tor Vergata, Via della Ricerca Scientifica, 1, Roma, Italy\\
\and
Discovery Center, Niels Bohr Institute, Blegdamsvej 17, Copenhagen, Denmark\\
\and
European Southern Observatory, ESO Vitacura, Alonso de Cordova 3107, Vitacura, Casilla 19001, Santiago, Chile\\
\and
European Space Agency, ESAC, Planck Science Office, Camino bajo del Castillo, s/n, Urbanizaci\'{o}n Villafranca del Castillo, Villanueva de la Ca\~{n}ada, Madrid, Spain\\
\and
European Space Agency, ESTEC, Keplerlaan 1, 2201 AZ Noordwijk, The Netherlands\\
\and
Haverford College Astronomy Department, 370 Lancaster Avenue, Haverford, Pennsylvania, U.S.A.\\
\and
Helsinki Institute of Physics, Gustaf H\"{a}llstr\"{o}min katu 2, University of Helsinki, Helsinki, Finland\\
\and
INAF - Osservatorio Astrofisico di Catania, Via S. Sofia 78, Catania, Italy\\
\and
INAF - Osservatorio Astronomico di Padova, Vicolo dell'Osservatorio 5, Padova, Italy\\
\and
INAF - Osservatorio Astronomico di Roma, via di Frascati 33, Monte Porzio Catone, Italy\\
\and
INAF - Osservatorio Astronomico di Trieste, Via G.B. Tiepolo 11, Trieste, Italy\\
\and
INAF Istituto di Radioastronomia, Via P. Gobetti 101, 40129 Bologna, Italy\\
\and
INAF/IASF Bologna, Via Gobetti 101, Bologna, Italy\\
\and
INAF/IASF Milano, Via E. Bassini 15, Milano, Italy\\
\and
INFN, Sezione di Bologna, Via Irnerio 46, I-40126, Bologna, Italy\\
\and
INFN, Sezione di Roma 1, Universit\`{a} di Roma Sapienza, Piazzale Aldo Moro 2, 00185, Roma, Italy\\
\and
IPAG: Institut de Plan\'{e}tologie et d'Astrophysique de Grenoble, Universit\'{e} Joseph Fourier, Grenoble 1 / CNRS-INSU, UMR 5274, Grenoble, F-38041, France\\
\and
IUCAA, Post Bag 4, Ganeshkhind, Pune University Campus, Pune 411 007, India\\
\and
Imperial College London, Astrophysics group, Blackett Laboratory, Prince Consort Road, London, SW7 2AZ, U.K.\\
\and
Infrared Processing and Analysis Center, California Institute of Technology, Pasadena, CA 91125, U.S.A.\\
\and
Institut N\'{e}el, CNRS, Universit\'{e} Joseph Fourier Grenoble I, 25 rue des Martyrs, Grenoble, France\\
\and
Institut Universitaire de France, 103, bd Saint-Michel, 75005, Paris, France\\
\and
Institut d'Astrophysique Spatiale, CNRS (UMR8617) Universit\'{e} Paris-Sud 11, B\^{a}timent 121, Orsay, France\\
\and
Institut d'Astrophysique de Paris, CNRS (UMR7095), 98 bis Boulevard Arago, F-75014, Paris, France\\
\and
Institute for Space Sciences, Bucharest-Magurale, Romania\\
\and
Institute of Astronomy and Astrophysics, Academia Sinica, Taipei, Taiwan\\
\and
Institute of Astronomy, University of Cambridge, Madingley Road, Cambridge CB3 0HA, U.K.\\
\and
Institute of Theoretical Astrophysics, University of Oslo, Blindern, Oslo, Norway\\
\and
Instituto de F\'{\i}sica de Cantabria (CSIC-Universidad de Cantabria), Avda. de los Castros s/n, Santander, Spain\\
\and
Jet Propulsion Laboratory, California Institute of Technology, 4800 Oak Grove Drive, Pasadena, California, U.S.A.\\
\and
Jodrell Bank Centre for Astrophysics, Alan Turing Building, School of Physics and Astronomy, The University of Manchester, Oxford Road, Manchester, M13 9PL, U.K.\\
\and
Kavli Institute for Cosmology Cambridge, Madingley Road, Cambridge, CB3 0HA, U.K.\\
\and
LAL, Universit\'{e} Paris-Sud, CNRS/IN2P3, Orsay, France\\
\and
LERMA, CNRS, Observatoire de Paris, 61 Avenue de l'Observatoire, Paris, France\\
\and
LESIA, Observatoire de Paris, CNRS, UPMC, Universit\'{e} Paris-Diderot, 5 Place J. Janssen, 92195 Meudon, France\\
\and
Laboratoire AIM, IRFU/Service d'Astrophysique - CEA/DSM - CNRS - Universit\'{e} Paris Diderot, B\^{a}t. 709, CEA-Saclay, F-91191 Gif-sur-Yvette Cedex, France\\
\and
Laboratoire Traitement et Communication de l'Information, CNRS (UMR 5141) and T\'{e}l\'{e}com ParisTech, 46 rue Barrault F-75634 Paris Cedex 13, France\\
\and
Laboratoire de Physique Subatomique et de Cosmologie, Universit\'{e} Joseph Fourier Grenoble I, CNRS/IN2P3, Institut National Polytechnique de Grenoble, 53 rue des Martyrs, 38026 Grenoble cedex, France\\
\and
Laboratoire de Physique Th\'{e}orique, Universit\'{e} Paris-Sud 11 \& CNRS, B\^{a}timent 210, 91405 Orsay, France\\
\and
Lawrence Berkeley National Laboratory, Berkeley, California, U.S.A.\\
\and
Max-Planck-Institut f\"{u}r Astrophysik, Karl-Schwarzschild-Str. 1, 85741 Garching, Germany\\
\and
McGill Physics, Ernest Rutherford Physics Building, McGill University, 3600 rue University, Montr\'{e}al, QC, H3A 2T8, Canada\\
\and
National University of Ireland, Department of Experimental Physics, Maynooth, Co. Kildare, Ireland\\
\and
Niels Bohr Institute, Blegdamsvej 17, Copenhagen, Denmark\\
\and
Observational Cosmology, Mail Stop 367-17, California Institute of Technology, Pasadena, CA, 91125, U.S.A.\\
\and
Optical Science Laboratory, University College London, Gower Street, London, U.K.\\
\and
SB-ITP-LPPC, EPFL, CH-1015, Lausanne, Switzerland\\
\and
SISSA, Astrophysics Sector, via Bonomea 265, 34136, Trieste, Italy\\
\and
School of Physics and Astronomy, Cardiff University, Queens Buildings, The Parade, Cardiff, CF24 3AA, U.K.\\
\and
Space Research Institute (IKI), Russian Academy of Sciences, Profsoyuznaya Str, 84/32, Moscow, 117997, Russia\\
\and
Space Sciences Laboratory, University of California, Berkeley, California, U.S.A.\\
\and
Special Astrophysical Observatory, Russian Academy of Sciences, Nizhnij Arkhyz, Zelenchukskiy region, Karachai-Cherkessian Republic, 369167, Russia\\
\and
Stanford University, Dept of Physics, Varian Physics Bldg, 382 Via Pueblo Mall, Stanford, California, U.S.A.\\
\and
Sub-Department of Astrophysics, University of Oxford, Keble Road, Oxford OX1 3RH, U.K.\\
\and
Theory Division, PH-TH, CERN, CH-1211, Geneva 23, Switzerland\\
\and
UPMC Univ Paris 06, UMR7095, 98 bis Boulevard Arago, F-75014, Paris, France\\
\and
Universit\'{e} de Toulouse, UPS-OMP, IRAP, F-31028 Toulouse cedex 4, France\\
% ...
}
}

\date{Received XX, 2013; accepted XX, 2023}

\abstract {This paper describes the methods used to produce photometrically calibrated maps from  the \Planck\
  High Frequency Instrument (HFI) cleaned, time-ordered information.  HFI observes the
  sky over a broad range of frequencies, from 100 to 857\,\GHz. To obtain the best calibration accuracy over such a large range, 
  two different photometric calibration schemes have to be used. The 545 and 857\,GHz\ data are calibrated by comparing flux-density measurements of Uranus and Neptune with models of their atmospheric emission.
 The lower frequencies (below 353\,\GHz) are calibrated using the Solar dipole.
A component of this  anisotropy is time-variable, owing to the orbital
motion of the satellite in the Solar System. Photometric
  calibration is thus tightly linked to mapmaking, which also addresses low-frequency 
noise removal. 
By comparing observations taken more than one year apart in the same configuration, we have identified apparent gain variations with time. {  These variations are induced by non-linearities in the read-out electronics chain}.  
We have developed an effective correction to limit their effect on calibration.
We present several methods to estimate the precision of the photometric calibration. We distinguish  relative uncertainties (between detectors, or between frequencies) and absolute uncertainties. Both these uncertainties {  lie in the  range from 0.3\%  to 10\% } from 100 to 857\,\GHz. We describe the pipeline used to produce the maps from the HFI timelines, based on the photometric calibration parameters, and  the scheme used to set the zero level of the maps a posteriori. We also discuss the cross-calibration between HFI and the SPIRE instrument on board {\it Herschel}. 
 Finally we summarize the basic characteristics of the set of  HFI maps included in the 2013 \Planck\ data release. 
}
   \keywords{Cosmology: observations -- Cosmic background radiation --
   Surveys -- methods: data analysis}

\authorrunning{Planck Collaboration}
\titlerunning{\Planck-HFI calibration and mapmaking}

\maketitle

\clearpage
\section{Introduction}
This paper, one of a set associated with the 2013 release of data from the \Planck\ mission\footnote{\Planck\ (\url{http://www.esa.int/Planck}) is a project of the European Space
Agency (ESA) with instruments provided by two scientific consortia funded by ESA member
states (in particular the lead countries France and Italy), with contributions from NASA
USA) and telescope reflectors provided by a collaboration between ESA and a scientific
consortium led and funded by Denmark.}~\respaps, describes the processing applied to  \Planck\  High Frequency Instrument (HFI) cleaned time-ordered information (TOI) to produce photometrically-calibrated sky maps.

CMB experiments can be calibrated using the dipole anisotropy induced by the motion of the instrument relative to the cosmological frame. This anisotropy
is naturally separated into two components: we refer to the component generated by the motion of Planck around the sun as the \emph{orbital dipole}, and that
generated by the sun's motion relative to the CMB as the \emph{solar dipole}. 

 In principle, the orbital dipole is the most precise calibrator,
as it depends on the very well known orbital parameters and the temperature of the CMB, measured  precisely by 
the \COBE-FIRAS experiment \citep{mather1999_firas}.  However, calibration using the orbital dipole involves comparison of data taken at large time separation (typically 6 months),
and the precision one can achieve using this calibrator is thus directly linked to that of the time stability of the data, and to the precision reached in addressing any time variable systematics. 
We have identified one such systematic, induced by non-linearities in the analogue-to-digital converters of the bolometers' read-out electronic chain, and for the present release have chosen
to use the solar dipole, based on the measurement of the solar dipole parameters from \WMAP ~\citep{hinshaw2009}, as the main calibrator for the 100 to 353\,\GHz\ channels.
These parameters are summarized in Table~\ref{tab:wmap_dipole}. 

At high frequency ($\nu\,\ge$500\,\GHz), the dipole becomes too faint with respect to the Galactic foregrounds to give an accurate calibration. 
Although we used the Galactic emission as measured by FIRAS for the calibration of the \Planck\ early papers~\citep{planck2011-1.5}, 
we have now obtained  a better accuracy using planet measurements. Thus, the absolute calibration of the two high-frequency
channels is done using Uranus and Neptune. \\

At all frequencies, the zero levels of the maps are obtained by assuming
no Galactic emission at zero gas column density, and adding the Cosmic
Infrared Background (CIB) mean level. \\

The paper is organized as follows. 
We first summarize the mapmaking procedure (Sect.~\ref{sec:mapmaking}).
We outline the calibration method used for the CMB-dominated channels (100 to 353\,\GHz) in Sect.~\ref{sect_LF_calib}. We discuss in this section unexpected response variations with time, and present an effective correction.  We then 
detail the calibration for the 545 and 857~\GHz\ channels (Sect.~\ref{sect_HF_calib}) and describe how the zero level of the maps can be fixed (Sect.~\ref{sec:zero_levels}). We finally quantify the accuracy of the photometric 
calibration, and give basic characteristics of the delivered maps in Sect.~\ref{calib_accuracy}. Conclusion are given in Sect.~\ref{sec:conclusion}.

\begin{table}
\caption{Parameters of the solar dipole, as measured by \WMAP\ \citep{hinshaw2009}}
\begin{center}
\begin{tabular}{ @{} l r@{}l  @{} }
\hline
\hline

\noalign{\vskip 2pt}

Amplitude  [mK$_{\mathrm{CMB}}]$ & $3.355$&$\pm0.008 $   \\
Galactic longitude  [\deg] &   $263.99$&$\pm0.14 $ \\
Galactic latitude  [\deg] &  $41.74$&$\pm0.03$ \\
\hline
\end{tabular}
\end{center}
\label{tab:wmap_dipole}
\end{table}

\section{Pipeline for map production}
\label{sec:mapmaking}

The products of the HFI mapmaking pipeline are maps of $I$, $Q$ and $U$, together with their covariances, pixelized according to the \healpix\ scheme~\citep{Gorski05} with a resolution parameter $\Nside=2048$. For a given channel, data sample $i$ may be described as  
\begin{equation}
 d_i \ = G\left( I_p + \frac{1-\eta}{1+\eta}\left( Q_p\ \cos{2\psi_i}
 + U_p\ \sin{2\psi_i} \right)\right)   + \ n_i  ,
 \label{eq:MM1}
\end{equation}
where $p$ denotes the sky pixel with Stokes parameters $I_p$, $Q_p$ and $U_p$, $n_i$
is the noise realization, $\eta$ is the cross-polarization parameter
(equal to 1 for an ideal spider-web bolometer  and 0 for an ideal polarization sensitive bolometer), $\psi_i$ is the
detector orientation on the sky, at sample $i$, and $G$ is the detector's gain.  Given \Planck's scanning strategy, reconstructing $I$, $Q$ and $U$ requires combining measurements from several detectors for most pixels. According to bolometer models, and given the stability of the HFI operational conditions during the mission, $G$ is not expected to vary significantly, 

In order to deal efficiently with the large HFI data set and the large number of maps to be produced, we use a two-step scheme to make maps from the HFI TOIs. The first step takes advantage of the redundancy of the observations on the sky. For each detector, we average the measurements in each \healpix\ pixel visited during a stable pointing period (hereafter called {\em ring}), into
an intermediate product, called an HPR for \healpix\ Pixels Ring.
Subsequent calibration and mapmaking operations use the HPR
as input. As we produce \healpix\
maps with the resolution parameter  \Nside\ set to 2048 we use the same internal resolution for building the HPR.

The in-flight noise of the HFI detectors, after TOI processing, is
mostly white at high frequency, with a ``$1/f$'' increase at low
frequency~\citep{planck2011-1.5}. In such a case, a destriping
approach is well suited for the mapmaking~\citep{ashdown2009}. In this approach, the noise in a ring $r$ is represented by an offset, denoted  by $\vec{o}_r$, and a white noise part $\vec{n}$,
which is uncorrelated with the low-frequency noise.
We may then reformulate Eq.~\ref{eq:MM1} as 
\begin{equation}
\vec{d}_i 
= 
G \times \tens{A}_{ip}\cdot\tens{T}_p + \tens{\Gamma}_{ir}\cdot\vec{o}_r + \vec{n}_i, 
\label{eq:MM2}
\end{equation}
where $\tens{T}$ represents the sky (which may be a 3-vector if polarization is accounted for) in pixel $p$, $\tens{A}$ is the  pointing matrix (which makes the link between data samples and  their positions on the sky) and $\tens{\Gamma}$ is the matrix folding the ring onto samples. From the above equation, $\vec{o}_r$ are derived through maximum likelihood. As there is a degeneracy between the average of the offsets and the zero level of the maps, we impose the constraint  $\langle  \vec{o} \rangle\ = \ 0$. \cite{tristram2011} have shown that with scanning and noise like those of HFI, an accurate reconstruction of the offsets $\vec{o}_r$ requires a precise measurement of $G$  for each channel. 

In addition, some signal components vary with time, adding more complexity to Eq.~\ref{eq:MM2}. Such components include the zodiacal light emission,  the CMB dipole anisotropy component induced by the motion of the satellite with respect to the Solar System, and the far sidelobe (FSL) pick-up signal. Time variability of the former comes from the variation of the observation angle of the Solar System region emitting this radiation, due to the ellipticity and cycloid modulation of the satellite's orbit. The FSL are discussed in \cite{planck2013-p03c} and \cite{planck2013-pip88}.
Accounting for these components in the mapmaking process requires an accurate calibration. Moreover, we need to take into account the low-frequency noise in the calibration process, so both operations (mapmaking and calibration) are interleaved.

For the production of the maps of the 2013 HFI data  release, we followed a four-step process.
\begin{enumerate}
\item We first build the HPR for all detectors, for three data sets: all the data for each ring,  and (for null tests) the data from just the first or just the second  half of each ring.
\item We then apply the following calibration operations to the HPR:
\begin{itemize}
\item solar dipole calibration, which sets the overall calibration factors for the 100--353\GHz\ detectors,
\item planet calibration (Uranus and Neptune), which is used to get the calibration factors for the 545--857\GHz\ detectors,
\item determine the relative gain variations over time of the 100--217\GHz\ detectors, using the \bogopix\ tool (see
  Sect.~\ref{subsec:bogo_gain_variation}). 
\end{itemize}
\item For each data set we then do the destriping and projections, using the \polkapix\ tool that was thoroughly validated in \cite{tristram2011}. We compute one set of offsets using the whole mission data set, and then use these offsets to compute the maps for the whole mission, as well as for restricted time intervals (corresponding to each individual survey, and to the nominal mission). Maps are built by simple co-addition in each pixel of the destriped, calibrated, and time varying component-subtracted signal. We subtract the \WMAP\ measured CMB dipole from all our maps, using the non-relativistic approximation. 
\item The zero-levels for the maps are set a posteriori. 
\end{enumerate}
We have produced single-detector temperature maps, as well as temperature and polarization maps using all the detectors of a single frequency and some detector subsets.
We have also produced hit-count maps  and variance maps for the $I$, $Q$ and $U$ values computed in each pixel. 
Overall, a  total of about $6500$ sky maps have been produced. We  used this data set to evaluate the performance of the photometric calibration.  Note that the HFI pipeline we have described is quite similar to that used for the Low Frequency Instrument (LFI)~\citep{planck2013-p02}.

In order to take into account the  Galactic signal integrated in the FSL and zodiacal light (hereafter called \zodi) components, which vary in time, we have constructed templates for the combination of these components at frequencies where Galactic emission in the FSL matters, i.e., 545 and 857\,\GHz, and of \zodi\ only at lower frequencies, as described in \cite{planck2013-pip88}. These templates are used to build HPRs. 
We provide two sets of maps. The first set is built without removing these spurious components, while the second set is the differences 
between maps from the previous set and maps from which the \zodi\ and FSL have been removed. The  difference maps might be can be used to correct the HFI maps for specific applications. 

In the following sections we will describe the calibration procedures and then assess their performance, and present some characteristics of the resulting maps.


\section{Photometric calibration of the low-frequency channels: dipole-based calibration \label{sect_LF_calib}}
\subsection{ADC non-linearities and calibration}
\label{sec:gain_variation}
{  With a larger data set than that analyzed in \cite{planck2011-1.7}, we could ideally use an orbital-dipole-based calibration, as described in \cite{tristram2011}. However, the additional redundancies revealed new systematic effects, ADC non-linearities and very long time constants (of the order of a few seconds) with very low energy content in the system's response. The former 
induce apparent gain variations with time. The latter shifts the CMB dipole a few arcmin in the scan direction, and hence 
creates leaks from the Solar dipole into the orbital dipole signal. These systematic effects prevented us from using the orbital dipole calibration. The very long time constants were identified after correcting for  the ADC non-linearities,  and have not yet been fully characterized yet. Both corrections will be implemented in the Planck 2014 data release.}

\begin{figure}[htbp]
\centering

\includegraphics[width=0.465\textwidth,bb=280 131 790 410,clip]{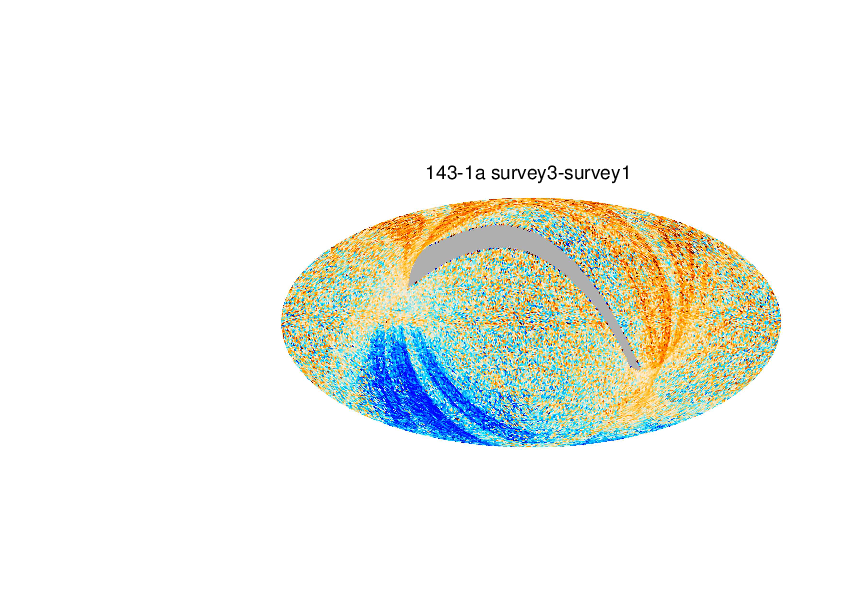}\\ 
\includegraphics[width=0.465\textwidth,bb=280 131 790 410,clip]{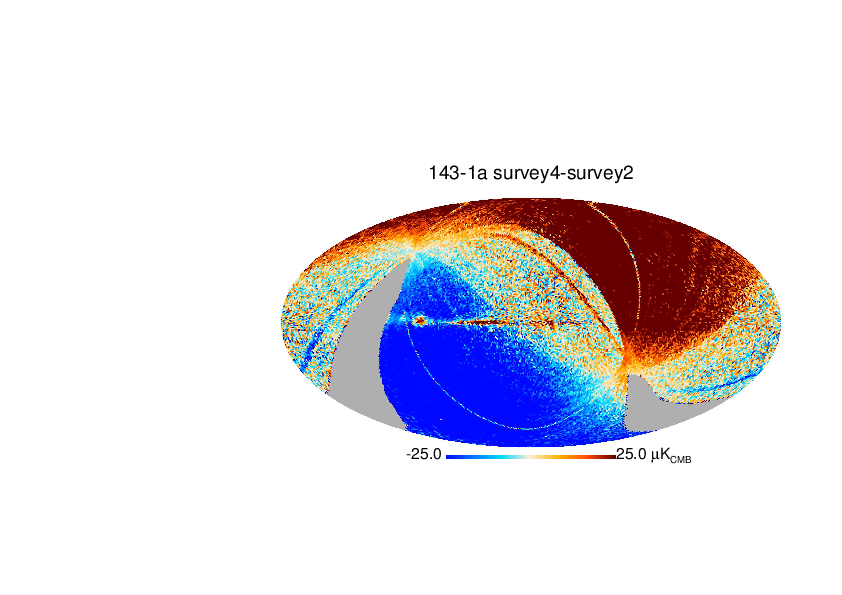} 
\caption{Differences between temperature maps built using data from
  detector 143-1a, for Surveys~1 and 3 (top) and 2 and 4 (bottom). In
  both cases, large-scale features appear. Their amplitude and
  disposition on the sky are compatible with residuals from the solar
  dipole, due to time variations of the detector gain, of the order of
  1 to 2\,\% .
  These residuals should be compared to the
  amplitudes of the solar dipole, 3.353~mK$_{\mathrm{CMB}}$ {  the orbital dipole, about 10 times lower.} }
\label{fig:143-1a_surv_diffs_raw}
\end{figure}

{  Effects of} such {  ADC-induced gain} variations are clearly visible when comparing Survey~3 with Survey~1 or Survey~2 with Survey~4. 
{  As an example, in Fig.~\ref{fig:143-1a_surv_diffs_raw} we show survey difference maps for one 143\,\GHz\ detector, built  using the calibration and mapmaking scheme presented in \cite{planck2011-1.7}. 
Large-scale dipolar features, aligned with the solar dipole,  are prominent in these maps. This shows that the constant gain assumption used to build these maps is incorrect.} 

Intrinsic bolometer sensitivity variations cannot explain such gain variations.  The HFI bolometers have been precisely characterized in flight using a dedicated sequence of $V(I)$ measurements, during the post-launch verification phase  and end-of-life periods. 
The static bolometer models predict that changes of their background during the observations could not explain response variations larger than 0.1\,\%. In addition, such variations are corrected for within the HFI DPC pipeline. In our present understanding,  these apparent response variations are
the result  of imperfections in the linearity of the analogue-to-digital
converters (ADC) used in the bolometer read-out units.  The variation of the bolometer background with time and the 
{  unevenness in} the ADC quantization steps leads, at first order, to an
apparent gain variation in the electronic chain. These non-linearities may also affect signals differently depending on their amplitude, for example the solar and orbital dipoles. 
%

\begin{figure}[htbp]
\centering

\includegraphics[width=.5\textwidth]{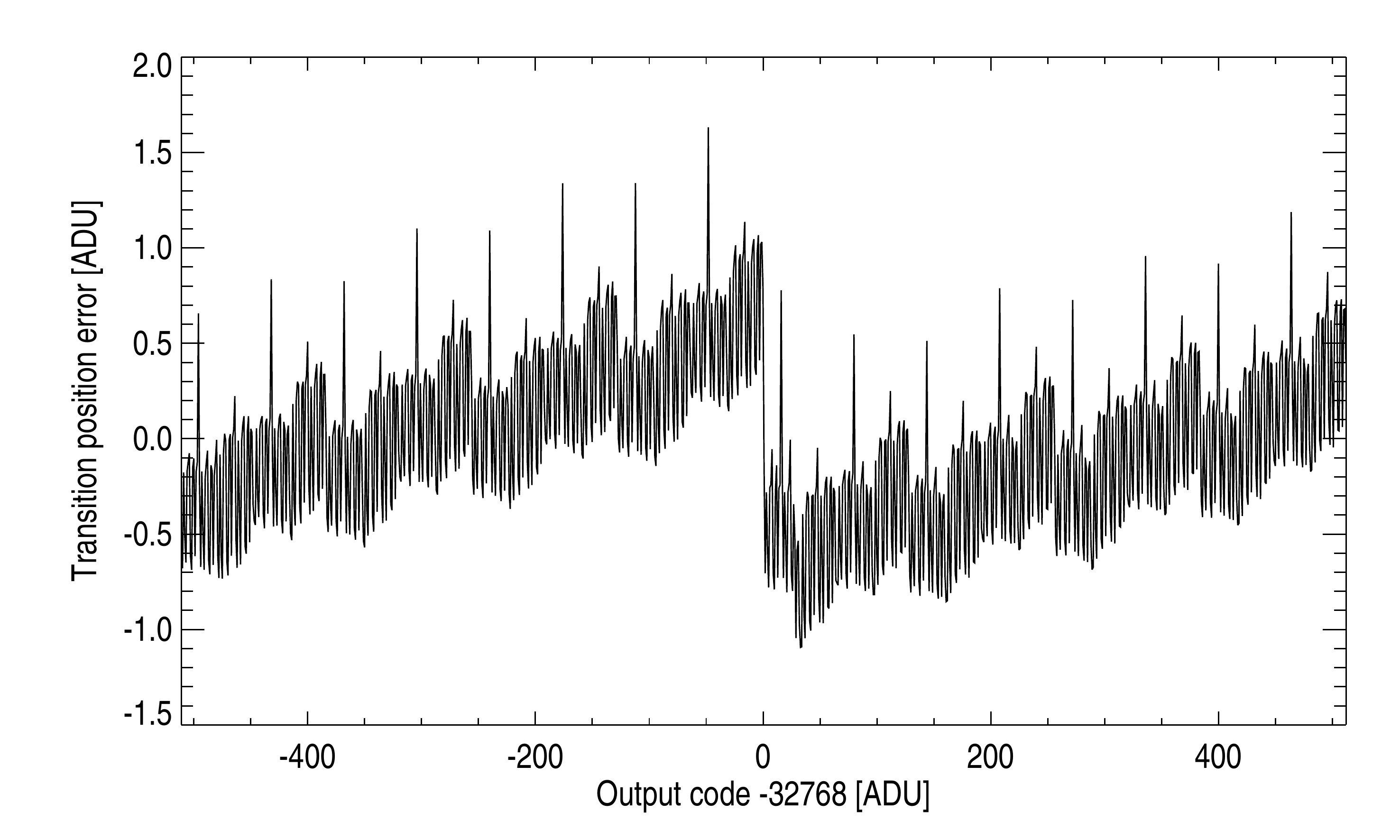}

\caption{Error on transition code positions measured on one chip around the ADC mid-scale, on the ground on a spare ADC. The largest error occurs at the sign transition, but  errors of  about 1 ADU also occur regularly every 64 {  steps}.}
\label{fig:ADC_groud_error}
\end{figure}

Figure \ref{fig:ADC_groud_error} shows the errors on the transition code positions measured on a spare ADC chip around the mid-scale, which is the most populated area. These ``integrated non-linearities'' (INL) present a prominent feature in all channels: the central step is always too narrow.  In addition to this, the 64-code, nearly periodic patterns contribute to the apparent gain variations, making it difficult to predict the consequence of such errors on the reconstructed, demodulated bolometer signal. 
Such an INL effect has however been included in full mission simulations, and it reproduces qualitatively the gain variation features observed in real flight data, {  with an amplitude of about $\pm 1\%$. This is larger than the required
calibration precision of the 100 to 217\,\GHz\ channels.}

In order to precisely correct all the data for this effect,  we need  accurate measurements of all the ADC INLs, together with a good model for the bolometer raw signal (including systematics).  Mapping the ADC response required more data than were acquired before the end of the HFI cold lifetime, so a dedicated campaign has been conducted over several months, at a focal plane temperature of about $4$\,K, to obtain a clean ADC characterization on Gaussian noise. Correcting this effect needs to be carried out prior to the TOI processing steps, and will require thorough checks of any products.  At the time of writing, correction procedures are being intensively tested {  but they have not been included in the 2013 \Planck\ data release. 

In the absence of a full correction procedure, we had to develop an effective method to address the apparent bolometer gain variations that arise from the ADC non-linearities.  
In this method, the absolute scale is fixed by the solar dipole, to ensure a better robustness against higher-order non-linearities, as described in Sect.~\ref{subs:soldip}. Relative gains are determined using 
the scanning redundancies, as explained in Sect.~\ref{sec:corr_gain}}.


\subsection{Solar dipole calibration \label{subs:soldip}}
The photometric calibration of the 100--353\,GHz bolometers is based on the CMB dipole.  

We estimate one value of the detector gain for each ring through a template fit of the HPR data. We fit {  the coefficients} of a linear combination of dipole, Galactic signal, and noise, neglecting the CMB and the polarization:
\begin{equation}
\vec{d} \ = \ g_r^{\mathrm D} . \vec{t}_{\mathrm D} + g_r^{\mathrm G} . \vec{t}_G + c_r+ \vec{n} \, .
\end{equation}
Here $\vec{d}$ represents the HPR samples from ring $r$, $\vec{t}_{\mathrm D}$ is the value of the total (Solar and orbital) kinematic dipole, $\vec{t}_{\mathrm G}$ is a model for the Galactic emission, and $\vec{n}$ is the white component of the noise. 
For simplicity, we used a non-relativistic approximation, as explained in Appendix~\ref{ssec:cmb_dip_conv}. 
We do not take into account the smearing of  the dipole by the instrumental beam in our procedure, as justified in Appendix \ref{ssec:fsl_conv}. We simultaneously  fit three parameters: 
$g_r^{\mathrm D}$, the gain of the kinematic dipole;
$g_r^{\mathrm G}$, the gain of the Galactic model; and
$c_r$, a constant accounting for the low-frequency noise.

As the satellite scans circles on the sky, the ratio of the dipole and Galactic signal amplitudes varies. We use a Galactic model to obtain a measurement of the dipole gain, even in rings where the dipole amplitude is low. However, imperfection of that model may lead to bias in the dipole gain. 
To reduce this bias, we exclude  pixels with a Galactic latitude lower than 9\deg. Because we calibrate on the kinematic dipole, we do not use the gain $g_r^{\mathrm G}$ in what follows. Pixels contaminated by point sources listed in the \Planck\ Catalogue of Compact Sources~\citep{planck2013-p05} are also excluded. The best model we have for the sky emission at the HFI frequencies being HFI measurements themselves, we use HFI sky maps {  at the detector frequency} 
as a Galactic model, as shown in Appendix~\ref{app:sky_dipcal}.

\begin{figure}[htbp]
\centering

\includegraphics[width=0.5\textwidth]{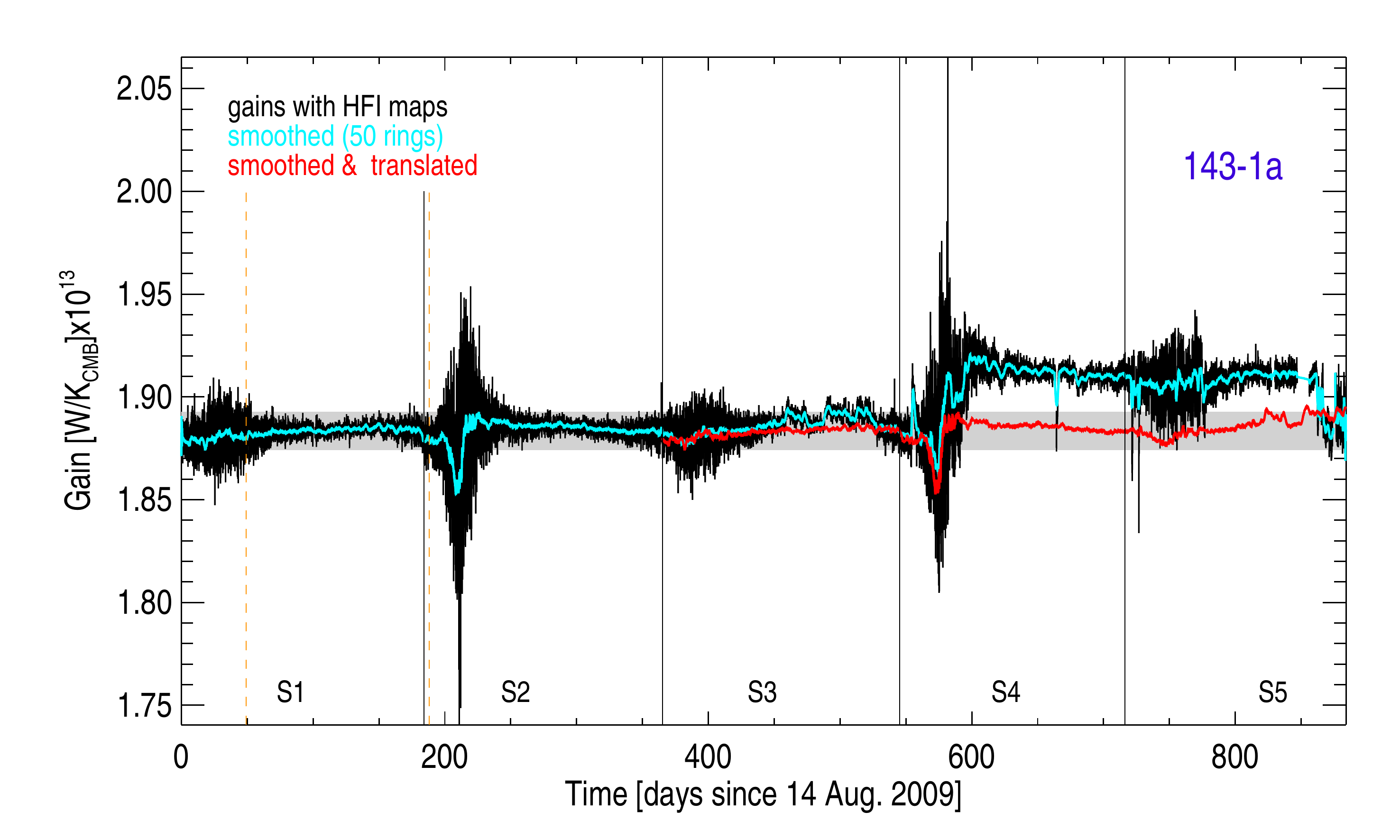}

\caption{Solar dipole gain reconstructed ring-by-ring for one HFI
  bolometer.  The thin black line represent the raw values, and the thick
  cyan line is a smoothed rendition with  a width of 50 rings (about 2 days). We have 
  indicated the conventional boundaries of the surveys as black vertical lines. The orange vertical dashed lines indicate the interval in which we compute the gain $\tilde{G}^{\mathrm{SD}}$  (computed between rings 2000 and 6000, or approximately days 60 and 190). The red curve shows  the smoothed gain variation shifted to match the repetition in Surveys 3 and 4 of the scan strategy followed in Surveys 1 and 2 (note that the scan strategy for Survey 5 differs from that of Survey 3).  The grey band highlights a $\pm 0.5\%$ excursion around the averaged gain $\tilde{G}^{\mathrm{SD}}$. The observed $\sim$1\% variations explain the large-scale residuals seen in Fig.~\ref{fig:143-1a_surv_diffs_raw}.
\label{fig:QD_var_shift}}
\end{figure}

Results of the gain estimation for each ring are shown in Fig.~\ref{fig:QD_var_shift} for one detector (143-1a) . 
We can see that the gain estimate is less accurate on some ring intervals. This is due to the \Planck\ scanning strategy:  these intervals correspond to {  epochs} when the \Planck\ spin axis is orthogonal to the dipole direction. 
We can also see the apparent ring-by-ring gain variations, of the order of $\pm 1\,\%$, explained in Sect.\ref{sec:gain_variation}. 
To show this more clearly, the figure compares the ring-by-ring variations reconstructed in Surveys 1 and 2 with those from Surveys 3 and~4.

The final gain value for each detector, hereafter denoted by $\tilde{G}^{\mathrm{SD}}$, is  defined as  the average of these estimates between rings 2000 and 6000, {  between which  the individual measurements for each ring have a dispersion of less than $1\,\%$}.


\subsection{Effective correction and characterization \label{sec:corr_gain}}
\label{subsec:bogo_gain_variation}
In order to handle time variation of the bolometer gains, we set up an effective correction tool, called \bogopix ~\citep{perdereau2006}. 
We start from  Eq.~\ref{eq:MM2}, but take explicitly into account the orbital dipole $ \vec{t}_{Do}$, which is time-variable, and also fit the gains $\vec{g}_r$  for each bolometer independently.  The problem finally reads 
\begin{equation}
\vec{d} 
= 
\vec{g}_r(\tens{A} \cdot \tens{T} + \vec{t}_{Do}) + \tens{\Gamma}\cdot\vec{o}_r + \vec{n}, 
\label{eq:MM3}
\end{equation}
where $r$ is the ring number. 
The unknowns are the offsets $\vec{o}_r$, the sky signal represented by $\tens{T}$, and the gains $\vec{g}_r$, sampled using one value per ring. 
Since the orbital dipole is an absolute calibrator, the solution for  $\vec{g}_r$ should also fix the absolute photometric calibration. 

\begin{figure}[htbp]
\centering
\includegraphics[width=0.5\textwidth]{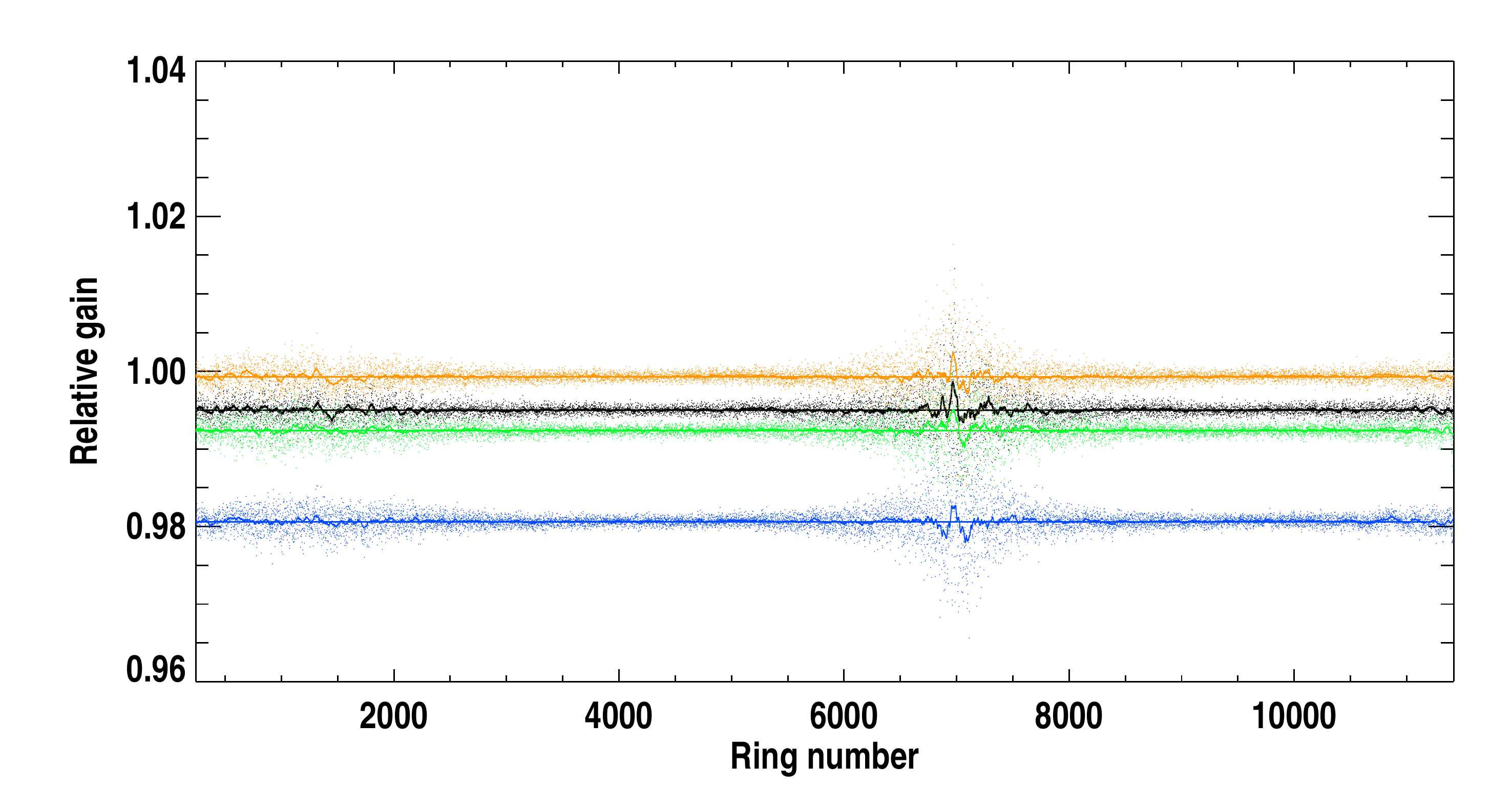}
\caption{Example of results obtained with \bogopix\ on the simulated data set used in \cite{tristram2011}, where constant gains biases were applied. The colours distinguish four different bolometers.  Dots correspond to individual measurements, and the thick line is a smoothed representation of these results with a 50 ring width. We plot relative reconstructed gains, with respect to their unbiased  value.  In this simulation, each bolometer's data was biased by factors 
of respectively 1.98 (blue) , 0.77 (green) , 0.50 (black) and 0.07 \% (orange) respectively which is precisely reflected by the recovered  \bogopix\ value.}
%
\label{fig:bogo_simu_results}
\end{figure}

We take advantage of the low amplitude of the observed gain variations to linearize this nonlinear problem, following an iterative approach. 
Starting from an approximate solution for the gains $\vec{g}_r$ and sky maps $\tens{T}$, we determine the variations with respect to these, $\vec{\delta g}_r$ and $\tens{\delta T}$, by solving : 
\begin{eqnarray}
\vec{d} =& (\vec{g}_r+\vec{\delta g}_r) (\tens{A} \cdot (\tens{T}+\tens{\delta T}) + \vec{t}_{Do})  + \tens{\Gamma}\cdot\vec{o}_r + \vec{n} \\
 \approx& \vec{g}_r (\tens{A}\cdot (\tens{T}+\tens{\delta T})+ \vec{t}_{Do}) + \vec{\delta g}_r (\tens{A}\cdot \tens{T}+ \vec{t}_{Do}) + \tens{\Gamma}\cdot\vec{o}_r + \vec{n}
\label{eq:MM3_lin}
\end{eqnarray}
The linearized Eq.~\ref{eq:MM3_lin} may then be solved for $\vec{\delta g}_r$, $\tens{\delta T}$ and $\vec{o}_r$ by a conjugate-gradient method. Using $\vec{\delta g}_r$ and $\tens{\delta T}$,
 the gains $\vec{g}_r$ and sky maps $\tens{T}$ can be
 updated. This process is iterated until a satisfactory solution is reached. 
To initialize the iterations, we start from the constant gain solution. We stop  when the relative change in the $\chi^2$  derived from  Eq.~\ref{eq:MM3} is low enough ({  in practice, when the change is less than $10^{-6}$}). 
This approach is similar to the one used for  the LFI calibration \citep{planck2013-p02b}. 
It was successfully tested using the data set of \cite{tristram2011}, derived from  simulated timelines with a \Planck-like scanning strategy, realistic noise 
(both for the white and $1/f$ components), Gaussian beams, and delta-function bandpasses, for four 143\,\GHz\ polarization-sensitive bolometers over about 12000 rings.  Figure~\ref{fig:bogo_simu_results} presents gains reconstructed  with \bogopix\ on simulated data, and compares them with the constant input gain values. From these results, we see that 
the precision of the gain value reconstructed for a single ring is
about $0.5\,\%$ (which is comparable with the global  precision of $5\times 10^{-5} $ for a constant gain for 12000 rings found in \citealt{tristram2011}).
We  computed the gain variations using single-detector data, thus neglecting polarization. 
{  As in destriping~\citep{tristram2011}, gradients within the sky pixels used for $\tens{T}$ will limit the accuracy of the gain determination. These gradients increase with frequency. Moreover, the ADC non-linearity will induce biases in the signal used for the gain determination. As this signal's dynamic range increases with frequency, we expect this bias also to increase with frequency. 
For these reasons, we used \bogopix\ to determine an effective correction for the apparent gain variations only for frequencies $\le217$\,\GHz. To avoid the central part of the Galactic plane and point sources, 
we used the  mask used for destriping  in the \Planck\ Early Results paper \citep[Figure~32]{planck2011-1.7}. }

\begin{figure}[htbp]
\centering

\includegraphics[width=0.5\textwidth]{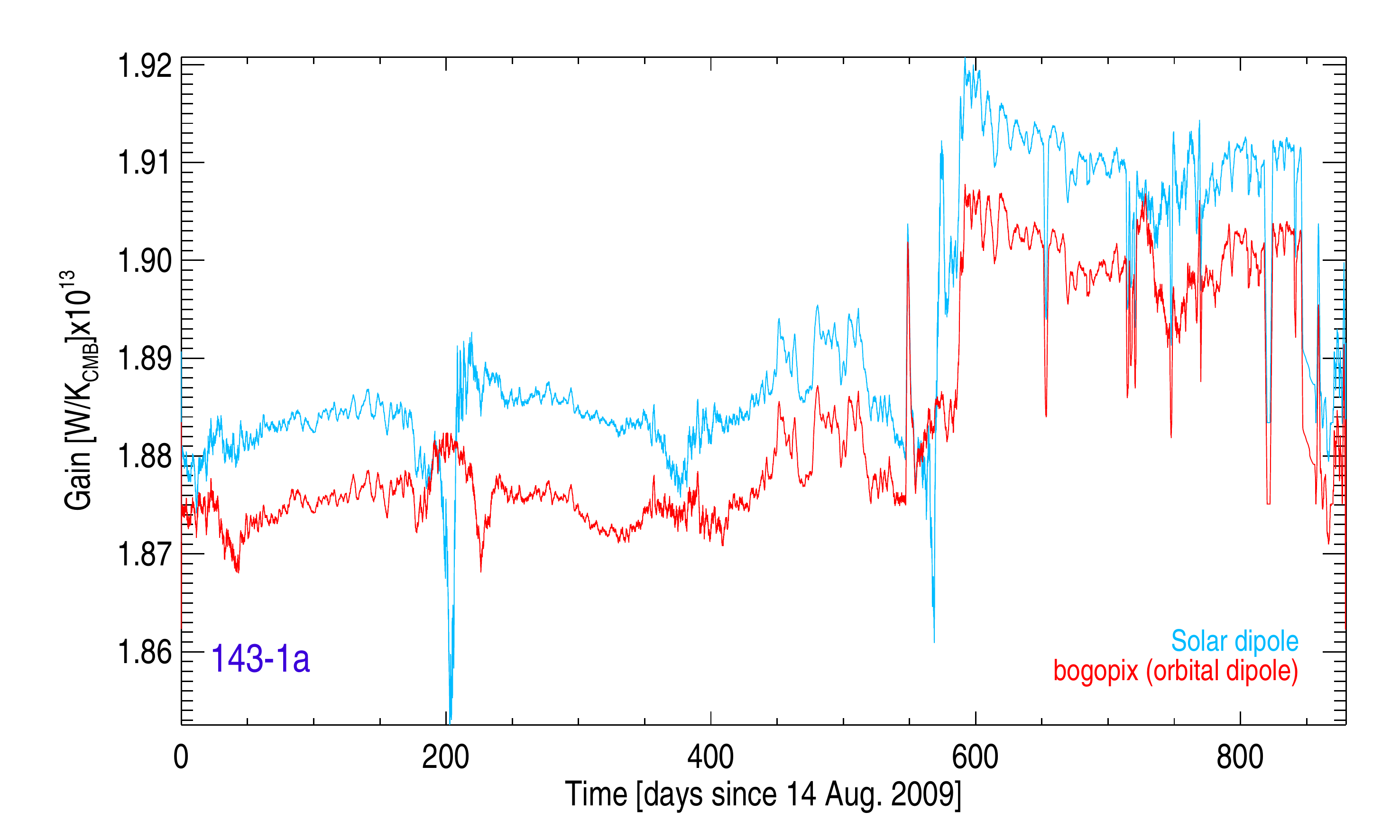}
\includegraphics[width=0.5\textwidth]{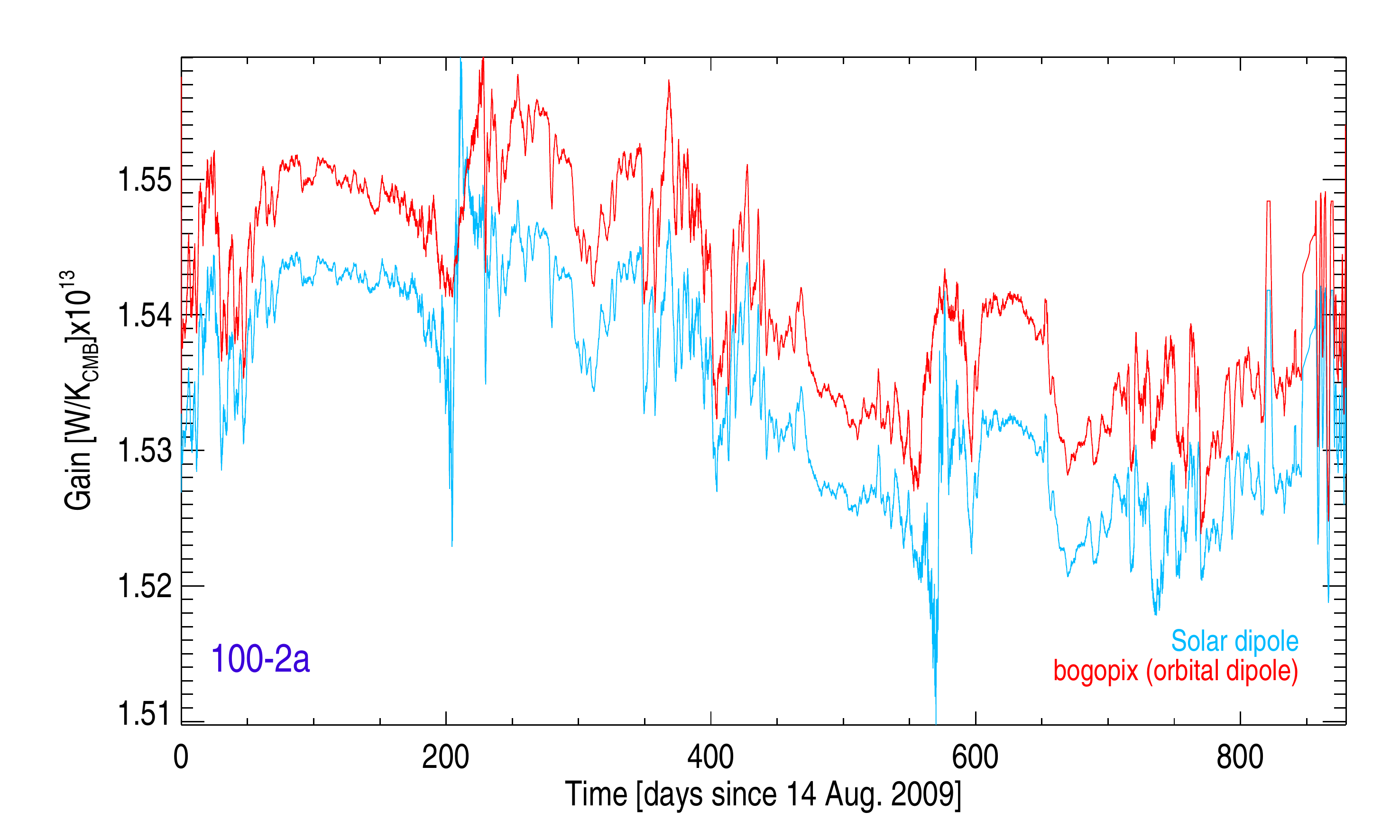}

\caption{\bogopix\ results for two HFI detectors, compared with those from the solar dipole calibration. Gain values for individual rings have been smoothed with a width of 50 rings (about 2 days), to increase the signal-to-noise ratio. There is good agreement  of the relative gain variations between the \bogopix\ results and those obtained from the HFI maps, except for the time intervals where the solar dipole amplitude is lower than the Galactic emission. The averaged value of the gains are, however, offset by factors (different from one detector to the other) of the order of 0.5 to 1\,\%. 
\label{fig:bogo_results}
}
\end{figure}

 As shown in Fig.~\ref{fig:bogo_results},  the variations of the gains $g_r$ found  with \bogopix\  follow nicely those from the solar dipole calibration ($g_r^{\mathrm{D}}$) in the regions where this signal is large. The lower level of fast variations from \bogopix\ in the time intervals where the scan lies close to the Solar dipole equator and at the same time close to the Galactic plane, indicates that the \bogopix\ results are less biased for these rings. We observe apparent gain variations on time scales of a few hour as well as months, with amplitudes of 1 to 2\,\% maximum, largely uncorrelated from one detector to another. 

The averaged gain level determined by the two methods are, however, different by 0.5 to 1\,\%, and the difference varies from one detector to another. We believe this is due to the different scales of the calibrating signals in the two methods: the absolute scale of \bogopix\  results is set by that of the orbital dipole, a factor of 5 to 10 lower in amplitude that the solar dipole used in the other method. These signals are thus affected to different degrees by the ADC non-linearities. In the simplest case, the effect of the  non-uniformity of the ADC digitization steps is a fixed offset (positive or negative) added on top of the signal, when this signal oversteps a given level, so  the resulting calibration bias will be lower for the largest calibration signal. 
\begin{figure}[htbp]
\centering
\includegraphics[width=0.45\textwidth]{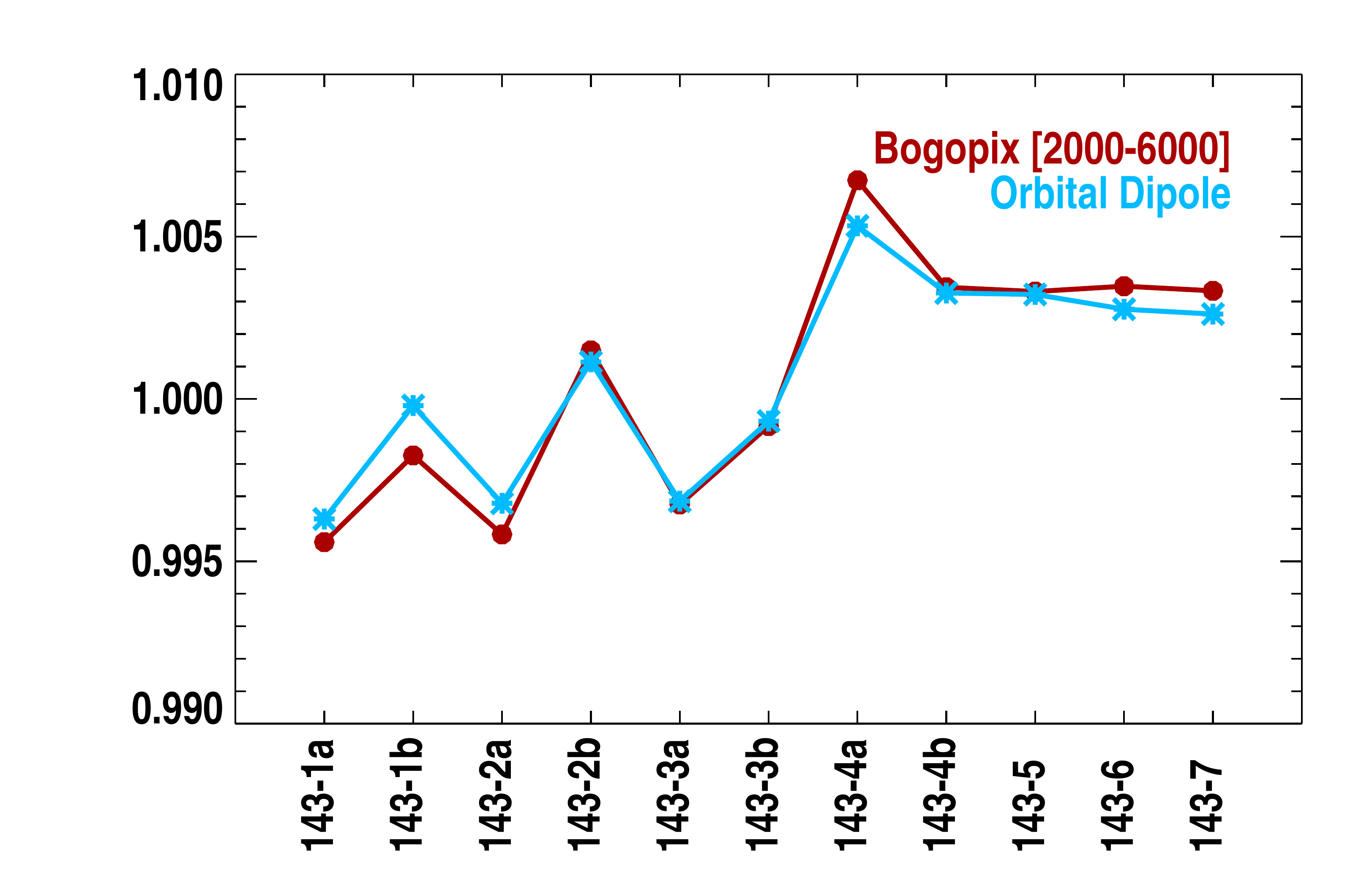}
\caption{Relative differences  of the orbital dipole calibration ($G^{\mathrm{OD}}$, blue) and  the average of \bogopix\ gains ($\tilde{G}^{\mathrm{bog}}$, red), with respect to the solar dipole calibration results($\tilde{G}^{\mathrm{SD}}$), for the 
143\,\GHz\ HFI detectors. Both schemes produce gains within 0.1\,\% of each other, which shows that they are both affected by the same systematics (the ADC non-linearities). 
\label{fig:calibs_bias_adc}}
\end{figure}

We study the difference between the averaged solar dipole gain, $\tilde{G}^{\mathrm{SD}}$, and the average of the \bogopix\ results, $g_r$, in the same ring  interval, denoted by $\tilde{G}^{\mathrm{bog}}$. 
We introduce another calibration process, based on the orbital dipole as  described in \cite{tristram2011}, together with \bogopix\ gains, renormalized so that they average to 1 between rings 2000 and 6000 (corresponding to days 60 and 190 approximately),  to correct
for the apparent relative gain variations. This produces another estimate of the absolute gain, $G^{\mathrm{OD}}$. 
The relative differences, $(\tilde{G}^{\mathrm{SD}}- G^{\mathrm{OD}})/\tilde{G}^{\mathrm{SD}}$, are shown in Fig.~\ref{fig:calibs_bias_adc} for each 143 \GHz\ detector. 
Both methods agree with each other within 0.05 to 0.1\,\%. We conclude that the difference between $\tilde{G}^{\mathrm{SD}}$ and  $\tilde{G}^{\mathrm{bog}}$ is genuine and it seems to be due to the use of the orbital dipole as the calibrator.
\begin{figure*}[htbp]
\centering
\begin{tabular}{@{}ccc@{}}
\includegraphics[width=0.32\textwidth]{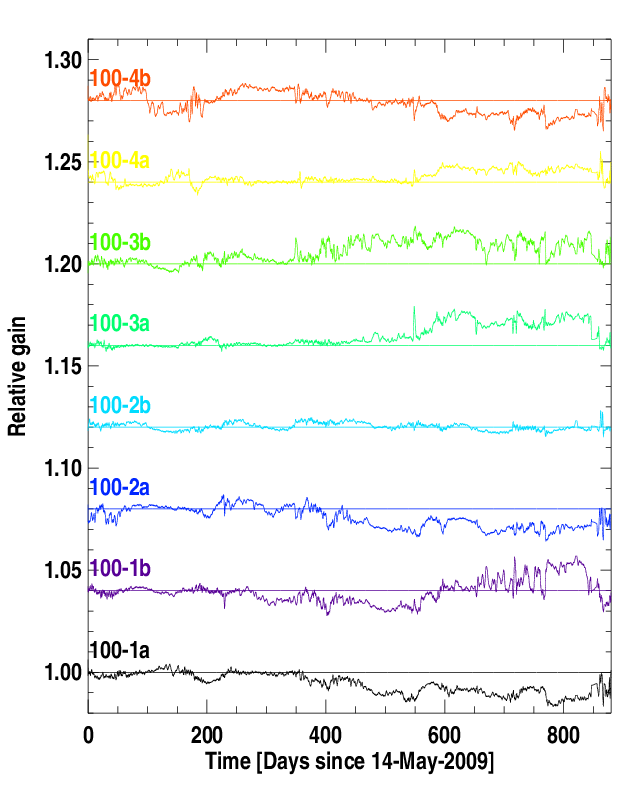}
&
\includegraphics[width=0.32\textwidth]{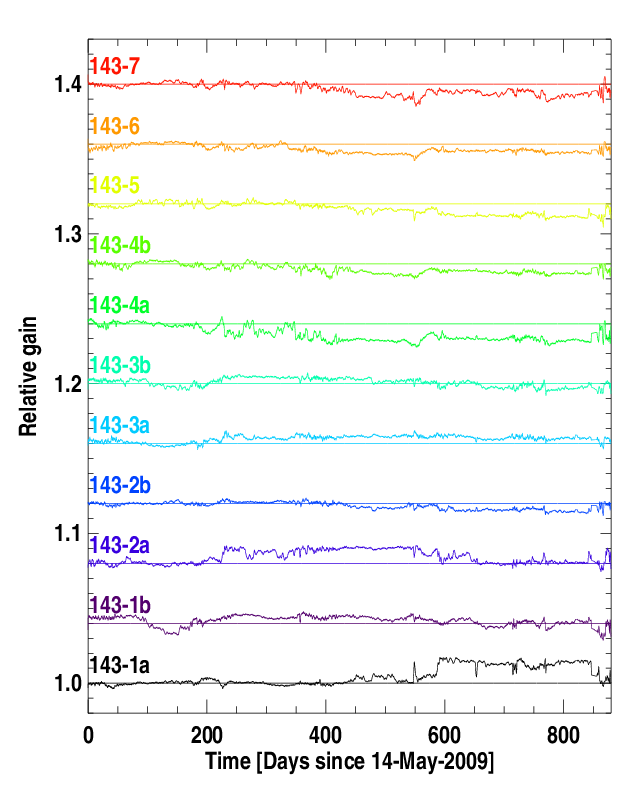}
&
\includegraphics[width=0.32\textwidth]{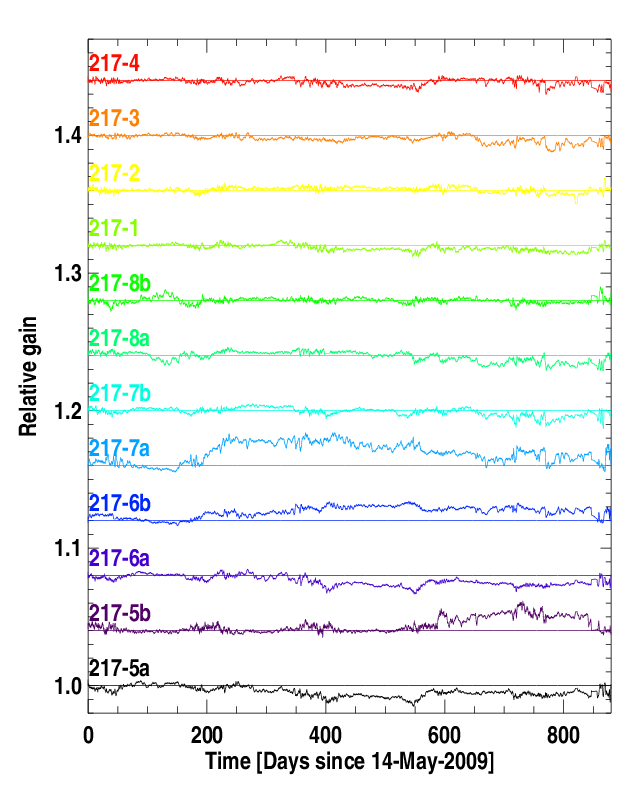}
\end{tabular}
\caption{Relative gains reconstructed by \bogopix\ for the 100, 143, and 217\,\GHz\  detectors as a function of time, smoothed with a width of 50 rings (about 2 days). Their overall amplitudes are of order 1 to 2\,\%, but both slow and fast (over a few tens of rings, i.e., a day) variations are observed. These variations are largely independent from one detector to the other.  Relative gains for each detector have been vertically displaced by 3\,\% for clarity. }\label{fig:bogo_res_summary}
\end{figure*}

We showed in \cite{tristram2011} that calibration errors induce large-scale features in the $Q$ and $U$ Stokes parameter maps. 
When using the orbital-dipole-based calibration factors to build these maps, we indeed observe such large-scale patterns, which is further evidence that the latter factors are biased. We also observed a noticeable residual dipole 
in the reconstructed detector maps, after subtraction of the \WMAP\ measured dipole, for the detectors where the difference between the solar and orbital dipole calibration was larger. 
We therefore conclude that, in the absence of an accurate correction for the ADC non-linearities, the orbital-dipole calibration scheme cannot be used to calibrate the HFI data. 

\subsection{Dipole calibration pipeline}

We used the \bogopix\ results only as to measure the {\em relative} gain variations, by normalizing to 1 on average between rings 2000 and 6000 (where the solar dipole calibration is computed). We show as an example a compilation of the relative gains reconstructed for the 100, 143, and 217\,\GHz\   detectors in Fig.~\ref{fig:bogo_res_summary}. The absolute calibration scale of the CMB channels (100--353\,\GHz) is set by the solar dipole calibration, as in the HFI early data release~\citep{planck2011-1.5}, which relied on \WMAP\ solar dipole measurements~\citep{hinshaw2009}.

\begin{figure}[htbp]
\centering

\includegraphics[width=0.465\textwidth,bb=280 131 790 410,clip]{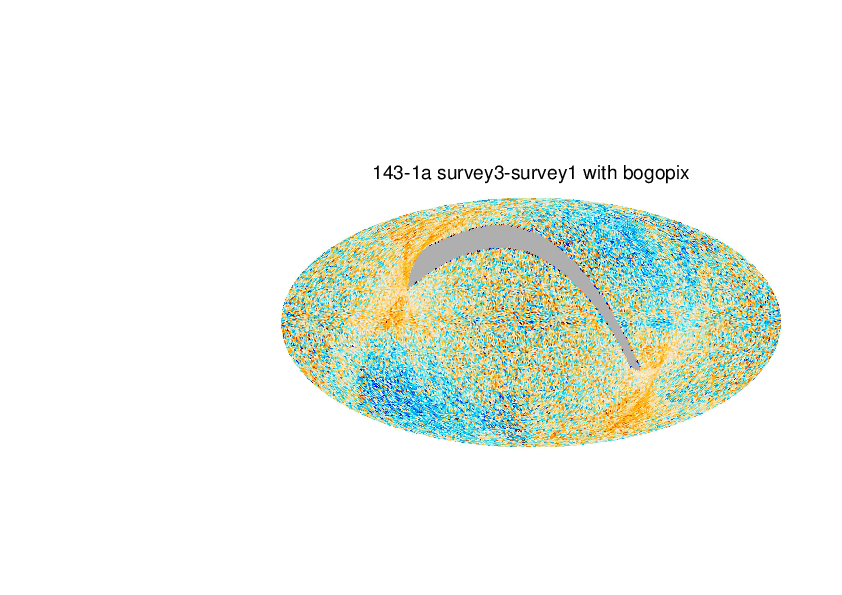}\\ 
\includegraphics[width=0.465\textwidth,bb=280 131 790 410,clip]{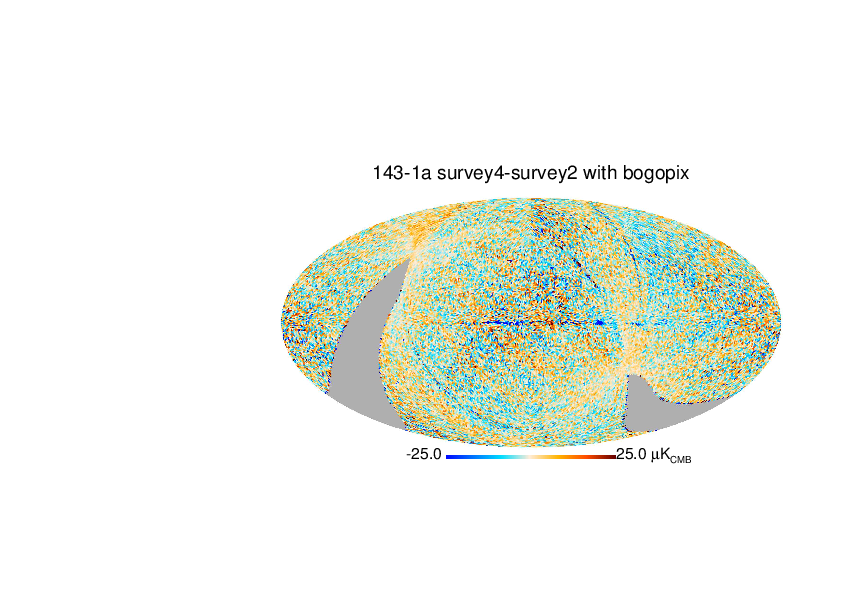} 

\caption{Residual differences between temperature maps built using data from detector 143-1a, for Surveys 1 and 3 (top) and Surveys 2 and 4 (bottom), derived using  the \bogopix\ results.  The level of  differences is much lower than in Fig.~\ref{fig:143-1a_surv_diffs_raw}. }
\label{fig:143-1a_surv_diffs_cor}
\end{figure}

As a first example of the improvements that \bogopix\ provides, we show in Fig.~\ref{fig:143-1a_surv_diffs_cor} the survey-difference maps (Survey 3 minus Survey 1, and Survey 4 minus Survey 2), for the detector used for Fig.~\ref{fig:143-1a_surv_diffs_raw}. The differences obtained using \bogopix\ are lower than $\sim 10\muK_{\mathrm{CMB}}$ outside the Galactic plane. The remaining
residuals in that region, in particular in the Survey3$-$Survey1 difference, can be attributed to the nonlinear nature of the systematic error, only the first-order linear part of which is handled by \bogopix. 

For frequencies $\ge 
353$\,\GHz, \bogopix\ results are not reliable, mainly because of the large spatial variation of the sky
emission inside a pixel  (we have used 1.72\arcm\ pixels here). Therefore, we do not correct  the highest-frequency channels for any gain variations. This leads to   calibration uncertainties of about 1\%  between maps from individual surveys.

\begin{table}
\caption{Statistical  and systematic uncertainties on the dipole calibration, for single detectors from the lower-frequency  HFI channels.   
The ``worst case'' column corresponds to a situation with a poorly matched sky template, whereas the third column is for the best case.  
In addition to each of these values, one has to take into account the \WMAP\ solar dipole amplitude uncertainty, 0.24\%, as this measurement is our primary calibrator.}
\begin{center}
\begin{tabular}{@{} c c c c @{}}
\hline
\hline
\noalign{\vskip 2pt}
Frequency & {Statistical error} & Systematic  & Systematic  \\
\noalign{\vskip 2pt}
$[$GHz$]$ &  {[\%]} & (worst case) [\%] & [\%] \\
\noalign{\vskip 2pt}
\hline
\noalign{\vskip 2pt}
100 & 0.004 & 0.64 & 0.37 \\
143 & 0.002 & 0.53 & 0.29 \\
217 & 0.002 & 0.69 & 0.41 \\
353 & 0.010 & 2.53 &  1.81 \\
\hline
\end{tabular}
\end{center}
\label{tab:calibration_syst_sat}
\end{table}


\subsection{Dipole calibration uncertainties for single detector}
\label{sec:abs_cal_unc}
For the dipole calibration scheme, the statistical uncertainties are
estimated by propagating the TOI sample variances (NET) to the
ring-by-ring gain estimation on the solar dipole, averaged between
rings 2000 and 6000, for each detector. These uncertainties are much
lower than the systematic uncertainties that dominate our calibration
measurement. We estimate the size of these systematic uncertainties on the calibration
of individual detectors by measuring the dispersion 
of these ring-by-ring gains. Both uncertainties are listed in
Table~\ref{tab:calibration_syst_sat}, which gives their average at
each frequency. The  \WMAP\  solar dipole amplitude uncertainty
 \citep[0.24\%,][]{hinshaw2009} is not included. The systematic errors
given here should be considered as upper limits on the real
systematics, as they have been derived from solar dipole ring-by-ring
gains prior to the \bogopix\  correction. We indicate the effect of
the choice of a Galactic template by indicating a ``worst case'' scenario (second
column of Table~\ref{tab:calibration_syst_sat}) in which a non-optimal template was used (see Appendix~\ref{app:sky_dipcal} for details). When
combining different detectors, some of these systematic errors
should partially average out for temperature. The gain variation part,
for example, is independent from one detector to another. To get a more
precise estimation of the calibration accuracy for the frequency maps
of this release, we have performed more elaborate tests, which are  presented in Sect.~\ref{sec:calib_charact}.


\section{Photometric calibration of the high-frequency channels}
 \label{sect_HF_calib}
 
\subsection{From FIRAS-based to planet-based absolute calibration} 
Since the early days of the \Planck\ data, it has been apparent that the ratio between HFI and FIRAS is not constant across the sky: we observe spatial gain variations, i.e., variation of the calibration coefficient $K$, and thus variation of the offset $O$ (see Eq.~\ref{eq_firas_HFI}), that mimic a decrease of  $K$ with brightness\footnote{These should not be confused  with the apparent gain variation with time, discussed in Sect.~\ref{sec:gain_variation}.}. Comparison with the dipole calibration at 353\,\GHz\ showed that the high-latitude gradients ($10\deg<|b|<60\deg$) give a better agreement. This was thus adopted for the calibration of the \Planck\ early results \citep{planck2011-1.7}.
We  studied this unresolved discrepancy with the FIRAS maps further while  preparing the first major release of \Planck\ data. As detailed in Appendix~\ref{A1}, numerous tests and checks have been conducted. However, we could not find any remaining HFI systematics, or any bias in our method of comparison of the two data sets, that could explain such a discrepancy. {  The only possibility comes from a systematic bias in the FIRAS ``pass4'' interstellar dust spectra. Indeed, \cite{liang2012} propose significant revisions to the FIRAS dust spectra that
that would reduce the discrepancy with HFI (see their figure~1). }

In parallel, indications have come to light of an overestimate of the HFI brightness at high frequencies, when calibrating using FIRAS (see Sect.~\ref{FIRAS_Syst_Effect}). These have led us to adopt a new photometric calibration scheme for the sub-millimetre channels. We now compare planet flux-density measurements at 545 and 857\,\GHz\ with models in order to set the absolute calibration. 
The ultimate scheme would be to intercalibrate the 545 and 857\,\GHz\ channels with the lower-frequency channels, using the planet models as relative calibrators. Indeed, for the Neptune and Uranus planet models used for the calibration, the absolute scale of the model is known to  about 5\,\%, whereas the  relative inter-frequency uncertainty is expected to be of order of 2\,\%~\citep{moreno2010}.  In this section we present our calibration procedure for the 2013 data release.

\subsection{Planet flux densities: measurements and comparison with the models}

\Planck\ observes the five outer planets: Mars, Jupiter, Saturn, Uranus, and Neptune. The 21 planet observations made by \Planck-HFI have been analyzed. For calibration purposes, only Neptune and Uranus are used because
\begin{description}
\item[(i)] Jupiter data lie partly in the nonlinear regime of the HFI read-out system, 
\item[(ii)] Jupiter and Saturn have strong absorption features that make  comparison with the broad-band measurements difficult,  
\item[(iii)] Mars's flux varies strongly from season to season.
\end{description}
 Even if this last issue can be handled precisely by the models, it complicates the analysis, so we defer the use of Mars to future work.
For the present data release we applied correction factors to the FIRAS-based calibration coefficients to match the Uranus and Neptune flux densities given by the \cite{moreno2010} model. 

\subsubsection{HFI beams and solid angles}
The HFI beam solid angles used in this analysis are those derived from Mars observations.  We do not use solid angles from the beams that are averaged over the scanning history (the so-called {\it effective} beams, \citealt{planck2013-p03c}), because we consider each observation of  Uranus and Neptune for each bolometer separately, and therefore we do not need to compute an average point spread function. We correct for the small response at large scales (more than 40\arcm\ from the beam centroid) that is due to incomplete deconvolution of the bolometer/readout electronics time response, as measured on Jupiter. Details of the beam solid angle measurements are given in \cite{planck2013-p03c}. 

The beam solid angle is frequency-dependent and its measured value thus depends on the SED of the source. The solid angle for a planet (with an SED roughly proportional to $\nu^2$) is different from that for the photometric convention $\nu I_{\nu}={\mathrm{constant}}$.  \cite{maffei2010} and \cite{tauber2010} investigated the variation of the beam size across the passband using a pre-launch telescope model.  For the lowest frequency HFI bands 100, 143, and 217\,\GHz, the beam size reaches a minimum near the centre of the band, making the solid angle a weak function of the source's SED.  The beam colour corrections for these bands are expected to be less than 0.3\,\%. At 353\,\GHz\ the solid angle increases with frequency across the band, but the beam colour corrections are expected to be less than 1\,\%.  The multi-moded horns at 545 and 857\,\GHz\ are more difficult to model because of uncertainty in the relative phase and amplitudes of the modes propagating through each horn. The models of \cite{murphy2010} give upper limits on the beam colour correction to the solid angles of 2\,\% at 545\,GHz and 1\,\% at  857\,\GHz.

The FWHM of the beam on the sky is defined by the energy
distribution at the entrance of the horn, which -- owing to the optical
design of the \Planck\ telescope -- does not depend much on the frequency
within the band. The throats of the horns are positioned very
close to the focal plane of the telescope, with a deviation 
varying from horn to horn. If the deviation is small, the variation of
the beam solid angle on the sky will be symmetric around the centre of
the band, and the combined correction will be small. On the other hand, if the horn is
significantly offset, the variation will not be symmetric, and the
correction will be larger. An estimate of the correction is presented
in \cite{planck2013-p03c}.

The variation of the solid angle inside the band is negligible
compared to the error we have on the photometry, and for now the beam
solid angle variation for the high-frequency channels has not been
taken into account in the calibration. 
The low variation of resolution across the passband  is unusual for a sub-millimetre experiment. By way of comparison,  in SPIRE
on {\it Herschel} the FWHM varies by $\pm$17\%.  The SPIRE
beam solid angles have been measured on Neptune; using the photometric
convention $\nu I_{\nu}={\mathrm{constant}}$, the corrections to the beam solid angles are
about 3.3\,\% at 350\,$\mu$m and and 5.9\% 500\,$\mu$m \citep{griffin2013}.

\subsubsection{Uranus and Neptune flux measurements and model comparison}
Our calibration procedure follows the following steps:
\begin{itemize}
\item A first photometric calibration was set using FIRAS at 545 and 857\,\GHz.
\item We created $2\deg\times2\deg$ maps with a 2\arcm\ pixel size around the planet positions by projecting the destriped and calibrated timelines, using the nearest grid point algorithm, from timelines scanning each planet.
\item We built maps of the same sky area, using
  observations taken at different epochs (when the planet was at a
  different position) to estimate the sky background, and
  subtracted them from the planet maps.  At $\nu\le 353$\,\GHz, the background is negligible. At 857\,\GHz, the astrophysical background is a few percent of the peak signal of Neptune. 
\item We measured the planet flux densities using aperture photometry on the background-subtracted maps. We integrated the flux up to 3 $\times$ FWHM. We corrected for the beam solid-angle difference between this scale and the  full solid angle. This correction amounts  to 0.8\,\% at 545\,\GHz\ and 1.5\,\% at 857\,\GHz.
\item At 545 and 857\,\GHz, we applied a correction factor to the FIRAS calibration to match the Uranus and Neptune flux densities given by the models. The factors were the same for all bolometers within a frequency channel, namely 1.07 at 857\,\GHz\ and 1.15 at 545\,\GHz.
\end{itemize}

The measurements are colour-corrected (using Eqs.~\ref{cc_2} and~\ref{cc_1}) and the flux densities are quoted for the two planet spectra. Colour corrections vary from about 0.92 (at 353\,\GHz) to 1.05 (at 143\,\GHz). In the sub-millimetre channels they are < 2\,\% at 857\,\GHz\ and about 5\,\% at 545\,\GHz. Errors on the colour corrections are estimated to be 0.25, 0.06, 0.01, 0.006, 0.003 and  0.002\,\% from 100 to 857\,\GHz\ \citep{planck2013-p03d}.

From the flux densities and the planet solid angles estimated for HFI at the date of the observations, we can compute the brightness temperatures $T_{\mathrm B}$. They are given in Table~\ref{tab:Tb_planets}, where we averaged the flux densities computed for all detectors in a channel, and all observations' {  epochs}  (four), prior to the computation of $T_{\mathrm B}$. The quoted error on $T_{\mathrm B}$ comes from the standard deviation of the flux-density measurements.

\begin{table}
\caption{Neptune and Uranus brightness temperatures measured by HFI. At 545 and 857\,\GHz\ the numbers are not independent measurements of the planet flux densities, since the data
 have been recalibrated to match the models.}
\begin{center}
\begin{tabular}{@{} c c c@{}}
\hline
\hline
\noalign{\vskip 2pt}
Frequencies & Uranus & Neptune \\ 
 $[$GHz$]$  & $T_B$ $[$K$]$ & $T_B$ $[$K$]$ \\ 
\noalign{\vskip 2pt}
\hline
\noalign{\vskip 2pt}
\begin{tabular}{c}
100 \\ 
143 \\ 
217 \\ 
353 \\ 
545 \\ 
857 
\end{tabular}

&
{  
\begin{tabular}{r@{}l} 
 124.3 & $\pm$5.0 \\
 108.4 & $\pm$2.9 \\
 97.0 & $\pm$2.5 \\
 83.3 & $\pm$2.3 \\
 (73.7 & $\pm$2.4) \\
 (67.5 & $\pm$1.3) 
\end{tabular}
}
&
{
\begin{tabular}{r@{}l@{}} 
129 & $\pm$15 \\
110 &$\pm$6 \\
 96.9 &$\pm$3.7 \\
81.1 &$\pm$2.7 \\
 (71.2 &$\pm$2.1) \\
(65.0 &$\pm$1.9) 
\end{tabular} 
} 
\\
\noalign{\vskip 2pt}
\hline
\end{tabular}
\end{center}
\label{tab:Tb_planets}
\end{table}

We use the models called ESA2 for Uranus and ESA3 for Neptune developed by R. Moreno for the {\it Herschel}-SPIRE absolute photometric calibration \citep{moreno2010}.  The millimetre and sub-millimetre spectra of Uranus and Neptune were modelled with a line-by-line radiative transfer code accounting for the spherical geometry of their planetary atmospheres, like that described for Titan by \cite{moreno2011}.
Atmospheric opacity due to the minor species CO (for Neptune only), and NH$_3$ far wings, as well as  collision-induced opacities of the main species
($\mathrm{H}_2$, $\mathrm{He}$, $\mathrm{CH}_4$)  were included. The thermal profiles in the troposphere, which is the atmospheric region probed between 90 and 900\,\GHz,  were taken from \cite{Lindal1992}. The uncertainty of the computed brightness temperature is mainly linked to the uncertainty on the thermal profile
with an absolute uncertainty value of 5\,\%. The relative calibration (between frequencies) is expected to be of the order of 2\,\%.

\begin{figure}
\begin{center}
\includegraphics[width=\columnwidth]{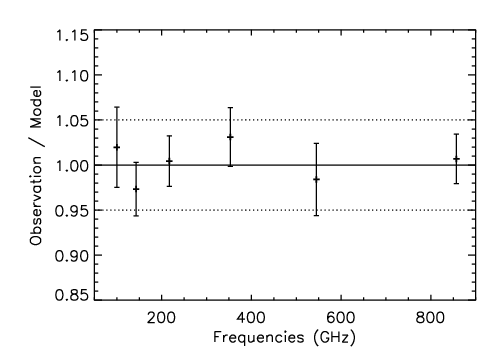}
\includegraphics[width=\columnwidth]{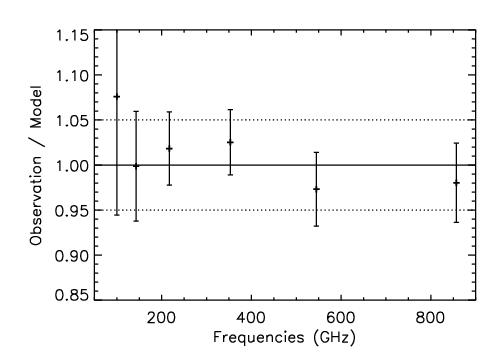}
\caption{Ratio of the flux densities measured by HFI and computed from
  the ESA2 (Uranus, {\it top}) and ESA3 (Neptune, {\it bottom}) models from \cite{moreno2010}. At 545 and
  857\,\GHz, the measurements are not independent measurements of the
  planet flux densities, since the 545 and 857\,\GHz\ channels have been
  re-calibrated to match the Uranus and Neptune
  flux densities given by the models.}\label{fig:planets_compa_model}
\end{center}

\end{figure}

\begin{figure}
\begin{center}
\includegraphics[width=\columnwidth]{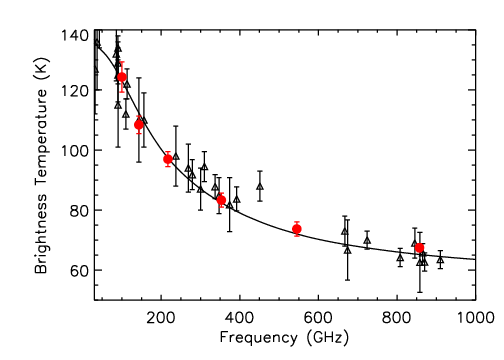}
\includegraphics[width=\columnwidth]{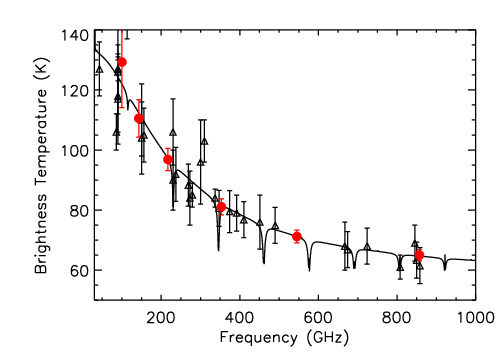}
\caption{Variation of brightness temperature with frequency for Uranus ({\it top}) and Neptune ({\it bottom}). The brightness temperatures derived from the flux densities measured using aperture photometry on HFI maps are the red points. The continuous lines are the ESA2 (Uranus) and ESA3 (Neptune) models from \cite{moreno2010}. The scale of the models is known at about the 5\,\% level, and the relative inter-frequency uncertainty is expected to be of the order of 2\,\%. At 545 and 857\,\GHz, the measurements are not independent determinations of the planet flux densities, since the 545 and 857\,\GHz\ channels have been re-calibrated to match the Uranus and Neptune flux densities given by the models. The other data points are extracted from the literature.}
\label{fig:planets_compa_model2}
\end{center}
\end{figure}

 We compute the flux densities using the brightness temperatures from the model and the planet solid angles estimated for HFI at the date of the observations. The model spectra (in Jy) are interpolated onto our bandpass frequencies, and convolved by our bandpass filters to obtain the flux densities as measured by HFI. In Fig.~\ref{fig:planets_compa_model}  we compare the flux measurements with the models. Error bars on the HFI data points correspond to the standard deviation of the measurements (for all bolometers and all {  epochs}).
For the two high frequencies (857 and 545\,\GHz), the agreement with the model has been forced by our calibration procedure. For the lower frequencies, calibrated using the dipole, we have an overall very good agreement with the model, the two being compatible within the error bars.
Fig.~\ref{fig:planets_compa_model2}  shows the same comparison, but on
spectra plotted in brightness temperature, and with other measurements from
the literature. Notice the high  accuracy of the HFI measurements over a wide range of frequencies.

\subsection{Planet  calibration uncertainties \label{subsect: planet_calib_uncert}}

At high frequencies, we estimate the error on the absolute calibration of the frequency maps to be  10\,\% (for both the 545 and 857\,\GHz\ channels). This uncertainty combines the statistical uncertainty in the flux-density measurements (5\,\%) with the systematic uncertainty in the Neptune and Uranus models, taken to be 5\,\%. Note that the latter is probably overestimated as we have a very good relative calibration between the low-frequency channels (143, 217, 353\,\GHz), which have a much more accurate absolute calibration, and the high-frequency ones.

\section{Setting the zero levels in the maps \label{sec:zero_levels}}
At this stage the zero levels of the maps are arbitrary. \Planck\ cannot fix them internally and we need to rely on the use of external data sets. The zero level comprises two parts.
\begin{enumerate}
\item A Galactic zero level: we estimate the brightness in the \Planck-HFI maps that corresponds to zero gas column density (zero gas column density means zero Galactic dust emission). As a gas tracer, we use the \hi\ column density (from 21\,cm emission), assumed to be a reliable tracer of the Galactic gas column density in very diffuse areas ({\bf column density lower than $2\times10^{20}\ \mathrm{cm}^{-2}$ to avoid any contamination by molecular gas}).
\item An extragalactic zero level: the cosmic infrared background monopole.
\end{enumerate}
The sum of the two offsets is appropriate for total emission analysis. For Galactic studies, only the Galactic zero level has to be set. 

\subsection{The Galactic zero level \label{sec:gal_zero_levels}}

Two methods have been combined to obtain reliable numbers. The first one
uses the correlation of the \Planck\ maps with \hi\ column density (following \citealt{planck2011-7.12}, \citealt{planck2011-6.6}, and \citealt{planck2013-pip56}). The basic idea is to estimate the brightness in the
\Planck\ maps that corresponds to zero column density by correlating
with \hi, which is assumed here to be a reliable tracer of the Galactic gas column
density in diffuse areas (it thus neglects any dust associated with
the diffuse {\sc Hii} gas that is not spatially correlated with the \hi). The model is simply
\begin{equation}
\label{eq_offset_1}
I_{\nu} = \alpha_{\nu} \times N_{\mathrm{HI}} + O_{\nu}\,.
\end{equation}
The correlation with \hi\ allows us to estimate $O_{\nu}$ independently for each frequency, but it relies on the assumption of a tight gas-to-dust correlation over relatively large areas of the sky.

The second method is based on the inter-frequency correlation of
\Planck\ maps, for which the model is
\begin{equation}
\label{eq_offset_2}
I_{\nu} = \alpha_{\nu} \times  I_{\nu_0}+ O_{\nu}\,,
\end{equation}
with $I_{\nu_0}$ being one the \Planck\ maps. Here the offsets are all relative to the offset of  $I_{\nu_0}$ that needs to be determined otherwise (by the first method for instance). The advantage of the second method is that no assumption is made on the phase in which the gas resides (we correlate dust emission with dust emission) and a larger area of the sky can be used to perform the correlation. 
All the data were smoothed to a common angular resolution of 1\deg. CMB anisotropies, as extracted in \cite{planck2013-p06}, were also removed from the data prior to the correlation.
 
For the correlation with \hi, we used the 21-cm all-sky data from the LAB survey \citep{kalberla2005}. The LAB data are a collection of close to 200\,000 spectra that were processed individually. The map of \hi\ column density used here is summed over velocities. 
The zero level of the LAB data (and of 21\,cm observations in general)
depends mostly on the baseline subtraction at the spectrum level. At
1420 MHz, the spectroscopic observation is the sum of the 21 cm line,
the synchrotron and free-free emissions (which are well
approximated by a power law at this frequency), and instrumental
baseline variations due to various effects, including ground radio
interference and system temperature variations. The 21 cm emission is
usually extracted by removing a baseline using a polynomial fit
constrained with velocity channels away from the \hi\ Galactic
emission. The two radiotelescopes used to build the LAB data have a
large velocity range coverage (from $-450$ to 400 km s$^{-1}$) allowing for a
good estimate of the baseline. In addition many sky positions
were observed several times, allowing improvement on the baseline
correction. Because the baseline correction is applied on each
individual spectrum, the noise on the zero level will be at the pixel
size on the final map and no bias at large angular scales should be
expected. Larger-scale zero-level variations could come from stray
(far sidelobe) radiation. However, the LAB data were constructed with the most
precise stray-radiation correction to date, leaving very faint
residual emission, at a level of 2\% (considering Galactic line emission). For the
gain calibration, strong radio sources are used (see \citealt{kalberla2005}). The calibration is performed regularly during observations to monitor any gain drift. The precision of the gain of the LAB data has no impact on the determination of the HFI zero level, as it is obtained through a correlation. 

The first method requires the use of a very strict mask, to include only regions where the gas is mostly in the neutral atomic form (no significant dark gas for example) and avoiding lines of sight with significant emission from clouds in the Galactic halo (intermediate velocity clouds and high velocity clouds), as they have slightly different dust emission properties. We select pixels where the local velocity cloud \hi\ column density is less than $2 \times 10^{20}$ cm$^{-2}$ and where no significant IVC emission is detected. This very strict mask includes 11.5\,\% of the sky.
For inter-frequency correlations (Eq.~\ref{eq_offset_2}), a second mask was built by including pixels where the local velocity cloud \hi\ column density is less than $3 \times 10^{20}$ cm$^{-2}$  (and no restriction on IVCs), increasing the sky fraction to 28\,\%. 

\begin{figure}
\centering
\includegraphics[width=6.cm]{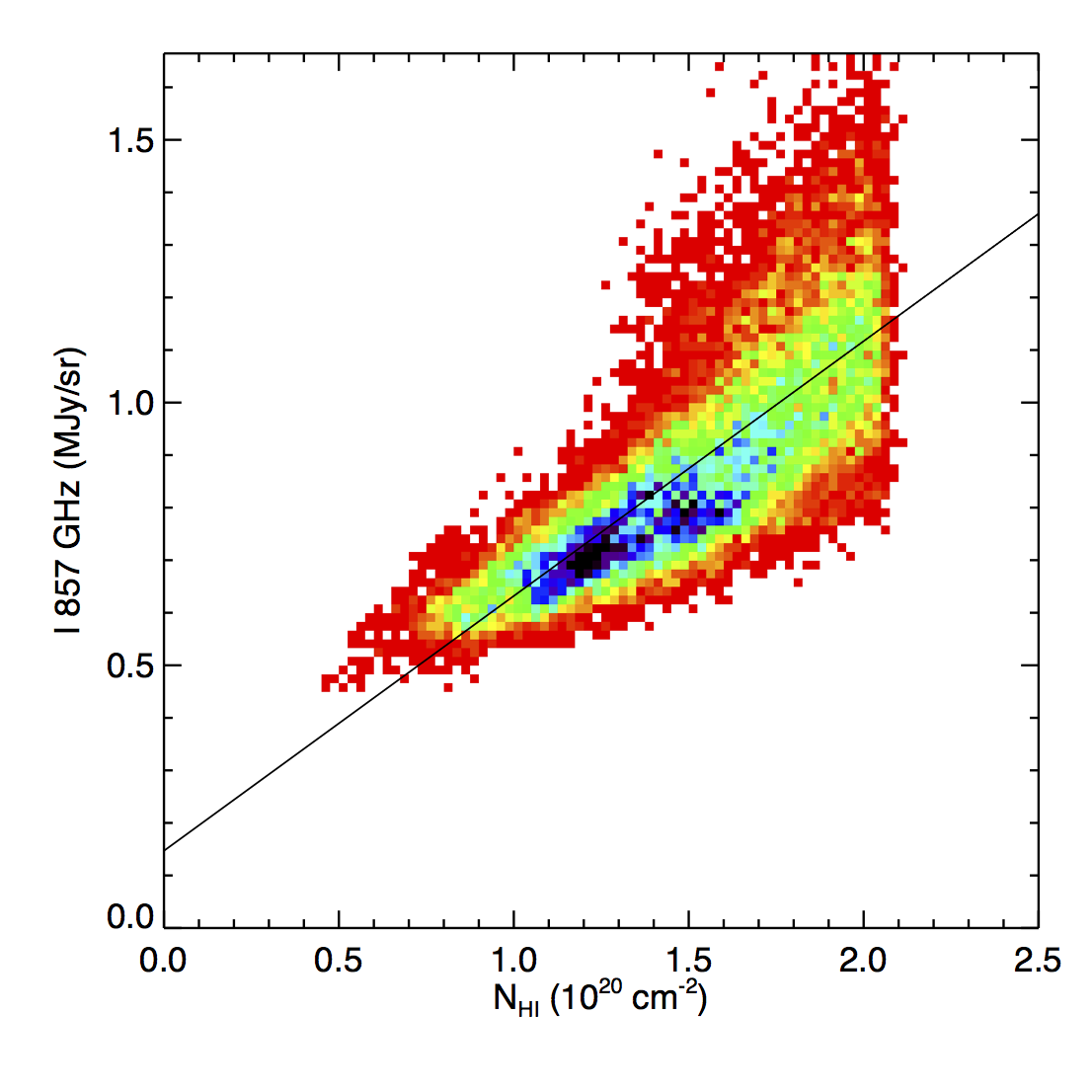}
\caption{857\,\GHz-\hi\ correlation over 11.5\,\% of the sky (N$_{\mathrm{HI}} < 2 \times 10^{20}$ cm$^{-2}$ -- smoothed to 1\deg-- and excluding intermediate velocity clouds). }
\label{fig:offset_hi}
\end{figure}

\begin{figure}
\centering
\includegraphics[width=9.cm]{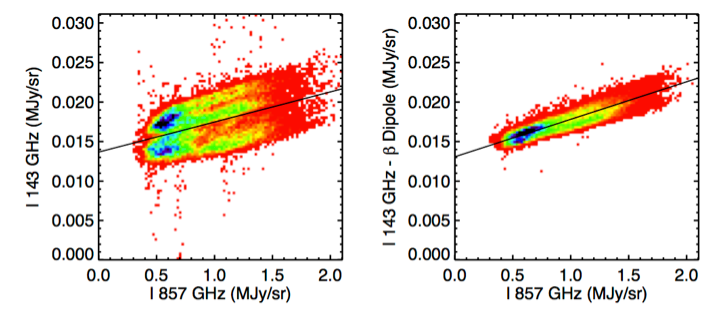}
\caption{Correlation between the 143 and 857\,\GHz\ frequency maps on 28\,\% of the sky (\hi\ column density smoothed at 1\deg\ lower than 3 $\times 10^{20}$ cm$^{-2}$). CMB anisotropies have been removed at 143\,\GHz. {\it Left:} Raw correlation. {\it Right:}  Correlation after a residual solar dipole has been removed at 143\,\GHz. The offset of this correlation sets the Galactic zero level of the 143\,\GHz\ map.}
\label{fig:offset_freq}
\end{figure}

To minimize the effect of the imperfect dust-to-\hi\ correlation and to obtain the highest possible signal-to-noise ratio, the Galactic zero levels were computed using Eq.~\ref{eq_offset_1} at 857\,\GHz\ and using Eq.~\ref{eq_offset_2} at the other frequencies, thus taking 
$I_{\nu_0}=I_{857}$.
In Fig~\ref{fig:offset_hi} we show the 857\,\GHz-\hi\ correlation. We
observe a significant dispersion in this correlation, possibly due to
variations of the dust-to-gas ratio or variations of the dust properties,
or due to the fact that \hi\ is not a perfect tracer of column density
(e.g., presence of dust in the warm ionized
medium). Figure~\ref{fig:offset_freq} shows an example of inter-frequency correlation, at 143\,\GHz. Their correlation plots are clearly split in two, revealing an effect unaccounted for in our model (Eq.~\ref{eq_offset_2}). Once projected onto the sky, the residual shows that the north and south parts of the mask have different offset values. This bi-modal structure is minimized once a residual solar dipole is removed from the data. To set the Galactic zero level, we therefore also fit for the amplitude of an additional residual solar dipole term in Eq.~\ref{eq_offset_2}. The amplitude of this residual dipole pattern is in accordance with the actual accuracy of the absolute calibration (discussed in Sect~\ref{sec:abs_cal_unc}). Indeed, with a given accuracy on the absolute calibration, a dipole with an amplitude of at least the given accuracy can be left in the map. There is no contradiction between the amplitude of the dipole left in the maps and the current absolute calibration uncertainties, listed in Table~\ref{tab:error_summary}. More precisely, we found residual dipole amplitudes compatible with a $0.3\,\%$ calibration error for 100--217\,\GHz\ and $1\,\%$ for 353\,\GHz. 

\subsection{The cosmic infrared background monopole}
The (isotropic) mean value of the CIB is computed using the \cite{bethermin2012} model. This is an empirical model based on the current understanding of the evolution of main-sequence and starburst galaxies. It reproduces the mid-infrared to radio galaxy counts very well. The values of the CIB (which is the integral of the emission from galaxies) have been computed using the HFI bandpass filters. They have then been converted into the convention $\nu I_{\nu}={\mathrm{constant}}$ using the CIB SED fit of \cite{gispert2000}. The values are given in Table~\ref{tab:cib}. They are consistent with those extracted from FIRAS data (see Table~\ref{tab:PZ_CIB}).  Errors are on the order of 20\,\%. The CIB has to be added to the maps for total emission analysis. 

\begin{table}
\caption{CIB monopole that has to be added to the maps.}
\begin{center}
\begin{tabular}{@{} c c @{}}
\hline
\hline
\noalign{\vskip 2pt}
Frequencies & CIB \\ 
 $[$GHz$]$  & $[$MJy\,sr$^{-1}]$ ($\nu I_{\nu}={\mathrm{constant}}$) \\ 
\noalign{\vskip 2pt}
\hline
\noalign{\vskip 2pt}
100 & 3.0$\times$10$^{-3}$ \\
143 & 7.9$\times$10$^{-3}$ \\
217 & 3.3$\times$10$^{-2}$ \\
353 & 1.3$\times$10$^{-1}$ \\
545 & 3.5$\times$10$^{-1}$ \\
857 & 6.4$\times$10$^{-1}$\\ 
\hline
\end{tabular}
\end{center}
\label{tab:cib}
\end{table}

\subsection{Set the appropriate zero levels of HFI maps}
For Galactic analysis, the Galactic zero levels, given in Table~\ref{tab:gal_offset}, have to be removed from the frequency maps in the 2013 data release .
For total emission analysis, the CIB monopole, given in
Table~\ref{tab:cib}, has furthermore to be added. As stated previously, for the CIB we estimate the error to be of the order of 20\,\%. For the Galactic zero level, errors are given in Table~\ref{tab:gal_offset}. The uncertainty on the 857\,\GHz\ Galactic offset is dominated by systematics. At lower frequencies, the uncertainties take into account the impact of the  CMB removal, the statistical uncertainty of the fit, and the error on the 857\,\GHz\ offset. 

\begin{table}
\caption{Table giving the offsets that have to be removed at each frequency to set the Galactic zero level. These offsets have been computed assuming zero Galactic dust emission for zero gas column density.}
\begin{center}
\begin{tabular}{@{} c c c @{}}
\hline
\hline
\noalign{\vskip 2pt}
Frequencies & Total maps & Zodi-removed maps \\ 
 $[$GHz$]$  &  $[$MJy\,sr$^{-1}]$  &
 $[$MJy\,sr$^{-1}]$  \\ 
  & ($\nu I_{\nu}={\mathrm{constant}}$) & ($\nu I_{\nu}={\mathrm{constant}}$) \\
\noalign{\vskip 2pt}
\hline
\noalign{\vskip 2pt}
100 & 0.0047$\pm$0.0008 & 0.0044$\pm$0.0009 \\
143 &0.0136$\pm$0.0010 & 0.0139$\pm$0.0010 \\
217 & 0.0384$\pm$0.0024 & 0.0392$\pm$0.0023 \\
353 & 0.0885$\pm$0.0067 & 0.0851$\pm$0.0058 \\
545 & 0.1065$\pm$0.0165 &  0.0947$\pm$0.0140 \\
857 & 0.1470$\pm$0.0147 & 0.0929$\pm$0.0093 \\ 
\hline
\end{tabular}
\end{center}
\label{tab:gal_offset}
\end{table}

\section{Characterization and checks of calibration \label{calib_accuracy}}
\label{sec:calib_charact}
In this section we present the various tests that have been carried out  to assess the precision and stability of the calibration of the HFI data. 

\subsection{Time stability of the calibration}
\label{sec:calib_time_stab}
\begin{figure*}
\centering
\includegraphics[width=0.75\textwidth]{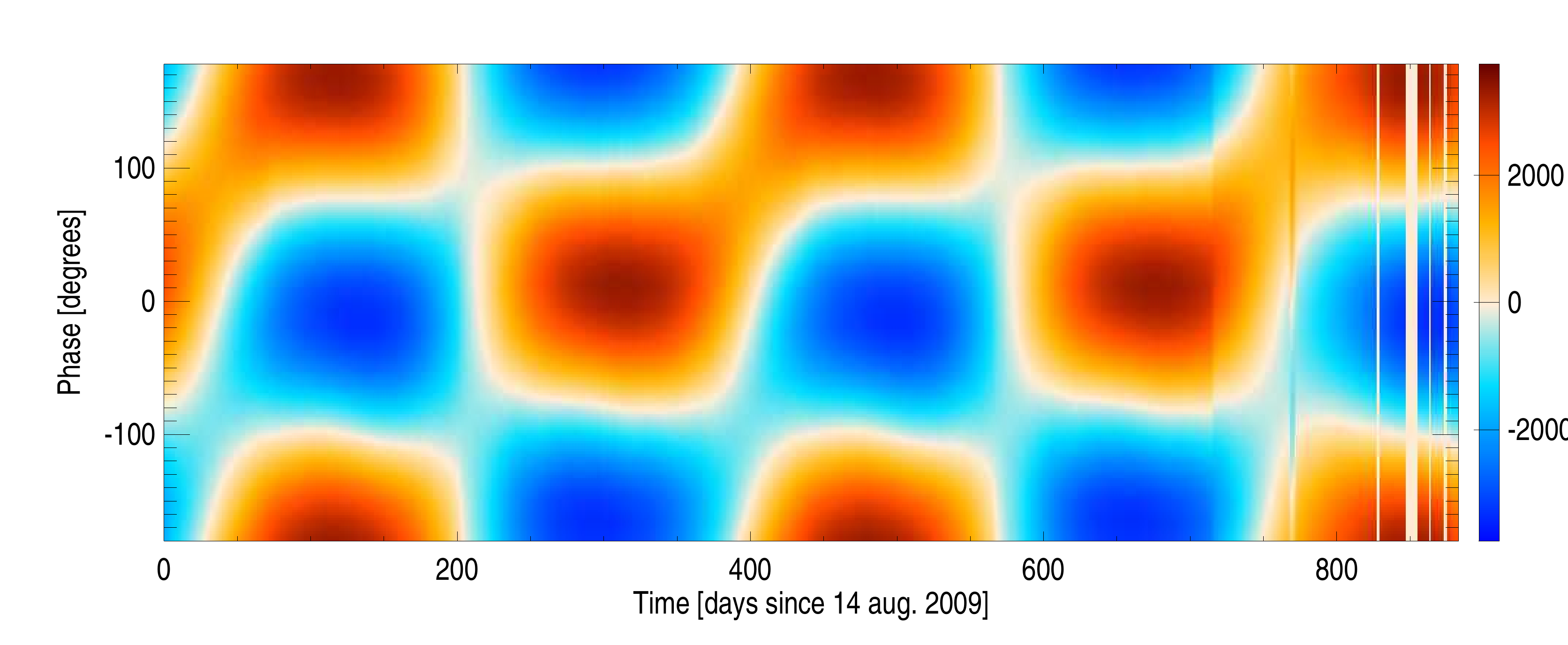}\\  %
\includegraphics[width=0.75\textwidth]{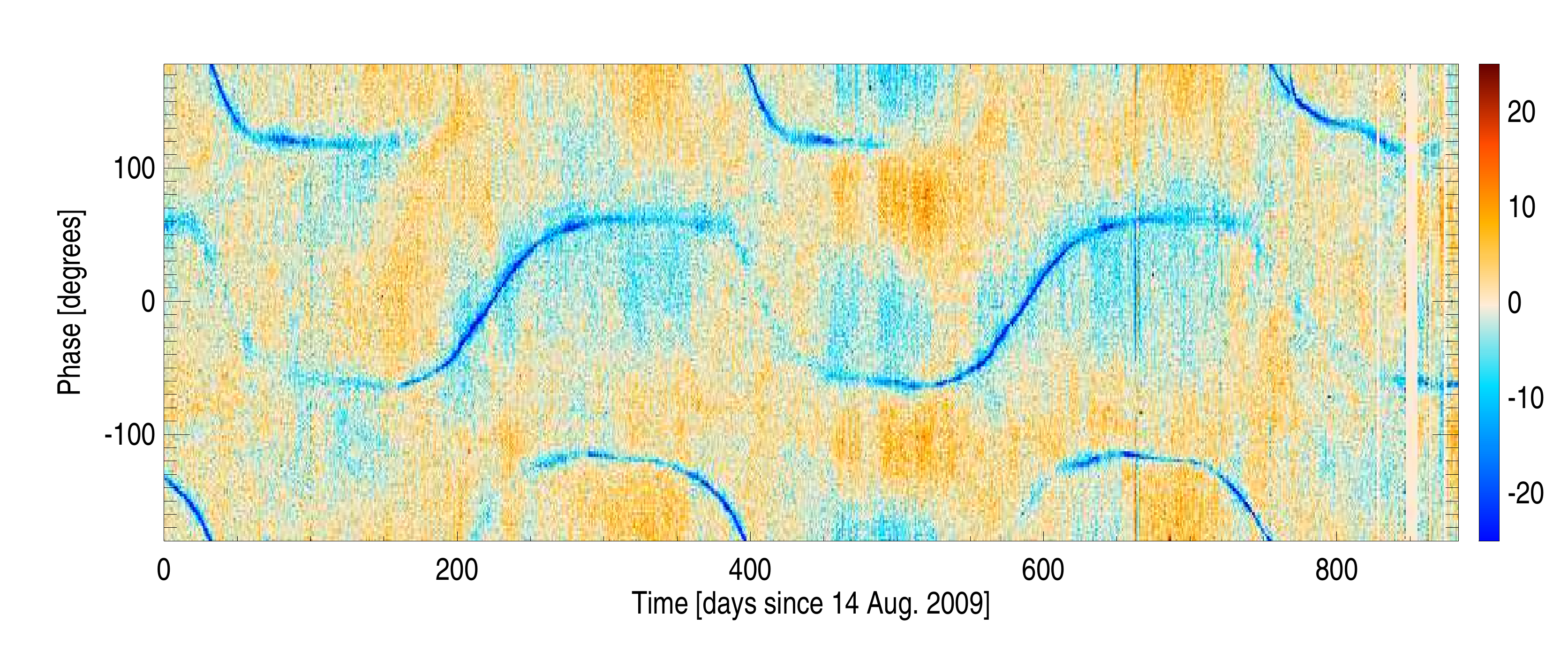}
\caption{{\it Bottom:} Distribution of the residuals in $\muK_{\mathrm{CMB}}$, computed using Eq.~\ref{eq:residual}, for detector 143-1a, plotted versus observation  date and satellite rotation phase. {\it Top}: Expected pattern for the solar dipole, in $\muK_{\mathrm{CMB}}$. Comparison of the two plots provides a check of the level of residual gain variation after applying the \bogopix\ gains. In the residuals the sharp features (dark blue) correspond to the Galaxy observations, where band-pass effects have not been corrected. 
\label{fig:resid_calib}}
\end{figure*}

To evaluate the accuracy of the apparent gain variation correction coming from \bogopix\ we compute, for each detector, the residual difference $R$ between the HPR data $d$ and a model including  
the destriping offsets $o_r$, the HFI $I$, $Q$ and $U$ maps, the  dipoles $t_D$ (orbital and solar) and the calibration parameters (relative ring-by-ring gains $g_r$ from \bogopix,  overall gain  $\tilde{G}^{\mathrm{SD}}$ based on the solar dipole, and zero point $z$ derived as described above). This corresponds, for each HPR sample $i$ of each ring $r$, and pixel $p$ to:
\begin{multline}
R_i\ =  (d_i-o_r)/(g_r \cdot{\tilde{G}^{\mathrm{SD}}}) -t_D -I_p \\
          - \frac{1-\eta}{1+\eta}\left( Q_p\ \cos{2\psi_i}  + U_p\ \sin{2\psi_i} \right)  -z 
\label{eq:residual}
\end{multline}
We display these residuals as a function of the rotation phase, i.e., the angle between the direction of the pixel in the HPR and the satellite velocity, in Fig.~\ref{fig:resid_calib}. In this representation, the orbital dipole extrema will be found at fixed phases 0 and $\pm\pi$. The solar dipole will present a modulated pattern, also  illustrated in Fig.~\ref{fig:resid_calib}.
As the solar dipole is the brightest component of the sky emission, its pattern in the residuals is a good indication of inaccuracy of the gain variation correction. This may also capture additional time variable signals that would not be accounted for in our processing, for example the primary spillover pick-up. 
The areas where the Galactic emissions dominate show up as outliers in
these residuals, for several reasons. First, they correspond to regions
were intra-pixel gradients are large, and will leave some imprint due
to the individual scanning trajectories of each detector. More
importantly, they present emission spectra different from that of the
CMB, on which we calibrate. Integrated over each detector's bandpass,
this will translate into an apparent brightness difference. At this
stage, we do not apply colour corrections to get rid of such effects,
considering that they can be minimized by a proper selection of the
sky area (i.e., avoiding the Galactic plane). 
Finally, imperfections in the time response of the detectors and in the pointing reconstruction will also induce larger residuals in the Galactic plane. Masking these regions, using a 40\,\% Galactic mask, we checked that, for all the 100--217\,\GHz\ detectors, the maximum level of the residuals we observe would correspond to a remaining gain variation lower than 0.3\,\% (i.e., residuals  lower than $10\ \mu {\mathrm{K}}_{\mathrm{CMB}}$).

\begin{figure}
\begin{center}
\begin{tabular}{@{}cc@{}}
\includegraphics[width=.95\columnwidth,bb=70 10 539 345]{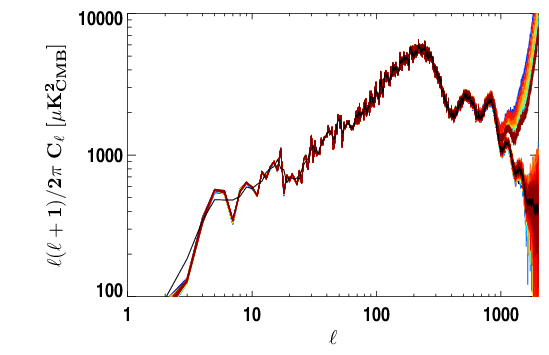}\\
\includegraphics[width=.95\columnwidth,bb=70 10 539 345]{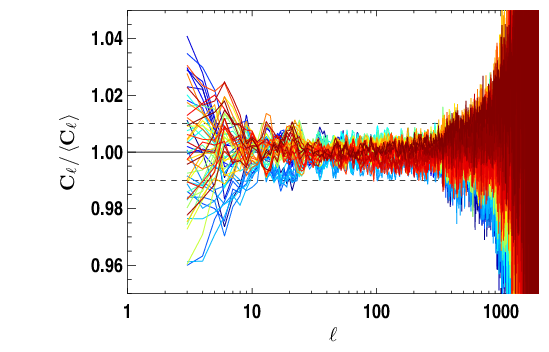}
\end{tabular}
\end{center}
\caption{Auto- and pseudo-cross-spectra obtained from the eleven
  143\,\GHz\ HFI detectors corrected for the beam  (top) and their ratio
  with respect to the average of the pseudo-cross-spectra
  (bottom). This average is indicated in black in the top panel. Each
  detector pair is shown in  a different colour. Note the
  noise suppression in the cross-spectra above $\ell\sim800$.}
\label{fig:rel_calib_spectra}

\end{figure}

\subsection{Intra-frequency calibration checks}
\label{sec:rel_calib_xspec}

We have checked the relative calibration of the detectors within a given frequency channel  using 
pseudo-cross-power spectra.  We start from the single-detector temperature maps, neglecting polarization. We mask sky areas where the Galactic emissions are large, keeping 40\,\% of the sky for frequencies lower than 300\,\GHz\ and 30\,\% above. 
We build the pseudo-cross spectra of this set of maps, using {\tt Xspect}~\citep{tristram-xspec2005}. 
We correct each pseudo-spectrum for its beam window function \citep{planck2013-p03c}. 
We then focus on the location of the first acoustic peak,  
so that results are not biased by beam uncertainties. 
For example, the set of
spectra we obtain for the 143\,\GHz\  HFI detectors is shown in
Fig.~\ref{fig:rel_calib_spectra}. Finally, we fit the recalibration
coefficients that minimize the differences between these spectra, for
$\ell$ in the range [25, 300]. For 545 and 857\,\GHz\ we apply a colour
correction for the band-pass mismatch between
detectors, assuming the {\it IRAS} spectral convention.
The relative calibration coefficients found  with this method should be considered as upper limits on the relative calibration precision of HFI, as we neglect polarization in this analysis.
They are given for all frequencies in Table~\ref{tab:max_relcalib_coeff}. For frequencies below 217\,\GHz\ the relative calibration accuracy is better than 0.4\,\%. These relative accuracies are consistent with the systematic uncertainties estimated in the previous section.

\begin{table}[h]
\caption{Maximum absolute value of the relative calibration coefficients fitted on pseudo-spectra similar to those of Fig.~\ref{fig:rel_calib_spectra}, between detectors of each frequency. These values are upper limits on the relative calibration errors within each channel (i.e., between all bolometers of a given channel).}
\begin{center}
\begin{tabular}{@{} l c c c c c c @{}}
\hline
\hline
\noalign{\vskip 2pt}
Frequency $[$GHz$]$ & 100 & 143 & 217 & 353  & 545 & 857 \\
Calibration [\%] & 0.39 & 0.28 & 0.21 & 1.35 & 1.3 & 1.4 \\
\hline
\end{tabular}
\end{center}
\label{tab:max_relcalib_coeff}

\end{table}

\begin{figure}[htbp]
\begin{center}
\begin{tabular}{@{}c@{}}
\includegraphics[width=0.45\textwidth,bb= 80 40 890 600 ]{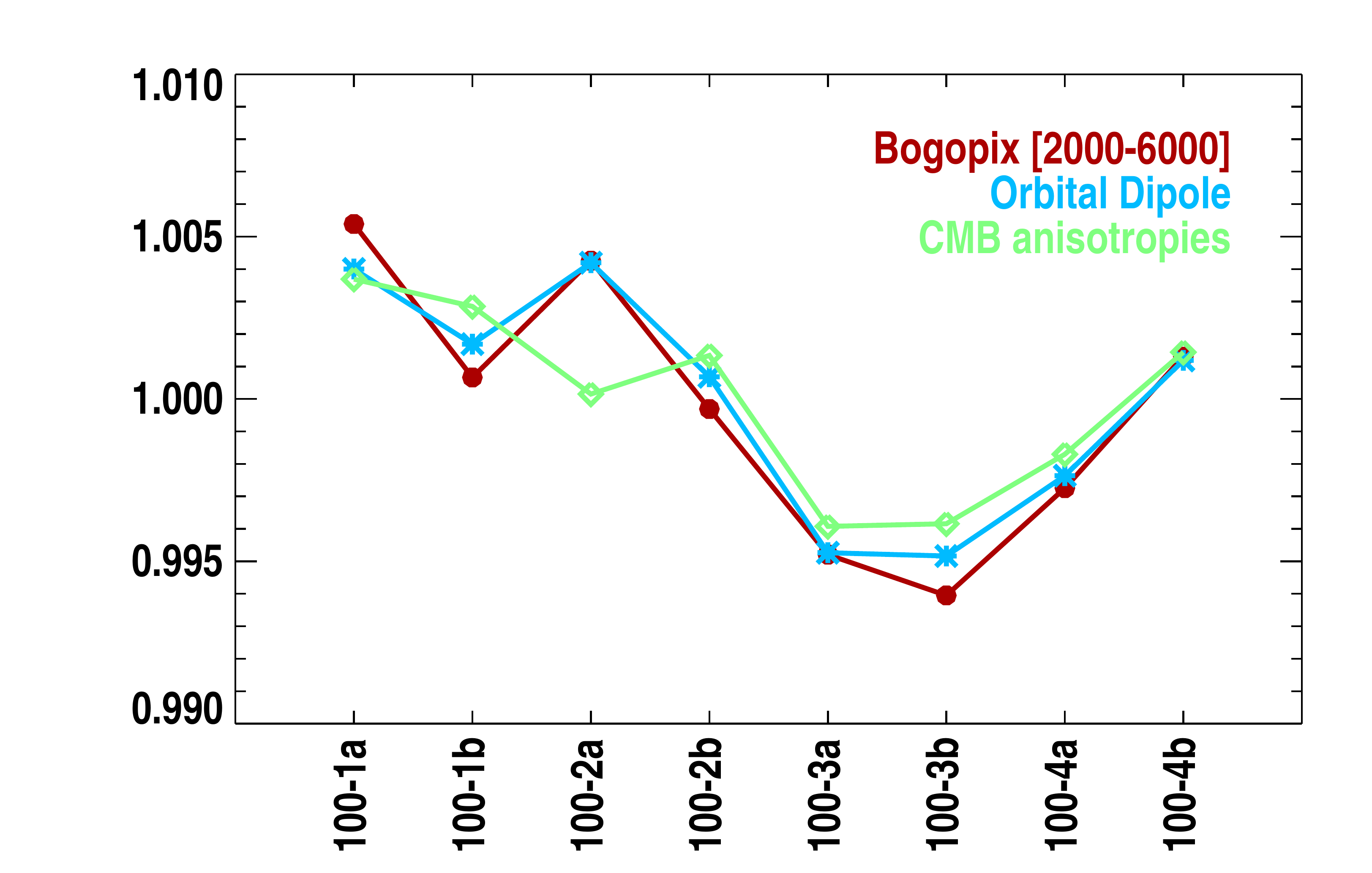}\\
\includegraphics[width=0.45\textwidth,bb= 80 40 890 600 ]{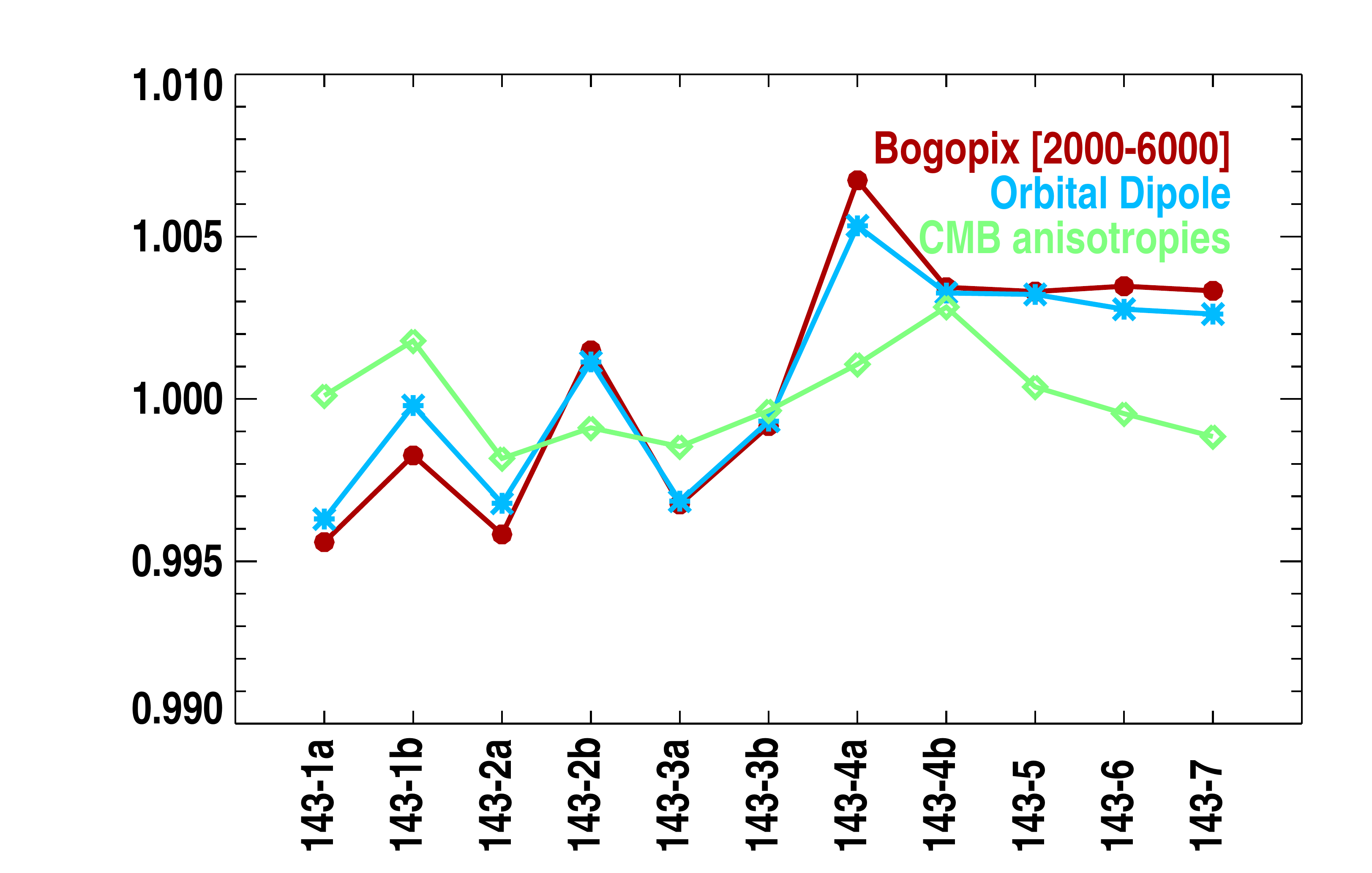}\\
\includegraphics[width=0.45\textwidth,bb= 80 40 890 600 ]{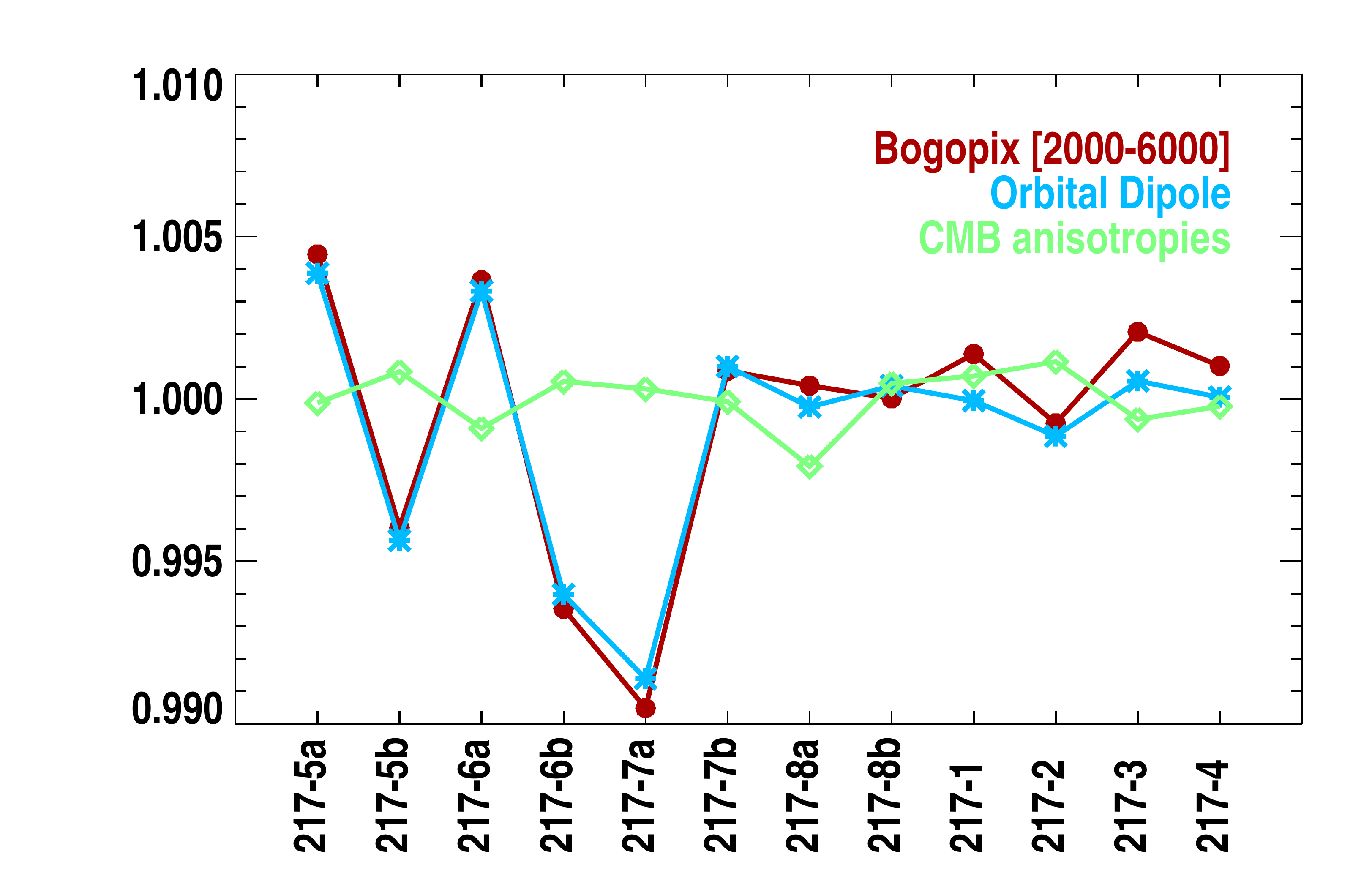}
\end{tabular}
\end{center}
\caption{Relative calibration coefficients found when calibrating on the orbital dipole (constant gain, light blue, with \bogopix\ in red) and using CMB anisotropies (see Sect.~\ref{sec:rel_calib_xspec}, in green), with respect to the Solar dipole gains, used to build the HFI maps. As ADC non-linearities are not corrected for, calibration systematics depend on the amplitude on the signal used to check for them. The amplitudes of such effects are within the systematic uncertainties quoted in Table~\ref{tab:max_relcalib_coeff}.} 
 \label{fig:adc_syst_error_summary}

\end{figure}

In Fig.~\ref{fig:adc_syst_error_summary}, we  compare the relative calibration coefficients derived from the pseudo-cross spectra, for all 100, 143 and 217\,\GHz\ detectors, with the relative differences between gains based on solar and orbital dipole calibration methods  (see Sect.~\ref{sec:corr_gain}). Both orbital dipole methods are affected by the same systematics, namely the ADC non-linearities. 
CMB anisotropies are well intercalibrated between detectors, using solar dipole calibration. This reinforces the choice of the solar dipole calibration.

\subsection{Inter-frequency and absolute calibration checks for CMB-dominated channels}
\label{sec:cmb_chan_calib_checks}
In this section, we describe the checks performed to study the calibration accuracy for CMB channels. 

\subsubsection{Pseudo-cross-power spectrum analysis\label{subs:interfre_xsp}}

We applied a technique similar to that presented in Sect.~\ref{sec:rel_calib_xspec} to assess the HFI inter-frequency relative calibration for combined maps, at frequencies where the CMB dominates at high Galactic latitudes.
We built pseudo-power spectra from the temperature maps for 100, 143, and 217\,\GHz,  applying the beam  correction described in \cite{planck2013-p03c}. 
As above, we determined the cross-calibration coefficients 
that minimize the difference between the pseudo-cross-power spectra of the HFI maps for $\ell$ in the range [25, 300]. 
Results from this analysis are shown in
Table~\ref{tab:xpol_freqcalib_coeff}. We see from these numbers that the internal relative calibration precision within HFI is better than $\pm0.15\,\% $. 

\begin{table}[h]
\caption{Cross-calibration coefficients that minimize the dispersion of the HFI temperature cross-spectra around their common mean.}
\begin{center}
\begin{tabular}{@{} l c c c @{}}
\hline
\hline
\noalign{\vskip 2pt}
Frequency [GHz] & 100 & 143 & 217 \\
\noalign{\vskip 2pt}
Calibration	 & 1.002 & 	0.999	  & 0.999 \\
\hline
\end{tabular}
\end{center}
\label{tab:xpol_freqcalib_coeff}
\end{table}

\subsubsection{Solar-dipole parameter fits}

We also studied the calibration accuracy using fits of the CMB dipole parameters on HFI maps. To perform this test, we used maps built without dipole subtraction. Such fits are likely to be biased in the presence of foregrounds, in particular due to the intrinsic dipole of the  Galactic emissions. We therefore used a template fitting
method to subtract dust emission; our dust template is based on
{\it IRAS} data~\citep{neugebauer1984}. We masked 10\% of the sky, based on Galactic dust and CO emission, as well as point sources, before fitting the amplitude and direction of the CMB dipole. We recover the \WMAP\ dipole amplitude measurement at the level of 0.1\,\% or better in all cases (Table \ref{tab:wmap_dipole_fits}). 
The direction, perhaps more affected by foreground residuals, is reconstructed within about $10\arcm$. These results are in agreement with the residual dipole measurements presented in Sect.~\ref{sec:gal_zero_levels}, which might be more sensitive to foreground removal and masking.

\begin{table}[h]
\caption{Differences between the CMB dipole parameters fitted on the HFI maps with those measured by \WMAP. The typical statistical errors on these fits are $\sim 0.01\,\% $ for the amplitude and 
less than 1\arcm\ for the direction.} 
\begin{center}
\begin{tabular}{@{}c c c c@{}}
\hline
\hline
\noalign{\vskip 2pt}
Frequency &  Amplitude  & Longitude   & Latitude \\
 $[$GHz$]$& [\%]  & [\arcm] & [\arcm] \\
\noalign{\vskip 2pt}
\hline
\noalign{\vskip 2pt}
100          	& $-$0.122	& ~2.30	& 11.09 \\
143	                & $-$0.074 &	~3.00  &		11.91 \\
217	                & $-$0.091 &	$-$5.10	 &	12.79 \\
\hline
\end{tabular}
\end{center}
\label{tab:wmap_dipole_fits}
\end{table}

\subsubsection{Calibration checks using component separation methods}

Finally, calibration consistency checks have been performed using
component separation tools. In particular, the
{\tt SMICA} component separation method \citep{cardoso2008} has been used to fit relative calibration coefficients for each frequency (including LFI data) on the CMB anisotropies \citep{planck2013-p06}. The foreground model is a non-parametric 4-dimensional model, meaning that the foregrounds are represented by four templates with arbitrary emission laws, arbitrary angular spectra, and arbitrary correlations (2- and 3-dimensional fits were also performed with compatible results). Relative-calibration coefficients between frequency power spectra obtained using this method are summarized in Table~\ref{tab:smica_freqcalib_coeff}.  They agree, within errors, with the results shown in Tables~\ref{tab:xpol_freqcalib_coeff} and \ref{tab:wmap_dipole_fits}. It should be noted that for frequencies $> 353$\,\GHz, Rayleigh scattering, not included in such studies,  will distort the CMB anisotropies used to derive such cross-calibrations, at the few percent level~\citep{yu2001}. Therefore cross-calibration coefficients found for 353 and 545\,\GHz, which are of the same order, should be considered as estimates of systematic cross-calibration uncertainties, rather than genuine corrections of our maps. 
 Such studies are routinely incorporated in \Planck\ likelihood minimizations\citep{planck2013-p08}, and more results are shown in \cite{planck2013-p08}. Comparisons with LFI and \WMAP\ are presented in \cite{planck2013-p11}.
 
\begin{table}[h]
\caption{Cross-calibration coefficients of the HFI sky maps at each frequency, with respect to the 143\,\GHz\ map, found with the {\tt SMICA} component
  separation method, with errors derived from a Fisher matrix
  analysis.}
\begin{center}
\begin{tabular}{@{} c c c @{}}
\hline
\hline
\noalign{\vskip 2pt}
Frequency &  Relative calibration  & Fisher errors  \\
 $[$GHz$]$&  --  & [\%] \\
\noalign{\vskip 2pt}
\hline
100 & 0.999 & 0.2\\
143 & 1 & 0.2\\
217 & 1.000 & 0.2 \\
353 & 0.993 & 0.3 \\
545 & 1.05 & 3.5 \\
\hline
\end{tabular}
\end{center}
\label{tab:smica_freqcalib_coeff}
\end{table}

\subsection{HFI/SPIRE cross-calibration on diffuse emission}
At high frequency, uncertainties in the SEDs of the  astrophysical components, together with their variation across the sky, make extensive calibration checks as performed in Sect.~\ref{sec:calib_time_stab} and \ref{sec:cmb_chan_calib_checks} more difficult. 
However, we can study the cross-calibration between HFI and other data sets, like  {\it Herschel}-SPIRE~\citep{griffin2010}.
HFI and SPIRE have two very close frequency channels: 857\,\GHz\ for \Planck\ versus 350\,$\mu$m for SPIRE, and 545\,\GHz\ versus 500\,$\mu$m. 
To compare the HFI and SPIRE brightness, we used nine large SPIRE public Galactic fields (for a total of about 75 deg$^2$), with mean brightness ranging from 1.8 to 285\,MJy\,sr$^{-1}$ at 857\,\GHz.

For each field we create SPIRE 350 and 500\,$\mu$m maps using the
Herschel Interactive Processing Environment pipeline {\tt HIPE} v9.1. We
applied the relative gain correction for extended emission and used
the destriper module. SPIRE data are calibrated using Neptune and are
given in Jy beam$^{-1}$ for the convention $\nu I_{\nu}={\mathrm{constant}}$. To convert the point source calibration to an extended emission calibration, we use the SPIRE beam solid angles of 822\,arcsec$^2$\ and 1768\,arcsec$^2$\ at 350 and 500~$\mu$m respectively.\footnote{\url{https://nhscsci.ipac.caltech.edu/sc/index.php/Spire/PhotBeamProfileDataAndAnalysis}} These values are also derived using the $\nu I_{\nu}={\mathrm{constant}}$ convention. SPIRE maps are then convolved with the HFI beam window function.

To compute the colour corrections, we use the SPIRE Relative Spectral Response Functions (RSRFs).\footnote{\url{https://nhscsci.ipac.caltech.edu/sc/index.php/Spire/PhotInstrumentDescription}} The SPIRE beam FWHM values vary as $\nu^\gamma$ with frequency, where $\gamma = 0.85$ at both 350 and 500\,$\mu$m~\citep{griffin2013}. To take this effect into account, we multiply the SPIRE RSRFs by $ \nu^{-2\gamma}$ and renormalize them.
Colour-correction factors, to convert SPIRE monochromatic flux
densities into HFI-like monochromatic flux densities, are computed
assuming the real source spectrum is a modified blackbody of a given
temperature and emissivity index: $B_{\nu}(T) \times \nu^{\,\beta}$.  The temperature $T$ and emissivity index $\beta$ are taken from the full-sky $T$ and $\beta$ maps available in the \Planck\ Legacy Archive. 

\begin{figure}
   \centering
   \includegraphics[width=9cm]{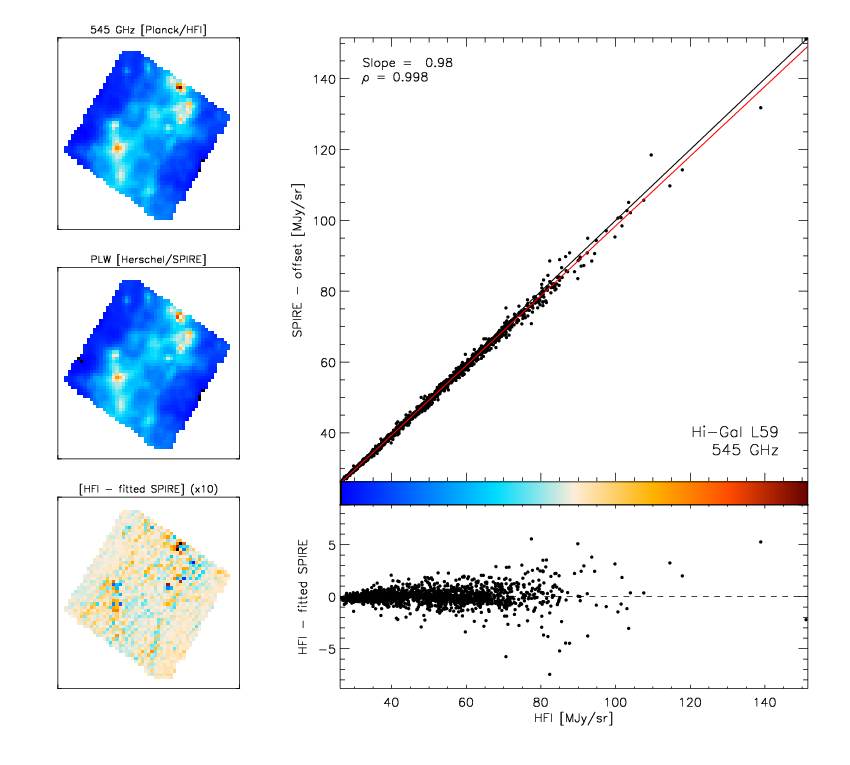}
   \caption{\label{spire_hfi}  SPIRE/HFI pixel-to-pixel comparison at 545\,\GHz\ in one 6.3 deg$^{2}$ field from the Herschel infrared Galactic Plane Survey \citep{molinari}. The red line is the result of a linear fit while the black line has a slope of unity. On the left are shown the HFI and SPIRE maps, together with the difference of the two. The difference is displayed between $-$6.3 and 6.3 MJy sr$^{-1}$.}
   \end{figure}

We estimate the agreement between the diffuse emission measurements from HFI and SPIRE by computing their correlation.
An example of  a scatter plot for one field and one frequency is shown in Fig~\ref{spire_hfi}. In all fields,
HFI and SPIRE measurements correlate very well, with an average Pearson
correlation coefficient of 0.998. The dispersion across the linear fit
ranges from about $2\,\%$ to $8\,\%$ of the mean brightness of the field. The SPIRE/HFI gain ratios are $0.972$ at 545 and 0.936 at 857\,\GHz. At 545\,\GHz, the agreement between the HFI and SPIRE absolute calibration is very good. At 857\,\GHz, we observe a systematic trend, with SPIRE being lower than HFI by 6.4\,\%. {  This difference is just within the joint uncertainties of the absolute calibration of the two instruments, which are 5\% for HFI (see Sect.\ref{subsect: planet_calib_uncert}), and 2\,\% for SPIRE (see SPIRE observer manual\footnote{\url{http://herschel.esac.esa.int/Docs/SPIRE/html/spire\_om.html}}), putting aside the 5\,\% systematic uncertainty of the Planet's models. We note however that the version of the SPIRE pipeline that
was used for this comparison was based on the Neptune ESA2 model
whereas HFI used the ESA3 model, which is of order 1.5\,\% higher 
over the frequency range of interest. Moreover, contrary to SPIRE, we
also use Uranus to calibrate. Considering these additional sources of
bias to the relative HFI-to-SPIRE calibration, we are not unduly worried about the current difference between
HFI and SPIRE at 857\,\GHz.}

\subsection{Map noise level assessment}

When combining detector data to build frequency maps, we apply an inverse noise weighting scheme. The weights we use are derived from the noise levels measured from clean TOIs together with the calibration coefficients. The resulting noise level in the combined maps is therefore a consistency check of the relative calibration between detectors, since a mis-calibration would result in additional noise, given the slightly different scanning path and redundancies of the detectors.  

\begin{figure*}[htbp]
\centering
\begin{tabular}{cccc}

\includegraphics[width=0.22\textwidth,bb=290 131 770 431]{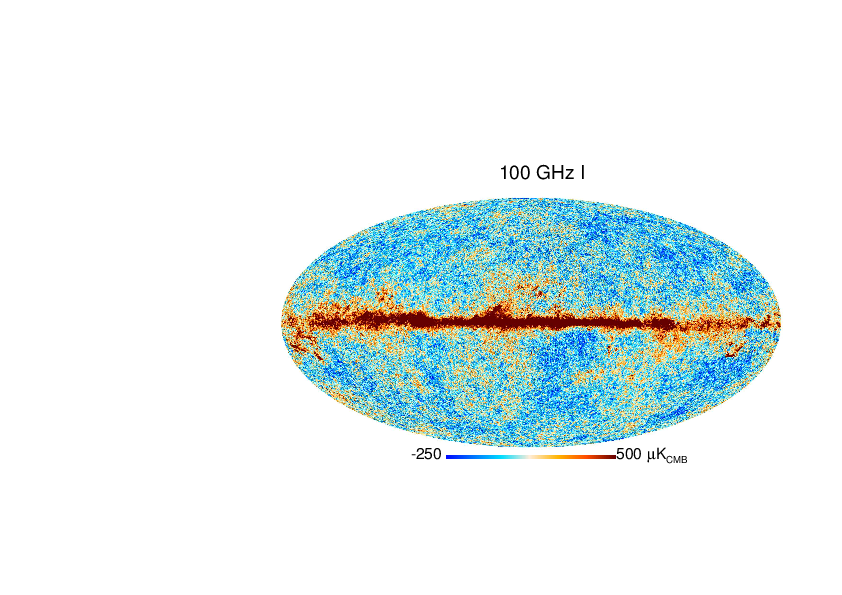}
& 
\includegraphics[width=0.22\textwidth,bb=290 131 770 431]{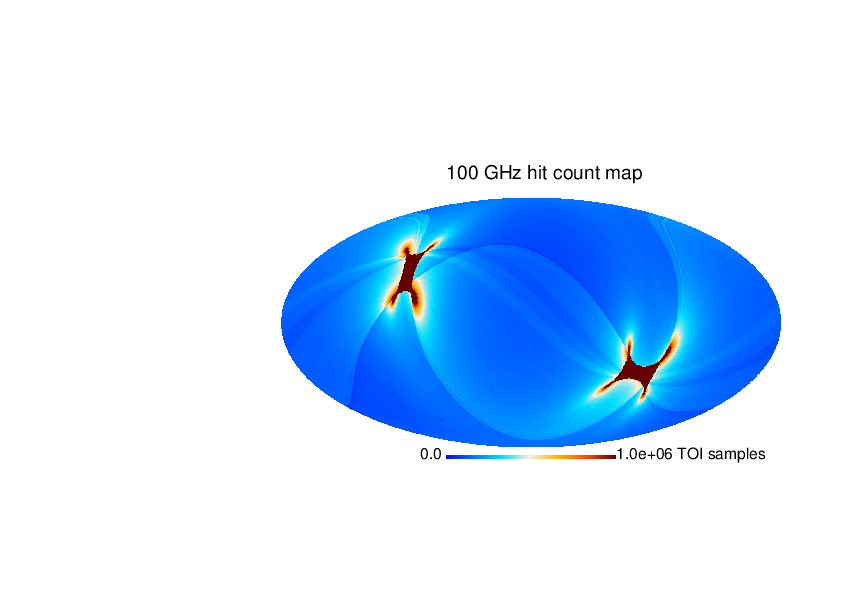}
& 
\includegraphics[width=0.22\textwidth,bb=290 131 770 431]{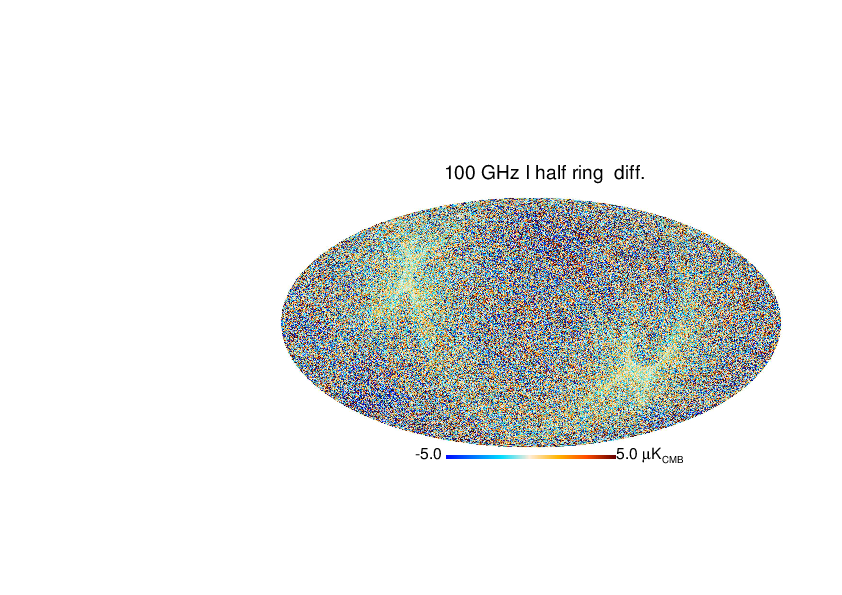}
& %
\includegraphics[width=0.22\textwidth,bb=290 131 770 431]{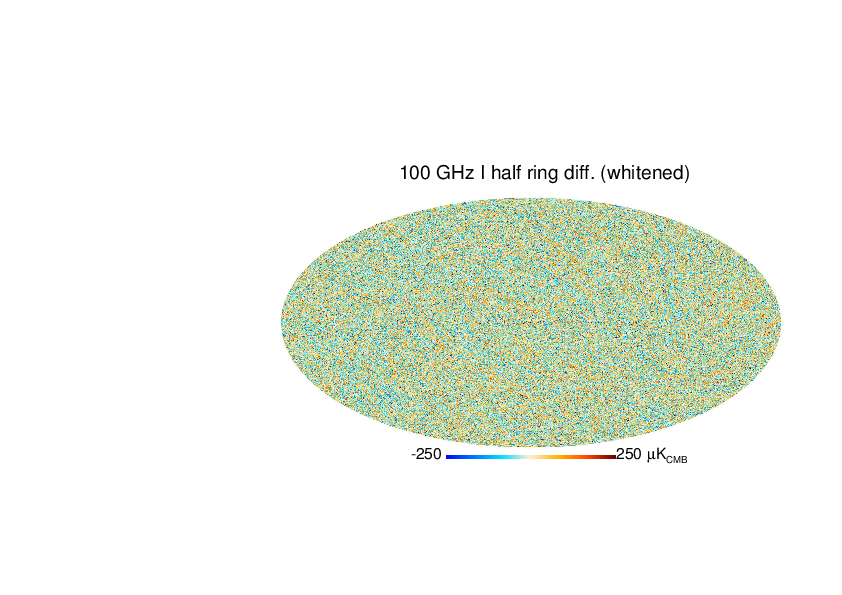}
\\
\includegraphics[width=0.22\textwidth,bb=290 131 770 431]{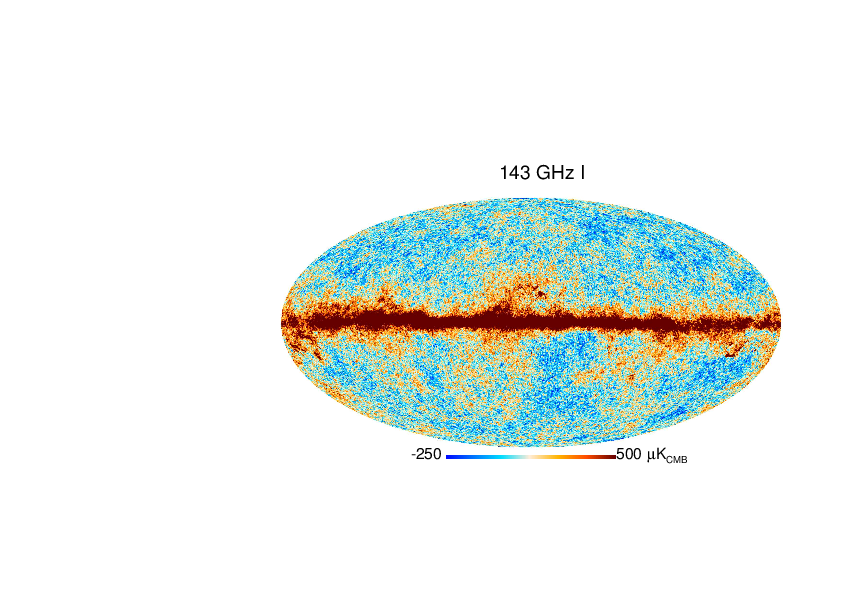}
& \includegraphics[width=0.22\textwidth,bb=290 131 770 431]{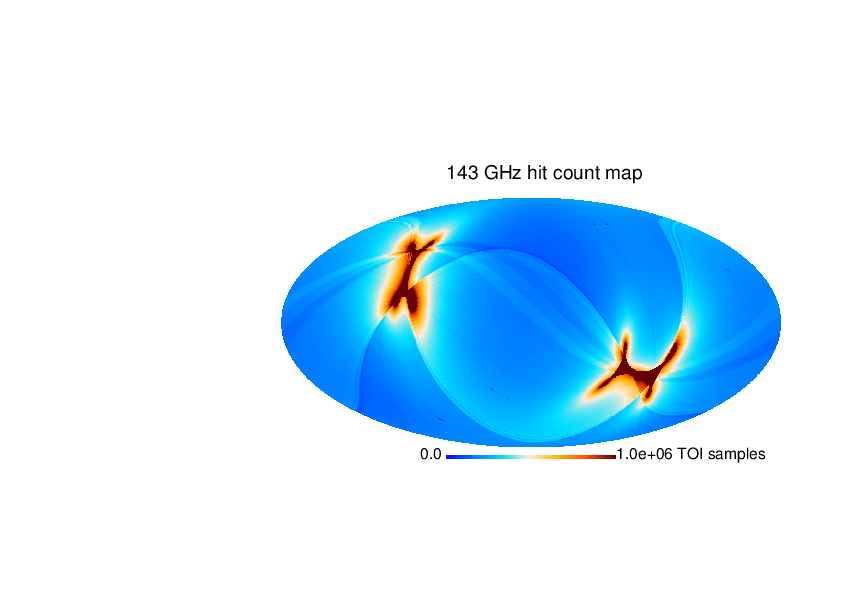}
& \includegraphics[width=0.22\textwidth,bb=290 131 770 431]{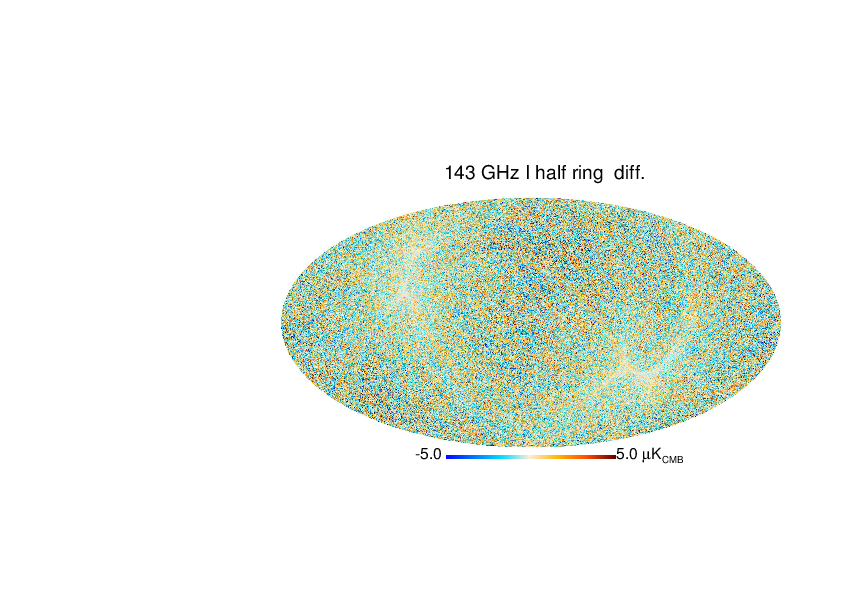}
& \includegraphics[width=0.22\textwidth,bb=290 131 770 431]{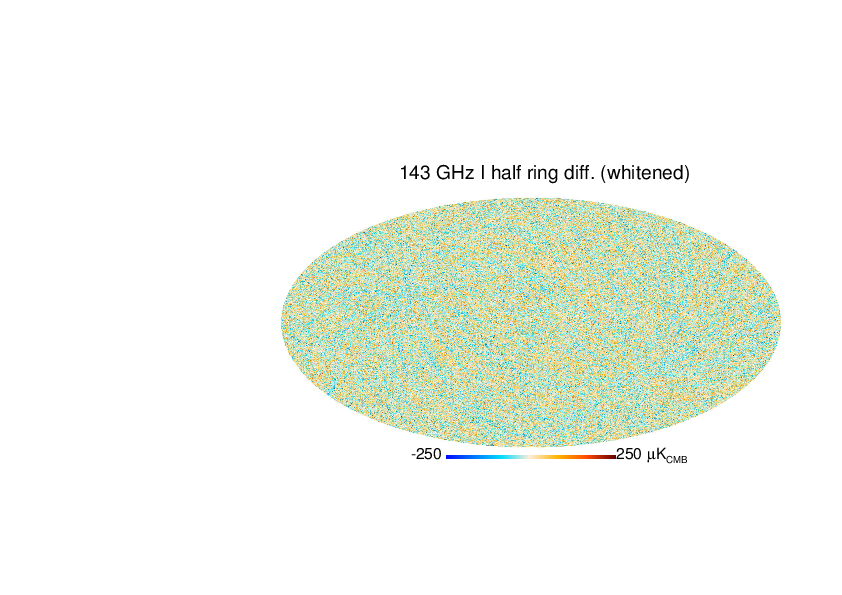}
\\
\includegraphics[width=0.22\textwidth,bb=290 131 770 431]{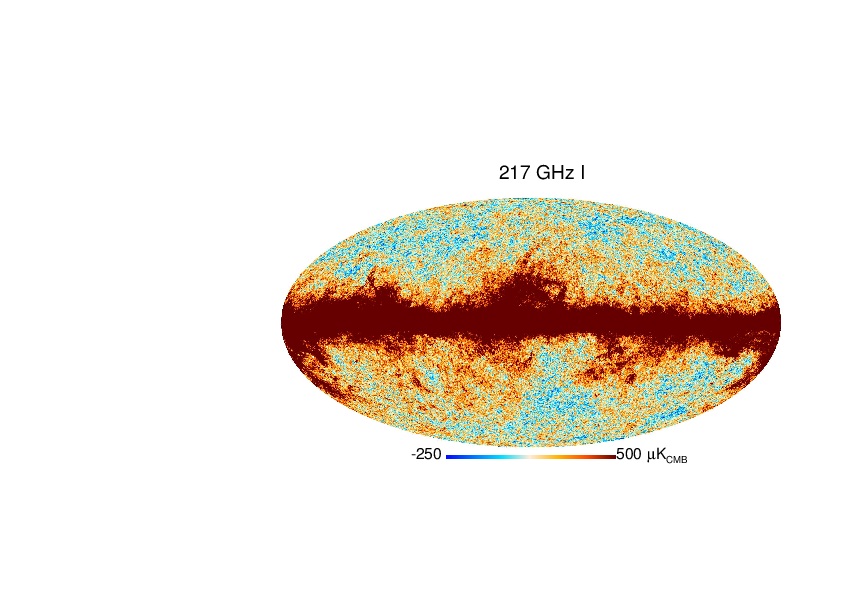}
& \includegraphics[width=0.22\textwidth,bb=290 131 770 431]{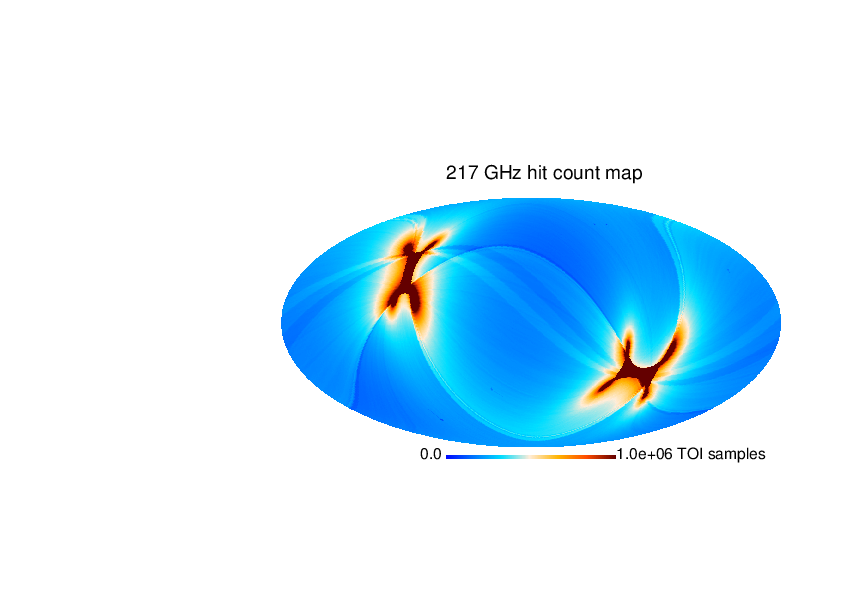}
& \includegraphics[width=0.22\textwidth,bb=290 131 770 431]{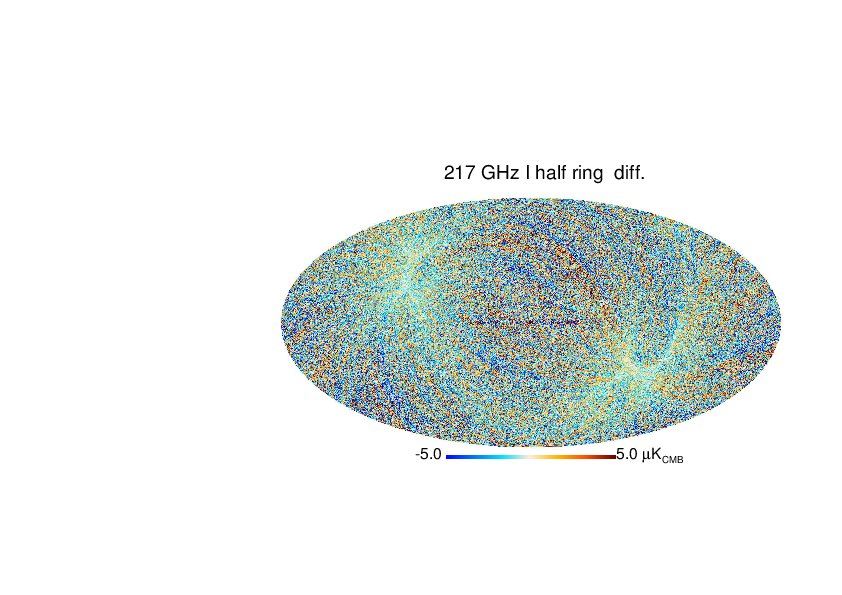}
& \includegraphics[width=0.22\textwidth,bb=290 131 770 431]{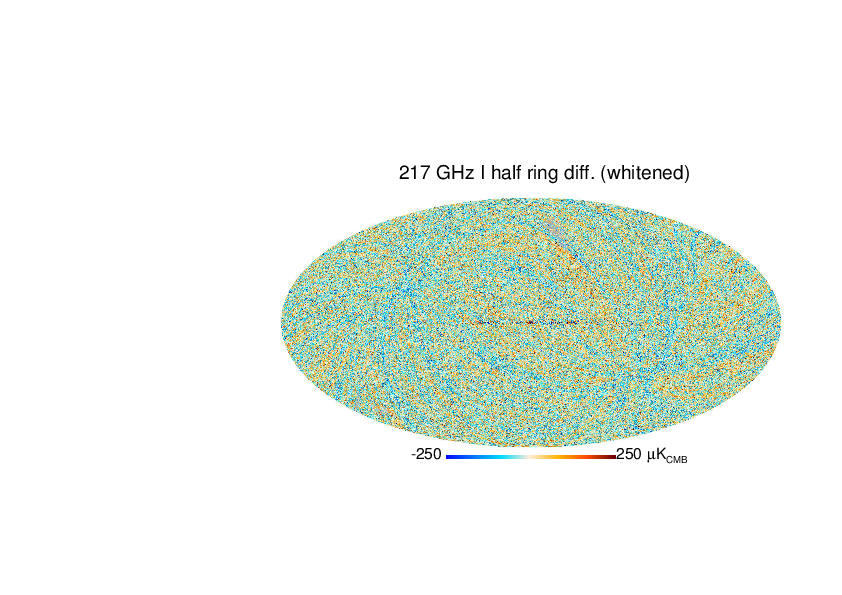}
\\
\includegraphics[width=0.22\textwidth,bb=290 131 770 431]{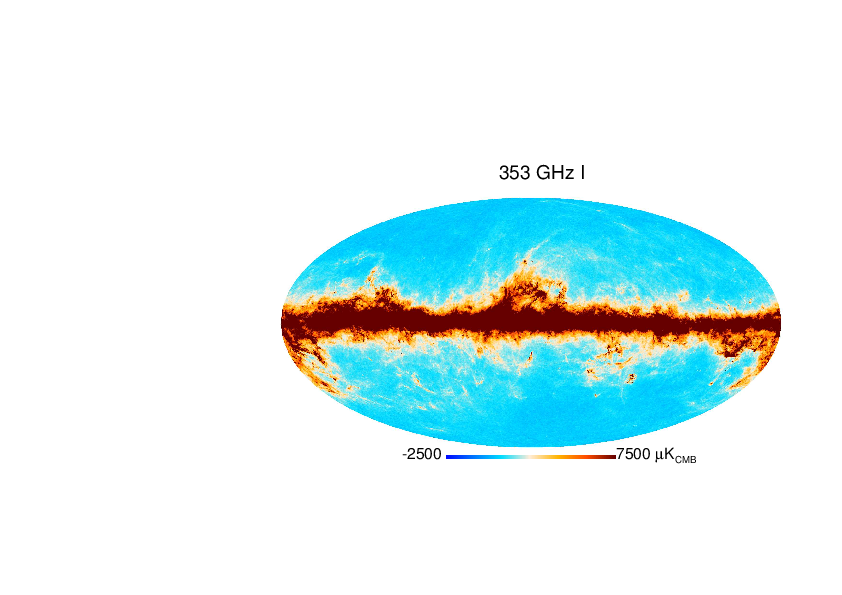}
& \includegraphics[width=0.22\textwidth,bb=290 131 770 431]{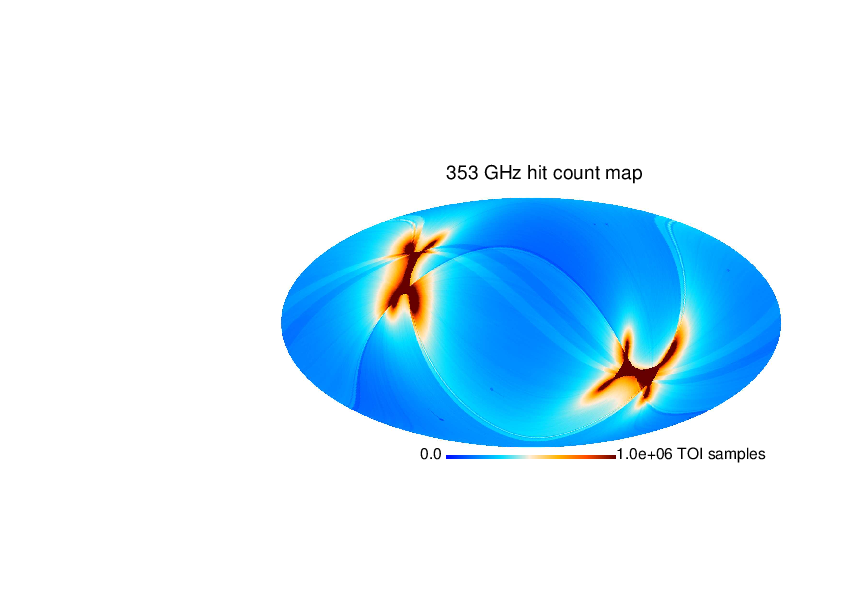}
& \includegraphics[width=0.22\textwidth,bb=290 131 770 431]{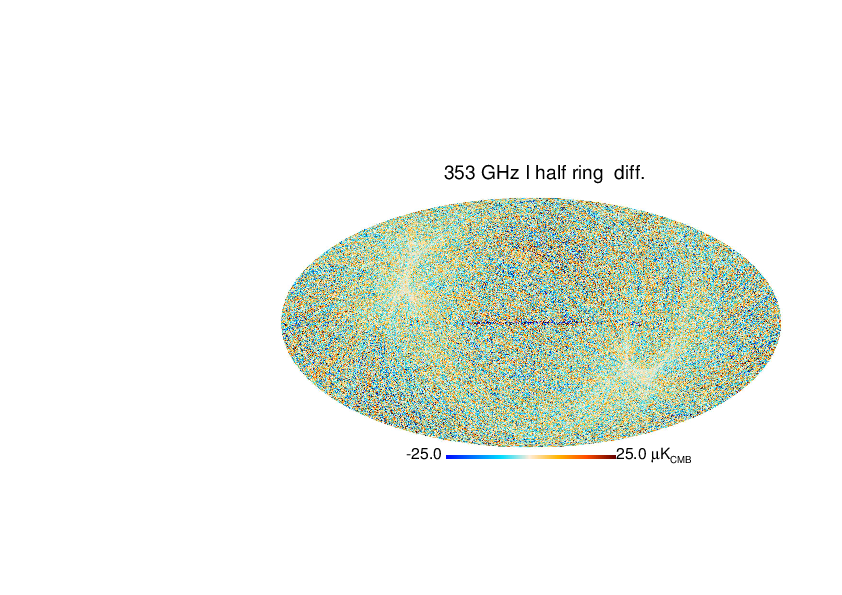}
& \includegraphics[width=0.22\textwidth,bb=290 131 770 431]{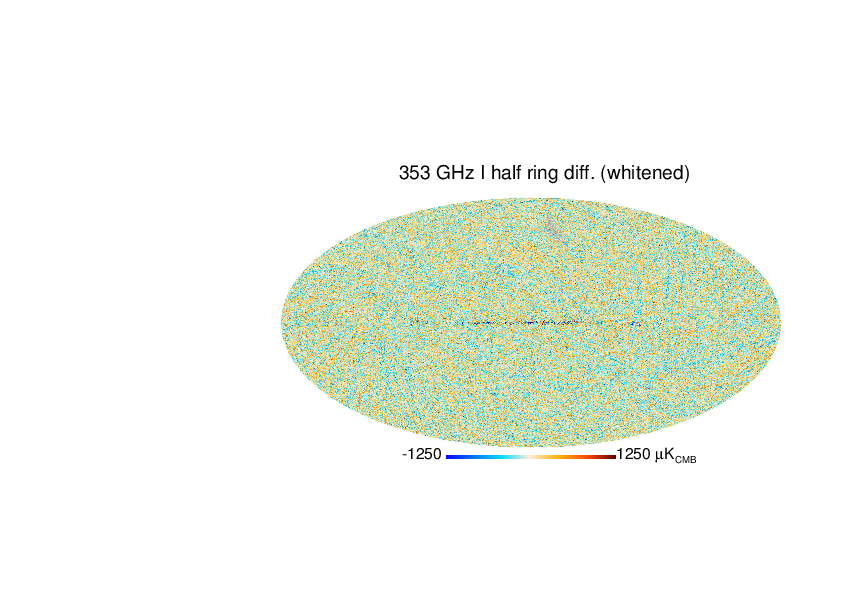}
\\
\includegraphics[width=0.22\textwidth,bb=290 131 770 431]{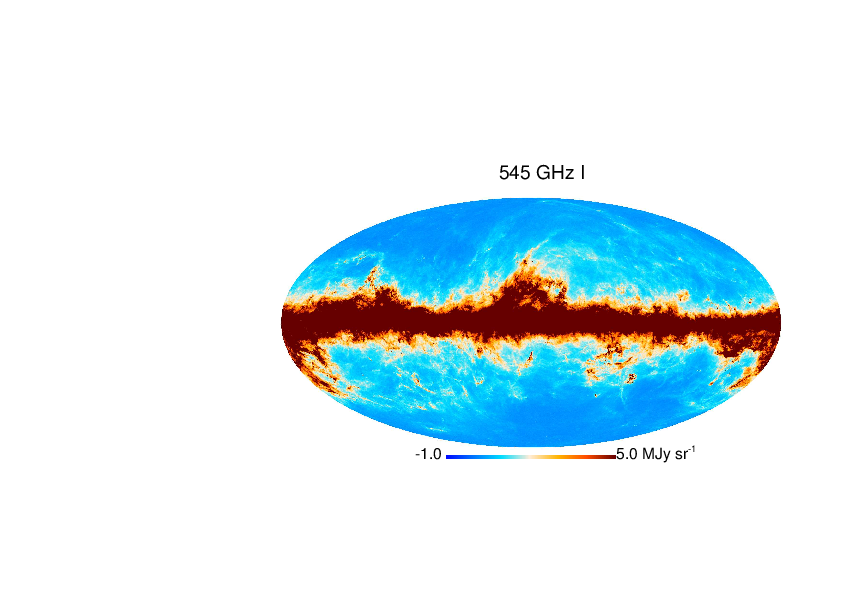}
& \includegraphics[width=0.22\textwidth,bb=290 131 770 431]{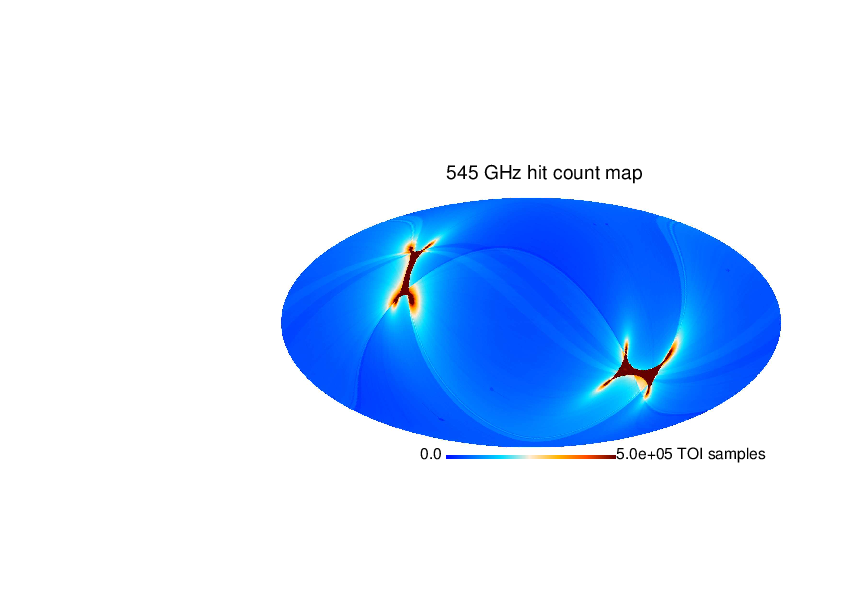}
& \includegraphics[width=0.22\textwidth,bb=290 131 770 431]{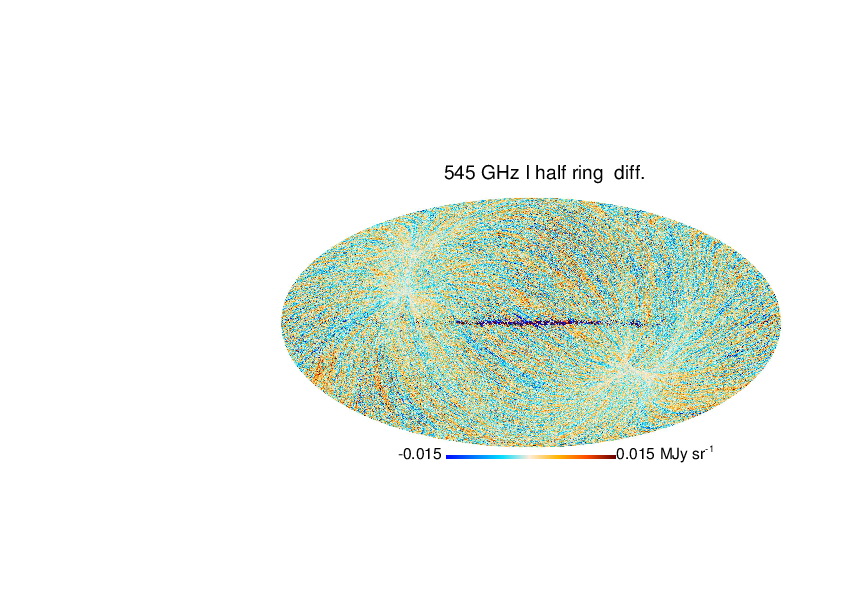}
& \includegraphics[width=0.22\textwidth,bb=290 131 770 431]{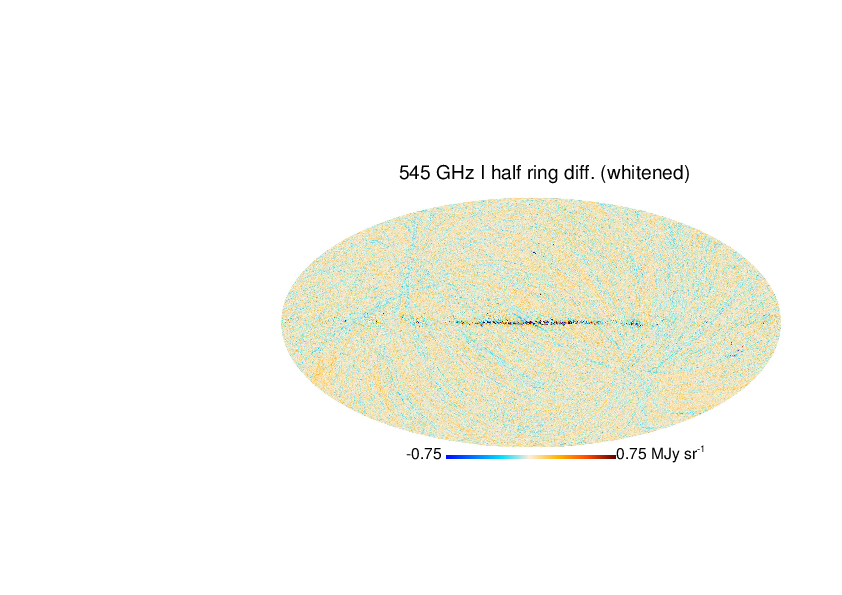}
\\
\includegraphics[width=0.22\textwidth,bb=290 131 770 431]{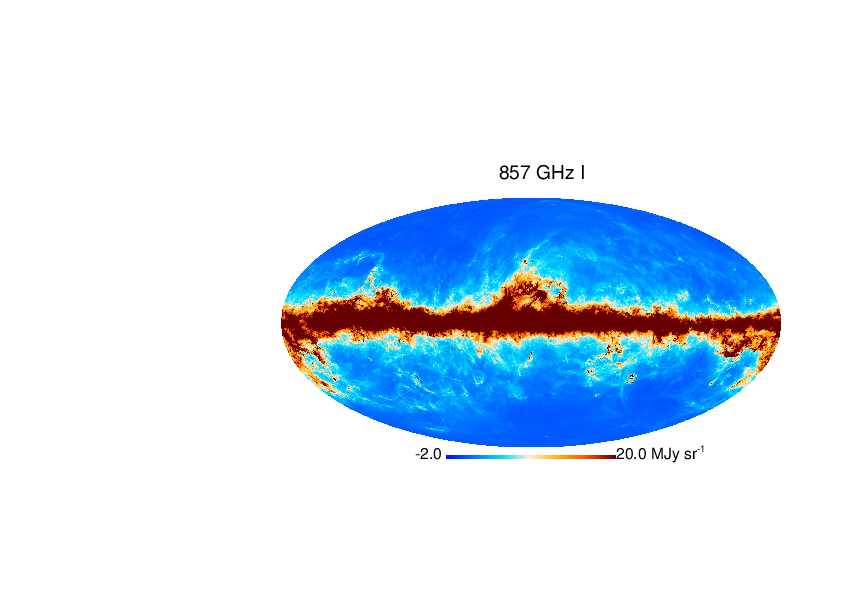}
& \includegraphics[width=0.22\textwidth,bb=290 131 770 431]{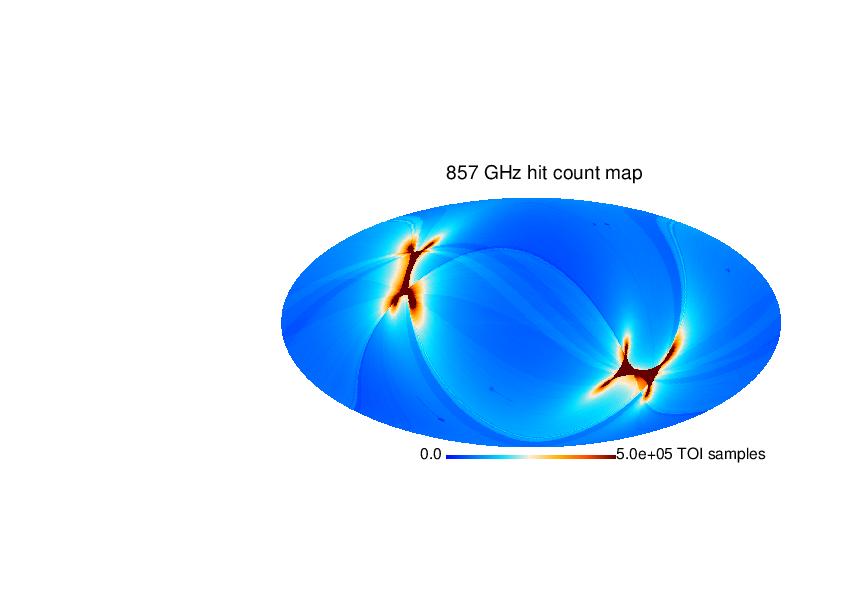}
& \includegraphics[width=0.22\textwidth,bb=290 131 770 431]{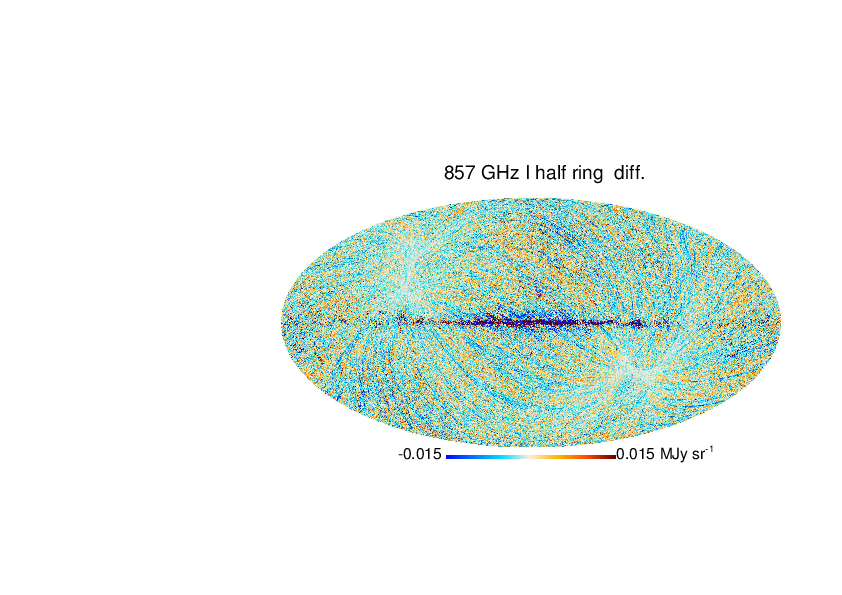}
& \includegraphics[width=0.22\textwidth,bb=290 131 770 431]{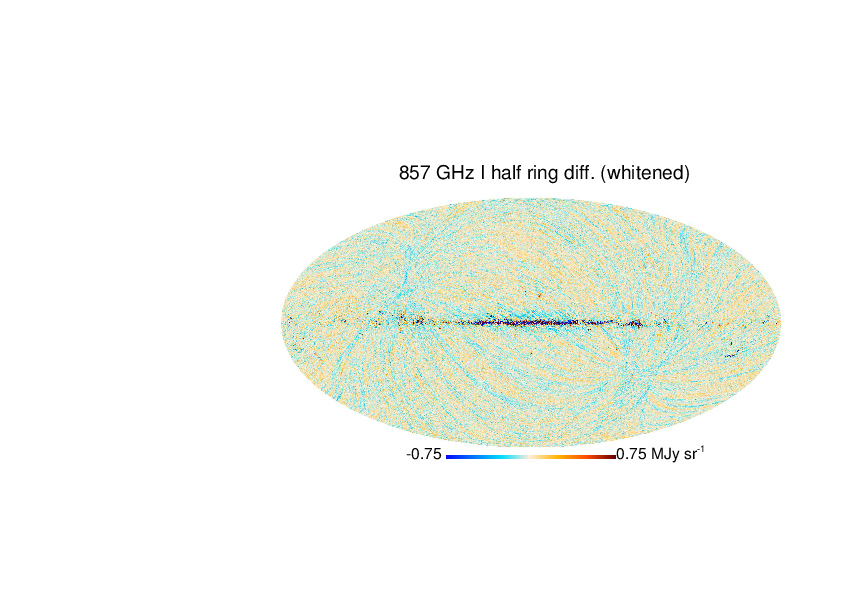}

\end{tabular}

\caption{Signal (left), hit counts (second column) and half differences between maps built with only the first and second half of each ring (third column) for all HFI frequencies. 
The half ring differences are clearly correlated with the hit count maps. 
The last column shows the half-ring difference maps, scaled by the square root of the number of TOI samples, which largely removes this correlation.
For the two highest frequencies, the differences show residual stripes and signal artefacts, at a low level (below 1\,\% of the sky signal). The difference maps have been degraded to $N{\mathrm {side}}=128$ \healpix\ resolution.}  
\label{fig:maps_hits_hrdif}
\end{figure*}

In Fig.~\ref{fig:maps_hits_hrdif} we show the intensity maps constructed for each of the HFI frequencies, together with the number of TOI  samples per pixel; and difference-maps built with the first and second half of each rings, both as the raw differences, and as differences scaled by the square root of the number of TOI samples to pre-whiten them.

The detector noise estimate used for the detector's data weighting is slightly different for the 2013 data release than for the previous release~\citep{planck2011-1.7}. 
As a consequence,  the pixel covariances we compute are now consistent with noise levels estimated from the difference maps built from the  first and second half of the rings.

\begin{figure*}[htbp]
\centering
\begin{tabular}{cc}
\includegraphics[width=0.45\textwidth,bb=33 597 425 828 ]{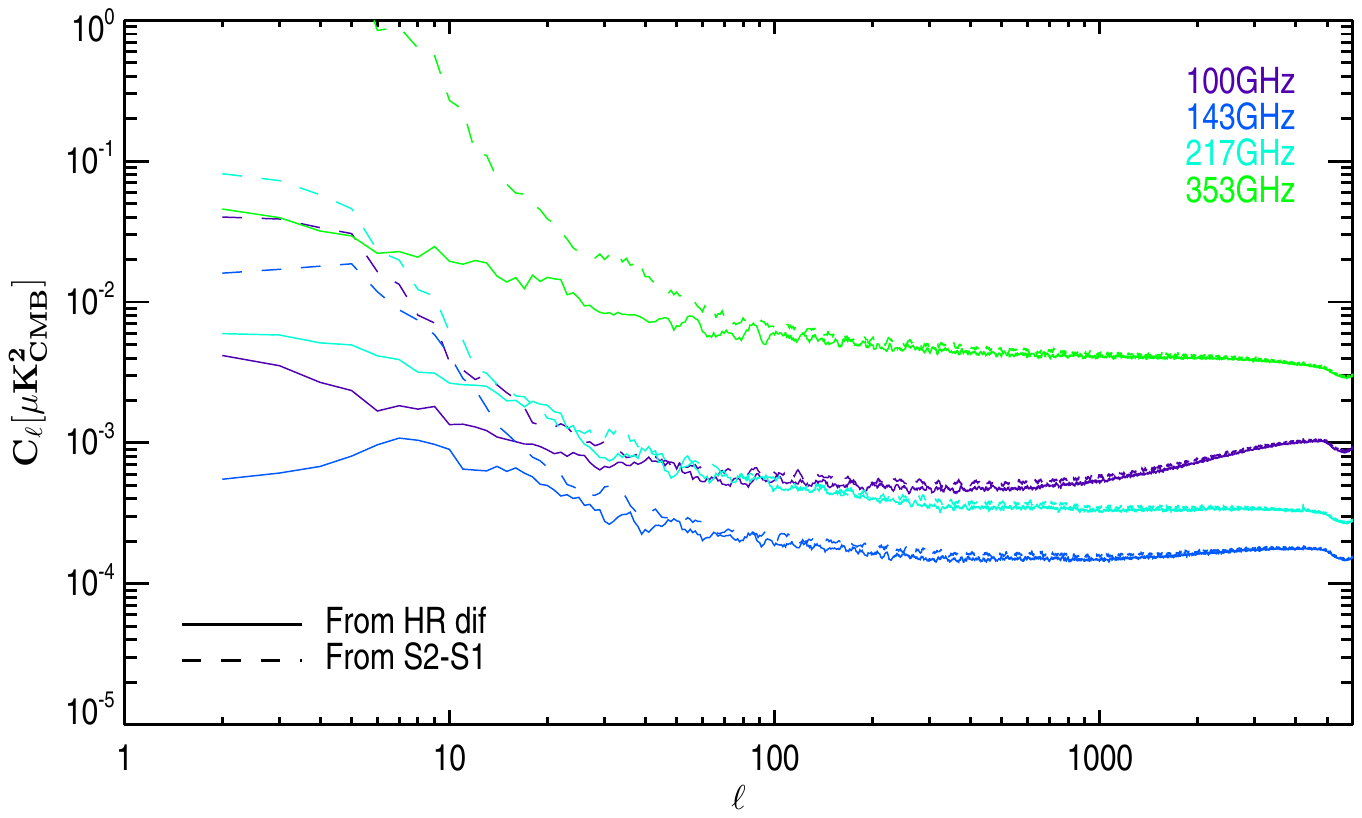}
&\includegraphics[width=0.45\textwidth,bb=33 597 425 828 ]{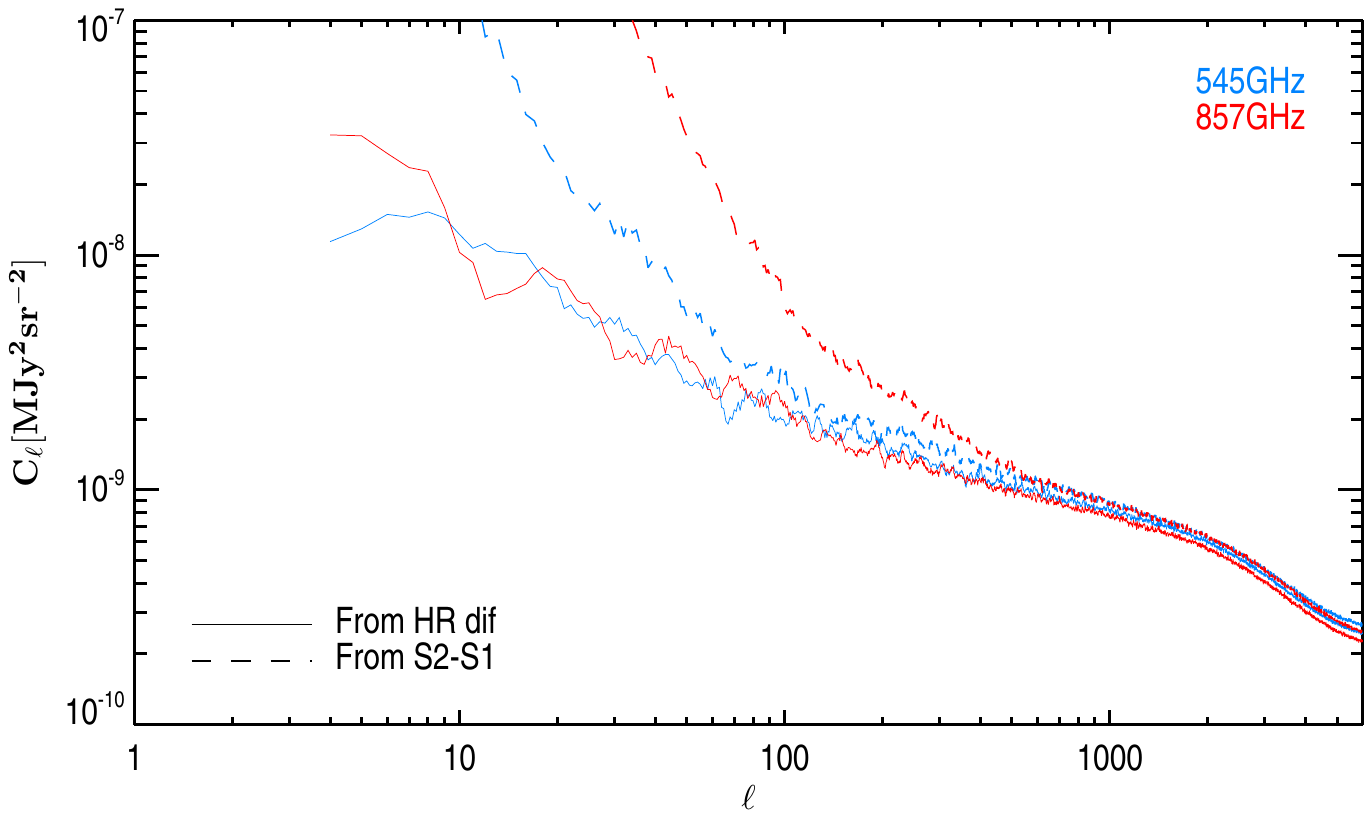}
\end{tabular}
\caption{Pseudo-power spectra reconstructed from the half-differences between maps from the first and the second half of each ring (continuous lines) and the half-differences between maps restricted to Survey 1 and 2 (dashed lines) for, respectively, the dipole calibrated channels (right) and the sub-millimetre ones (right). These pseudo-spectra were computed using Galactic masks, removing 15\,\%  ($\le353\,GHz$) or 40\,\% of the sky (sub-mm channels), combined with the \Planck\ point source mask.
For high-frequency channels, power spectra are dominated by signal and destriping residuals, due to gradients inside the pixels, which are not scanned at exactly the same positions in the two data sets. In the survey differences, other systematics like time response, pointing drifts, and residual gain variations also induce larger residuals.}
\label{fig:noise_in_maps}
\end{figure*}

Figure~\ref{fig:noise_in_maps} presents pseudo-spectra of the null test
difference maps, computed with a 15\,\% Galactic mask for frequencies
up to 353\,\GHz, or 40\,\% for the higher frequencies, combined with a point
source mask derived from the \Planck\ catalogue of compact sources \citep{planck2013-p05}.
We compare these spectra in Fig.~\ref{fig:noise_in_maps}  with those from the half-difference of the maps reconstructed from Surveys 1 and 2, properly normalized to compensate for the lower integration time. As illustrated previously, in Fig.~\ref{fig:143-1a_surv_diffs_raw}, such differences are sensitive to, among other things, time variations in the gains. As they compare observations made with roughly opposite scan directions, they may also exhibit residuals in regions where the sky signal is intense, and large gradients due to imperfect deconvolution of time response \citep{planck2013-p03}. As a consequence, their spectra, shown as dashed lines in Fig.~\ref{fig:noise_in_maps}, are higher at low multipoles than those of the half-ring differences. The fact that both half-difference spectra are very close to each other at high multipoles for frequencies lower than 353\,\GHz\  is an indication that these differences provide an estimate of the high spatial-frequency part of the noise included in the HFI 2013 data release. For the sub-millimetre channels, both spectra present a significant $\ell$ variation, showing that they are contaminated by systematic residuals. 

From these pseudo-spectra  we estimate the noise
level in the HFI maps by computing their average, after 
re-normalization by the sky coverage, in the $\ell$ range
100--6000. Using the averaged hit count per pixel, we convert these
averages into an equivalent rms per TOI sample. We compare this
estimate with two others: the rms of the half-ring map
differences, properly whitened using the hit counts; and the averaged
square-root of the variance computed in each pixel by the projection
module, scaled to a dispersion per TOI sample using the averaged hit
counts. These estimates are compared in Table~\ref{tab:noise_in_maps}. In general, they are in fair agreement for
the three lowest frequencies, indicating that they are a good 
estimate of the noise level in the maps. 
At higher frequencies,
however, signal residuals give a larger contributions. Therefore,  such methods only provide an upper limit on the high-frequency noise in the maps. 
%
\begin{table}[h]
\caption{Results of the three methods for deriving the TOI rms per
  sample from: (a) the variance maps; (b) the rms of the half ring
  difference maps; and (c) the pseudo spectra from
  Fig.~\ref{fig:noise_in_maps} (as explained in the text) for each
  frequency. Units are $\mu$K$_{\mathrm{CMB}}$ for 100 to 353\GHz,
  and MJy\,sr$^{-1}$ $(\nu I_{\nu}={\mathrm{constant}})$ for the sub-mm channels. These results should be considered as rough estimates only. The higher the frequency, the larger are the contributions of systematics residuals in the half-differences, e.g., time constants and signal gradients.}
\begin{center}
\begin{tabular}{@{} c c c c @{}}
\hline
\hline
\noalign{\vskip 2pt}
Frequency &  Variance maps   & Diff. maps  & Pseudo spectra   \\
 $[$GHz$]$&   (a)  & (b) & (c) \\
\noalign{\vskip 2pt}
\hline
\noalign{\vskip 2pt}
100 & 1569 & 1546 &1554\\
143 & 777 & 775 & 826\\
217 & 1109 & 1105 &1212\\
353 & 3671 & 3712 & 4101 \\
545 & 0.604 & 0.976 & 0.817 \\
857 & 0.695 & 2.58 & 0.920 \\
\hline
\end{tabular}
\end{center}
\label{tab:noise_in_maps}
\end{table}


\section{Conclusions \label{sec:conclusion}}
In this paper we have presented the mapmaking and calibration procedures used for the \Planck\ HFI data in the  2013 release. The calibrator for the CMB frequency data (100--353\,\GHz) is the 
solar dipole anisotropy as measured by \WMAP\ \citep{hinshaw2009}. This calibration is performed through a 
ring-by-ring template fit. Its limitations are largely a consequence of the non-ideal behaviour of the ADC from  the bolometer read-out electronics. Tiny deviations from linearity in these devices  cause apparent gain variation of the detector chain with time, which we have addressed using an effective gain correction, \bogopix. We showed that this scheme was able to reduce the apparent gain variation in time from 1--2\,\% to lower than 0.3\,\%, by studying the 
residuals of the map built for different time. Higher-order signal distortions induced by this systematic effect prevent us from using the more precise, orbital dipole-based calibration scheme presented in \cite{tristram2011}.

Correction for the ADC non-linearities should be made prior to any data reduction step. It requires precise measurements of each ADC response, which is currently taking place using data from the warm (4K) instrument. First tests of corrective software are also under way, with promising results.  
{  The time transfer functions used to deconvolve the data are built using planets and galactic plane observations. These are not sensitive to time constant longer than one second. However, the thermal behaviour of the bolometers/bolometer plate system  show longer time constants~\citep{planck2011-1.3} which have now been shown to shift significantly the dipoles axis and could thus affect the data at very low levels, below those of the ADC non-linearities correction, and contribute to the residual level of systematic inconsistencies in the calibration presented in this paper. }

The calibration for the 545 and 857\,\GHz\ channels is performed by comparing Uranus and Neptune flux densities with models of their emissivities. We had to switch to this scheme owing to apparent systematic effects  
in the FIRAS spectra we used in the HFI Early Data release. At those frequencies, time variations of the gain are lower than other systematic calibration uncertainties.

We  revised our zero level-setting method, which now relies on 
the CIB monopole and the zero of the Galactic emission, defined as zero dust emission for a null \hi\ column density.

At all frequencies, the statistical uncertainty of the calibration  is negligible compared to the systematic uncertainty. 
The systematic uncertainty has been evaluated  using several methods, presented in Sect.~\ref{sec:calib_charact}. We evaluated three types of systematic uncertainties:
\begin{description}
\item{(a)} {\it Residual apparent variations of gains with time}. For the 100 to 217\,\GHz\ maps, we showed in Sect.~\ref{sec:calib_time_stab} that, using \bogopix, these variations were lower than 0.3\,\%, for each individual detector. As 
shown in Fig.~\ref{fig:bogo_res_summary} the gain variations appear to be independent from one detector to the other, so such uncertainties should average out in the combined maps from this release. 
This 0.3\,\%  uncertainty is therefore a conservative upper limit on the level of residual gain variations in the frequency maps. At higher frequencies, no estimation, nor 
correction for the apparent gain variation, is available. We choose to quote the 
level of variations we observed in the single detector measurements
of \bogopix\ at lower frequencies, which is 1\,\%; this is again an upper limit for combined maps. Given \Planck's scanning strategy, such uncertainties might be relevant for point-like sources studies, as these are observed in general once per survey, or globally when comparing sky maps from individual surveys.

\item{(b)} {\it Relative calibration uncertainties}, which should be used when combining different frequency maps, e.g., when reconstructing the SED of an object. We presented in Sect.~\ref{sec:cmb_chan_calib_checks} several methods to evaluate such uncertainties for 100 to 217\,GHz channels. Both a direct comparison of pseudo-power spectra outside the Galaxy and results from the component separation method {\tt SMICA} show that the inter-calibration between the 100, 143, and 217\,\GHz\ channels is better than 0.2\,\% (we keep the more conservative estimate, using {\tt SMICA} for the reported errors). We complement these results with the upper limits extracted from {\tt SMICA} at 353 and 545\,\GHz, using the central value (1 and 5\%, respectively) as an upper limit on the uncertainty. For the relative calibration of the 857\,\GHz\ maps, we quote the 5\,\% uncertainty on the photometry used in the planet calibration.  

\item{(c)} {\it Absolute calibration uncertainties} that should be considered when comparing with other data sets. This involves comparing \Planck\ data with an external calibrator.  
Below 353\,GHz, such uncertainties have been evaluated by two complementary 
approaches: reconstructing the dipole and comparing it with the {\it WMAP} measurements (Sect.~\ref{sec:cmb_chan_calib_checks}); and evaluating the amplitude of residual dipoles in our maps, after foreground removal (Sect.~\ref{sec:gal_zero_levels}). From 100 to 217\,\GHz, both methods show
consistency with \WMAP\ at better than 0.3\,\%. The second approach shows agreement at 1\,\% for 353\,\GHz. As the data are calibrated  on the \WMAP\ dipole measurement, an additional uncertainty of $0.24\,\%$ has to be combined with the HFI intrinsic uncertainties.  {  Due to the nature of the calibrator, the absolute accuracies stated here only apply at very low $\ell$. When studying smaller angular scale anisotropies, transfer functions, including that resulting from the - yet unaccounted for - very long time constants (see Section \ref{sec:gain_variation}), should be taken into account.}

{  Indeed  the comparison of  HFI with WMAP $C_{\ell}$ measurements at the level of the first and second peak show a discrepancy of 2.4 \% in spectra, thus a possible calibration discrepancy of 1.2\% \citep{planck2013-p11}.  Considering the relative calibration accuracy discussed above this must come from a common systematic effect on either the HFI CMB channels or on the WMAP V and W data, affecting the dipole calibrations and/or the transfer functions. }

For the two highest frequencies, the absolute scale is limited by the accuracy of the planetary atmosphere models (5\,\%), combined with systematic uncertainties in our flux measurements (5\,\%), which results in a total uncertainty of 10\,\%. 
Such uncertainties are relevant for comparing \Planck\ data with other data sets. When comparing with data sets sharing the same calibrator as HFI, the \WMAP\ dipole or the planet models of \cite{moreno2010}, the uncertainty on these calibrators should therefore be omitted in the comparison. 
\end{description} 
 
We summarize the calibration uncertainties for the HFI  frequency maps  in Table~\ref{tab:error_summary}.
 
\begin{table}[h]
\caption{Summary of the HFI systematic calibration uncertainties for the frequency maps of the 2013 data release. Column (a) gives the residual 
relative variation of calibration with time, (b) gives the relative calibration uncertainty from one HFI channel to the other, and (c) the absolute calibration uncertainties of each HFI channel, including the uncertainty of the calibrators. 
We indicate in the last column the uncertainties of the calibrators (\WMAP\ dipole and models of planets) that are already included in the absolute uncertainties listed in column (c). These have to be taken into account when comparing with data sets relying on the same calibrators.}
\begin{center}
\begin{tabular}{@{} c c c c c @{}}
\hline
\hline
\noalign{\vskip 2pt}
Frequency & Time stability & Relative   &  Absolute & Model\\
 $[$GHz$]$& (a)  [\%]  & (b) [\%] & (c) [\%] &[\%]\\
\noalign{\vskip 2pt}
\hline
\noalign{\vskip 2pt}
100 & 0.3 & 0.2 & 0.54 & 0.24 \\
143 & 0.3 & 0.2 & 0.54 & 0.24\\
217 & 0.3 & 0.2 & 0.54 & 0.24\\
353 & 1.0 & 1.0 & 1.24 & 0.24 \\
545 & 1.0 & 5.0 & 10.0 & 5.0~ \\
857 & 1.0 & 5.0 & 10.0 & 5.0~ \\
\hline
\end{tabular}
\end{center}

\label{tab:error_summary}
\end{table}

\begin{acknowledgements}
The development of \Planck\ has been supported by: ESA; CNES and CNRS/INSU-IN2P3-INP (France); ASI, CNR, and INAF (Italy); NASA and DoE (USA); STFC and UKSA (UK); CSIC, MICINN and JA (Spain); Tekes, AoF and CSC (Finland); DLR and MPG (Germany); CSA (Canada); DTU Space (Denmark); SER/SSO (Switzerland); RCN (Norway); SFI (Ireland); FCT/MCTES (Portugal); and PRACE (EU). 
A description of the \Planck\ Collaboration and a list of its members
with the technical or scientific activities they have been involved
into, can be found at {\url{http://www.rssd.esa.int/index.php?project=PLANCK&page=PlanckCollaboration}.}
\end{acknowledgements}

\bibliographystyle{aa}

\bibliography{../calib_bib,../Planck_bib}

%

\appendix
\section{Calibration conventions \label{sec:conv}}

\subsection{Colour corrections}

Whatever the origin of the calibrator (on the sky or an internal blackbody), the calibration is performed with a source of known spectral energy distribution (SED). Except for CMB anisotropies, in general, the observed source will  have 
a SED different from the calibration source. 
Although the simplest way to express the calibration is to give the
response as a function of the power falling onto the detector,  we
usually use a secondary expression of the measurements as spectral
densities. This allows us to compare the measurements with others
and with models. Spectral densities are either an intensity
(W\,m$^{-2}$\,sr$^{-1}$\,Hz$^{-1}$) for brightness or flux densities
(W\,m$^{-2}$\,Hz$^{-1}$) for unresolved sources, expressed at a reference frequency such that the power integrated in the spectral bandpass 
is equal to the measured power.
The intensity (or flux density) is thus always linked to the choice
of both a reference frequency and an assumed SED.

CMB anisotropies are calibrated on the CMB dipoles (and inter-calibrated on higher-order CMB anisotropies). The CMB temperature gives a calibration only for the SED of the CMB anisotropies. CMB anisotropies are thus
expressed as $\delta$T in K$_{\mathrm{CMB}}$. For astrophysical
components with a different SED, this calibration has to be
re-expressed as intensity at the reference frequency, using an SED convention.
 Following the {\it IRAS} convention, the spectral
intensity data $I_{\nu}$, are often  expressed at fixed nominal
frequencies, assuming the source spectrum is $\nu I_{\nu}=$~constant (i.e., constant
intensity per logarithmic frequency interval, labelled ``ref''). The colour correction factor $\cal C$ is defined such that:
\begin{equation}
\label{cc_2}
I_{\nu_0}^{\mathrm{act}} = \frac{I_{\nu_0}^{\mathrm{ref}}}{\cal C} \, ,
\end{equation}
where $I_{\nu_0}^{\mathrm{act}}$ is the actual specific intensity
of the sky at frequency $\nu_0$, $I_{\nu_0}^{\mathrm{ref}}$ is the corresponding
value given with the {\it IRAS}~\citep{neugebauer1984} or DIRBE~\citep{silverberg1993} convention{\footnote{The DIRBE and {{\it IRAS} data products give $I_{\nu_0}(\nu I_{\nu}=constant)$}}
and $\nu_0$
is the frequency corresponding to the nominal wavelength of the 
band. With these definitions,
\begin{equation}
\label{cc_1}
{\cal C} = \frac{\int (I_{\nu}/I_{\nu_0})^{\mathrm{act}}{ R_{\nu} d \nu}}{\int (\nu_0/\nu)  R_{\nu} d \nu} \, ,
\end{equation}
where $(I_{\nu}/I_{\nu_0})_{\mathrm{act}}$ is the actual specific intensity of the 
sky normalized to the intensity at frequency $\nu_0$ and $R_{\nu}$
is the spectral response \citep[see][]{planck2013-p03d}.

\subsection{CMB dipole conventions\label{ssec:cmb_dip_conv}}
For HFI calibration at low frequency, unlike the LFI calibration~\citep{planck2013-p02b}, we used {  the non-relativistic} approximation of the dipole anisotropy:
\begin{equation}
\frac{\delta T}{T}  =  \beta \cos{\theta} ,
\label{eq:dipole}
\end{equation}
where $\beta = v/c$ is the ratio between the  observer velocity $v$ and the speed of light. For the CMB dipole, $\beta \simeq 1.2\times10^{-3}$. 
The leading-order term of the relativistic corrections is $\beta^2
(\cos^2{\theta}-1/2)$~(\cite{peebles_wilkinson1968}, \cite{kamionkowski2003}, \cite{planck2013-pipaberration}). The amplitude of this
correction is of $\pm 1/2\beta^2$, so this quadrupole term has a relative amplitude  of $0.6\times 10^{-3}$ with respect to the non-relativistic term of Eq.~\ref{eq:dipole}. However, this 
quadrupole is only coupled to the dipole when masking part of the sky,
which is small ($\sim$~$10\%$) in our calibration scheme, so the real
bias must be smaller. 
Indeed, when using the orbital dipole, which
is a factor of about $10$ smaller than the solar dipole, as the calibrator,
\cite{tristram2011} showed that using the non-relativistic approximation leads
to a relative bias as small as $6\times 10^{-6}$. Given the level of systematic uncertainties we estimate for our calibration, it is therefore legitimate, and much simpler, to use the non-relativist approximation of the solar dipole anisotropy.  

\subsection{Far sidelobes\label{ssec:fsl_conv}}
The impact of the far sidelobes (FSL) on HFI data is discussed in detail in \cite{planck2013-p03c}. We present only a summary of their impact on calibration in this appendix.

FSL  may affect the calibration determination in different ways. The power measured by our detectors $p_{\mathrm{mes}}$  may be schematically written as: 
\begin{equation}
p_{\mathrm{mes}}= g ( S_{\mathrm{ML}}+ S_{\mathrm{FSL}}) + \mathrm{noise}
\end{equation}
where $S_{\mathrm{ML}}$ denotes the sky signal coming through the main lobe and $S_{\mathrm{FSL}}$ that coming through the far sidelobes.
We also denote by $f_{\mathrm{FSL}}$ the fraction of power going into the FSL. 
For the planet photometry, some level of knowledge of $f_{\mathrm{FSL}}$ is needed to correctly compare the reconstructed flux with the planet brightness. However, the relative FSL power is lower than 0.3\,\%~\citep{tauber2010} for all HFI frequencies, which is well below the systematic uncertainties in the planet emission models that we are using, which are about $ 5\,\%$ (see Sect.~\ref{sect_HF_calib}). Therefore FSL can safely be ignored in the 545 and 857\,\GHz\ calibration analyses. 

For the diffuse emission calibration on the CMB dipole at low HFI frequencies, we use a fit of the observed data to the solar dipole as measured by the \WMAP\ team~\citep{hinshaw2009}, as detailed in Sect.~\ref{sect_LF_calib}, without convolving it with a beam model. Note that this  is different from the LFI calibration pipeline~\citep{planck2013-p02b}. To clarify the consequences of this choice, we have to examine  the FSL signal in detail. 
The FSL signal may be decomposed to first order into three main components, depending on their optical paths: the primary, named PR, and the secondary and baffle spillovers collectively named SR in ~\cite{tauber2010}. The primary spillover
originates mainly from directions on the sky close to the spin axis, so
is will be roughly constant for each fixed pointing period and will be removed by our destriping procedure. The baffle spillover corresponds to radiation reaching the detectors after reflection {  on material surrounding the mirrors,} including the telescope's baffles. Again, 
one may expect this component to be roughly constant over a ring, as a result of the averaging of many original directions. 
Finally, the secondary spillover is selected over a wide ($\sim
15\deg\times 30\deg$) area {  centered about} $10\deg$ away from the main beam (see Fig.~5 in \citealt{tauber2010}) and is formed by radiation reaching the HFI horns without  reflection on any telescope part. This will generate an image of the sky offset with respect to the main beam and integrated over a very broad area. In particular, it will contain a dipole component. 

Table~2 in \cite{tauber2010} gives the relative amplitudes of the different FSL components with respect to the main beam signal, combining secondary and baffle spillovers in the 'SR' column. One may see that, in this framework, PR and SR have comparable amplitudes. At maximum, the SR 
relative amplitude amounts to 0.2\% at 100\,\GHz\ and decreases fast for frequencies higher that 217\,\GHz. To clarify the importance of the baffle spillovers, we performed simulations with the actual \Planck\ scanning strategy and FSL models similar to those of 
\cite{tauber2010}. This indicated that the SR FSL signal after destriping has an amplitude of  about $70\%$ of {  the value listed in Table~2} of \cite{tauber2010}.  This is confirmed by the estimate of the FSL amplitudes presented in \cite{planck2013-pip88}, which gives confidence in the models used in \cite{tauber2010}. From this, as shown in \cite{planck2013-p03c}, we conclude that, at maximum, a spurious dipolar signal with a relative amplitude of $\sim 0.13\% $ could be present in our data. As the CMB is a Gaussian signal, 
neglecting the beam transfer function, i.e., at very low $\ell$ ($\ell \leq 30$ given the SR beam), one may then consider that
\begin{equation}
S_{\mathrm{FSL}}= \varepsilon_{\mathrm{SR}}S_{\mathrm{ML}} + \mathrm{constant}
\end{equation}
at ring level. Thus our calibration will determine an effective gain: 
$g_{\mathrm{eff}}= g(1+\varepsilon_{\mathrm{SR}})$. This will lead to a reconstructed sky
signal approximately equal to $S_{\mathrm{ ML}}$. Therefore, at large angular
scales, if we ignore the spurious SR signal remaining after destriping  in the calibration process,
this cancels to first order the effect of not accounting for it in further
analysis (like power spectra). At smaller scales, however, the SR
signal becomes negligible and this cancellation is not effective
anymore. To summarize, the HFI dipole calibration as performed for the 2013 data release may result in an approximately   $ 0.25\% $ overestimate of the band powers for $\ell$ above about $ 40$. This calibration 
systematic effect (roughly $ 0.13\% $) is lower than other sources of systematic uncertainties evaluated in this paper. 

\section{Choosing the sky template for the solar dipole calibration\label{app:sky_dipcal}}

\begin{figure}[htbp]
   	\centering
	\includegraphics[width=0.5\textwidth]{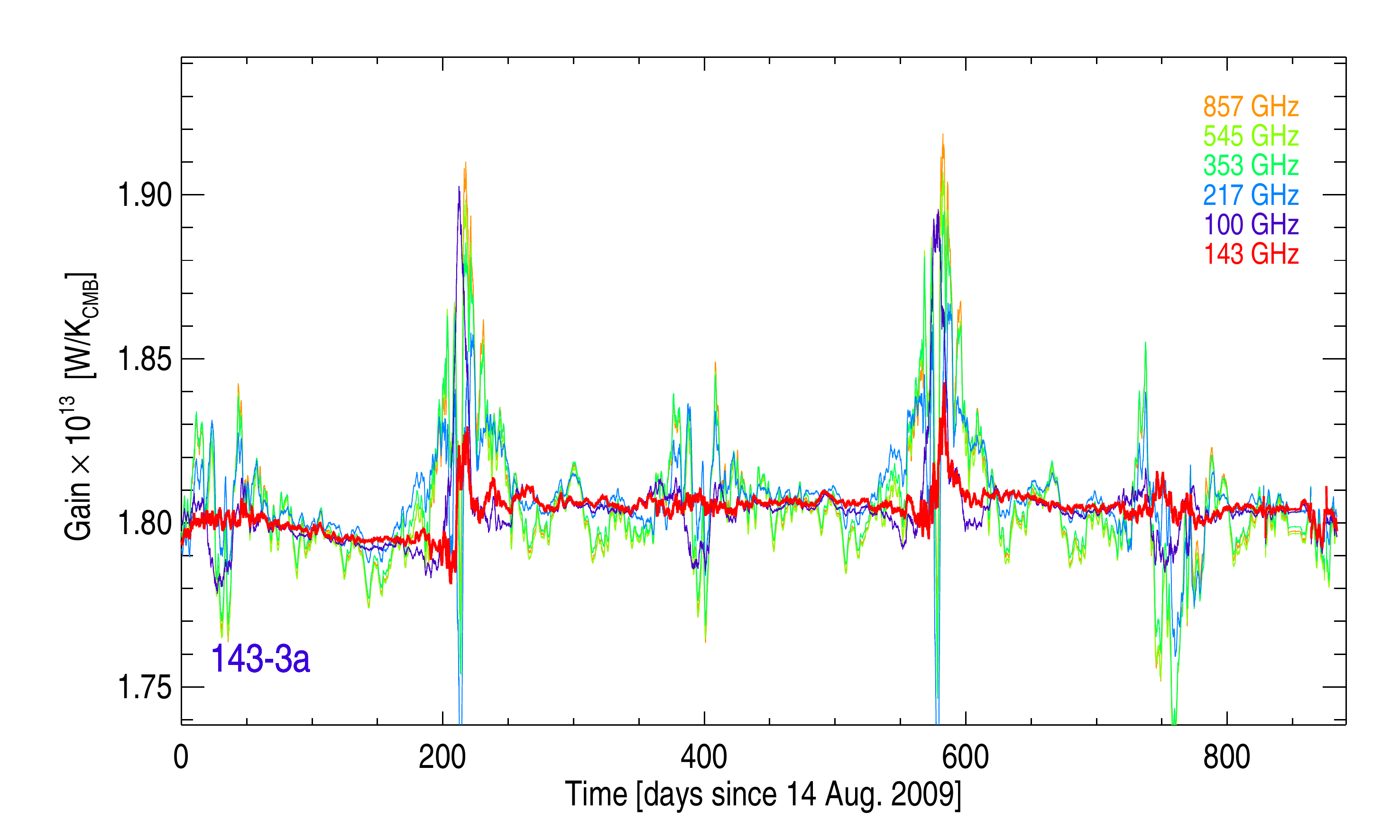}

    	\caption{Impact of the Galactic template on the solar dipole ring-gain measurements, for  one detector at 143\,\GHz. We compare the results obtained using the HFI temperature maps at each frequency (from, e.g., previous reconstructions). For this plot, ring-by-ring gains have been smoothed with a  width of 50 rings (which corresponds to about $2$ days). The  largest variations occur for rings for which the solar dipole amplitude is low relative to the Galactic emission  (around days 50, 200, 400, 550, and 700). Using the HFI map at the detector's frequency as a Galactic template minimizes the systematic ring-to-ring variations. 
    \label{fig:QD_template_effect}}
\end{figure}
\begin{figure}[htbp]
   	\centering
	\includegraphics[width=0.5\textwidth]{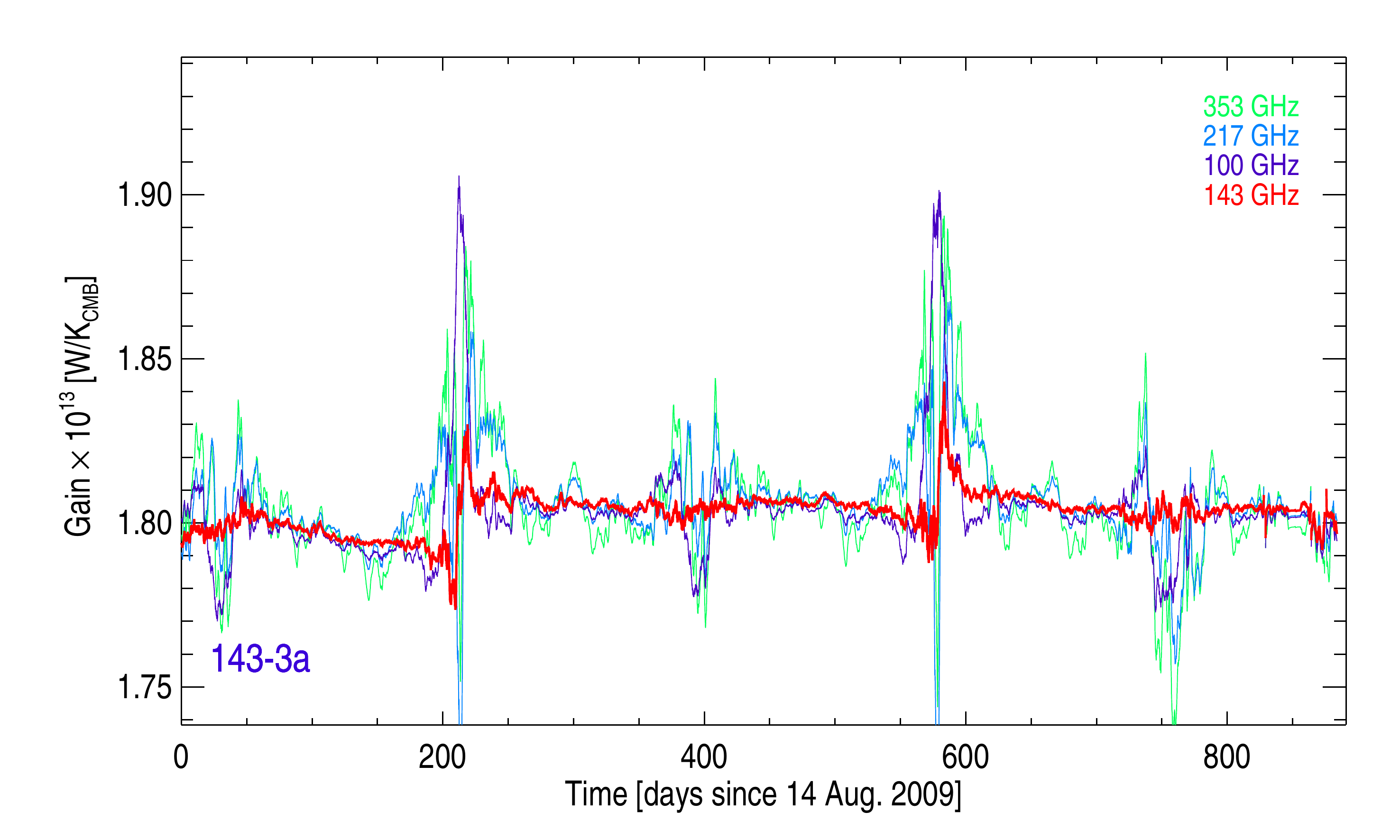}

    	\caption{Same as Fig.~\ref{fig:QD_template_effect}, but taking into account polarization as reconstructed in the HFI $Q$ and $U$ maps. This makes very little difference in the results, showing that polarization is not the main source of systematic variations for the rings where the  Galactic emission is large. 
    \label{fig:QD_template_effect_pol}}
\end{figure}
To determine  the solar dipole calibration factor, we  have to take into account the Galactic foreground. This is done with a template (see Sect.~\ref{subs:soldip}).
Fig.~\ref{fig:QD_template_effect} shows the impact of the choice of Galactic template. We compare the ring-by-ring gains obtained with each of the 
HFI frequency maps.
Using a Galactic template helps improves the estimation of the calibration factor for rings scanning regions close to the Galactic equator (e.g., around day 200).  
Unsurprisingly, the template producing the lowest apparent variations is that with the same frequency as the detector of interest. However, even in this case, apparent systematic gain variations are observed  at the times when the dipole amplitude is low,
In Fig.~\ref{fig:QD_template_effect_pol} we show results  obtained taking into account the polarization (from the HFI $Q$ and $U$ maps). The small difference this induces shows that polarization does not play a major role in the apparent ring-to-ring variations, even for rings where Galactic emission is larger.


\section{\label{A1} HFI and FIRAS data comparison}

The procedure we used for comparing  HFI and  FIRAS data is very similar to that adopted by the {\tt ARCHEOPS} collaboration~\citep{macias2007}. It was summarized in \cite{planck2011-1.7}; here we give full details.

\subsection{\label{FIRAS} FIRAS data: spectra and derived maps}

\paragraph{FIRAS spectra}
The FIRAS instrument, its operating modes, calibration, and the
data products are described  in the FIRAS Explanatory
Supplement (FIRAS team, 1997, \url{http://lambda.gsfc.nasa.gov/product/cobe/firas_exsupv4.cfm}).  FIRAS has a scanning,
four-port (two inputs; two outputs) Michelson interferometer that uses
polarizing grids as beamsplitters and creates an interferogram
(i.e., the Fourier transform of the source spectrum) by scanning a
movable mirror platform (the ``Mirror Transport Mechanism'', or MTM). A
dichroic splitter at each output port (designated ``left''
or ``right'') further splits each beam into low (30--630\,\GHz) and
high (600--2910\,\GHz) frequency bands. The MTM could be scanned at
either of two speeds: ``slow'' or ``fast''. Furthermore the MTM sweep could be set
to one of two scan lengths, ``long'' or ``short'', thus affecting the
spectral resolution. 
Most research applications call for one or more 
high-level products, such as the dust spectrum maps that we are using
here. In these high-level products, the different modes and detector signals
were combined to form the HIGH and LOWF frequency data-sets.
The two dust-spectrum maps (FIRAS$\_$DUST$\_$SPECTRUM$\_$HIGH.FITS and 
FIRAS$\_$DUST$\_$SPECTRUM$\_$LOWF.FITS) cover 98.7\,\% of the sky
and give the residual sky spectrum, from about 2970 to 68\,\GHz,
after modeled emission from the CMB, interplanetary dust, and interstellar lines
have been subtracted. The remaining signal is thus dominated by thermal 
continuum emission from Galactic interstellar dust (and the cosmic
IR background e.g., \citealt{puget96}, \citealt{fixsen98} and \citealt{lagache99}). 

\paragraph{Uncertainty estimates}
Uncertainties in the FIRAS data come from several different sources
and manifest themselves in several different ways.
Fortunately, many of the uncertainties are well
described by a few dominant terms that show the source of
the uncertainty.  All of them are fully detailed in the FIRAS
explanatory supplement. The errors are divided into groups:
the detector measurements; the calibration emissivities; the
bolometer model parameters; the temperature measurements
of all but XCAL (XCAL is the external Calibrator); 
and the temperature measurement of the XCAL.
For our purposes, only the detector noise, the uncertainties
in some parameters derived from the calibration, and
the uncertainty in the absolute temperature scale of
the external calibrator are of importance.
The covariance matrix can be written as:
\begin{equation}
\tens{V} = \tens{C} + \tens{J}+ \tens{P} \, .
\end{equation}
The $\tens{C}$ term (called the C-matrix by the FIRAS team) is the detector noise. It includes off-diagonal
terms due to frequency correlations introduced by the apodization
of the coadded interferograms, before Fourier transformation
into spectra.
The \tens{J} term (called JCJ by the FIRAS team) corresponds to uncertainties linked to
the bolometer model parameters (only the JCJ gain is important here).
This error is regarded as a systematic error.
The \tens{P} term (called PTP by the FIRAS team) is the absolute thermometry uncertainty; it
is not a statistical uncertainty. 
This \tens{P} term is included in the error budget
since it is the dominant error for the absolute temperature of
the CMB. It is thus important for
comparison of FIRAS measurements to other experiments (but only
for frequencies smaller than about 430\,\GHz). 
The levels of these three uncertainties are shown in Fig. \ref{FIRAS_error}.
For the detector noise, only the square root of the diagonal
(the ``C-vector'') is displayed. 
\cite{fixsen97} used a conservative estimate of the
gain uncertainty of 2\,\% for the 600--2400\,\GHz\ FIRAS data.
   \begin{figure}
   \centering
   \includegraphics[width=\columnwidth]{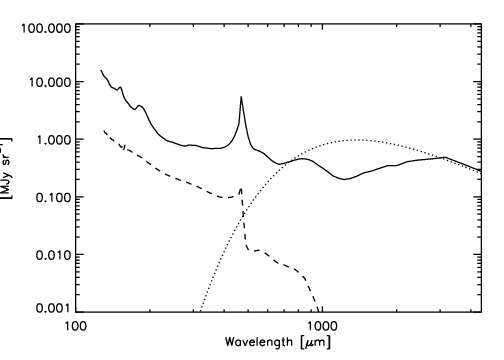}
   \caption{\label{FIRAS_error} FIRAS uncertainty summary. Only the uncertainties
     relevant for our purpose are displayed here: the C-vector ({\it continuous line}), 
     the JCJ gain multiplied by the average sky spectrum ({\it dashed line}), and the PTP
     uncertainty ({\it dotted line}).}
   \end{figure}

\paragraph{Building FIRAS maps}
For individual pixels at high Galactic latitude, the S/N ratio of 
FIRAS spectra is about 1. Of course, this is not the case in the
Galactic plane, where the S/N ratio is about 50 times higher (at high
frequencies). The FIRAS map at a selected frequency can be
obtained by convolving the FIRAS spectra with the \Planck-HFI 
bandpass filters.  However, this method gives very noisy FIRAS
maps at high Galactic latitude (especially for $\nu<
430$\,\GHz). Since we are interested in both the Galactic plane and
its surroundings, we have chosen to derive the FIRAS maps
together with their errors from fits of FIRAS spectra.  Each
individual FIRAS spectrum is fitted with a modified blackbody spectrum,
\begin{equation}
S_{\nu} = \tau \left( \frac{\nu}{\nu_0} \right) ^{\beta} P(\nu, T_{\mathrm{dust}})  \, ,
\label{eqn:mbbs}
\end{equation}
where $\tau$ is a measure of the relative dust column density for each
pixel, $\beta$ is the spectral index, and $P(\nu, T_{\mathrm{dust}})$ is the
Planck function.  Since we are searching for the best representation
of the data and not for physical dust parameters, we include the
contribution of the cosmic infrared background in the fit. Moreover, we restrict
the fit to the frequency range of interest -- this avoids the need for a
second dust component as in \cite{finkbeiner99}.

For each frequency used for the calibration, we find the best
values of $\tau$, $T_{\mathrm{dust}}$, and $\beta$ using a $\chi^2$
minimization method for each pixel. We include the correlations between
FIRAS frequencies, and fit in frequency intervals
related to the frequency of interest. We compute the error in each pixel
by considering the deviation of the emissivity induced by all models
(Eq.~\ref{eqn:mbbs}) allowed at 68\,\% confidence level by the FIRAS spectrum fit.
Only the C-matrix was considered in the fit. The JCJ and PTP terms are added as a
systematic error at the end of the process.

Ideally,the  fits would be be performed on independent frequency intervals so that
the maps derived for each of the \Planck-HFI frequencies are independent.
However reducing the frequency interval increases the noise,
so it was necessary to use overlapping frequency intervals.
Fortunately, the fitting results are not very sensitive to the choice of frequency interval;
there are no systematic effects, and the values derived
at the \Planck-HFI wavelengths are consistent. This is not the case for
the error bars, which can vary by factors of 2. Fits are performed for $560<\nu<1765$\,\GHz,
$400<\nu<1500$\,\GHz, $270<\nu<1000$\,\GHz, $170<\nu<860$\,\GHz\ and 
$75<\nu<670$\,\GHz\
for the 857, 545, 353, 217, and 143\,\GHz\ HFI frequencies, respectively.
Typically, for Galactic latitudes $|b|<5\deg$, the 1$\sigma$ uncertainty is  1.0, 1.5, 4.2, 
9.8, or 35.1\,\% at 857, 545, 353, 217 and 143\,\GHz, respectively. For $15\deg<|b|<20\deg$, they reach 5.5, 11, 29, 54, and 158\,\%.  Fig. \ref{fit_FIRAS} shows the results
for three different pixels. 

   \begin{figure}
   \centering
   \includegraphics[width=9cm]{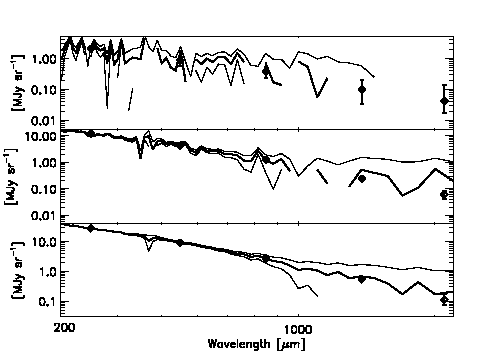}
   \caption{\label{fit_FIRAS} Example of FIRAS spectra (with their $\pm$1$\sigma$ error) with the
computed values at the HFI wavelengths.}
   \end{figure}

\subsection{\label{degrade} \Planck-HFI data: towards the FIRAS resolution}
We built uncalibrated \Planck\ maps for each detector, and convolved them with the FIRAS beam. 

\paragraph{FIRAS beam}
The FIRAS beam has been measured using the Moon.  Due to
imperfections in the sky horn antenna, the effective beam shows both
radial and azimuthal deviations from the nominal 7\deg\ top-hat beam
profile (Fig. \ref{beam_prof}).  Since \COBE\ rotates about the
optical axis of the FIRAS instrument, the average beam 
has circular symmetry; but a single
interferogram is acquired in less than a rotation period and can have an 
asymmetric beam. \cite{fixsen97} estimate that the assumption of beam symmetry
may produce residual beam shape errors of order of 5\,\%, and we take this into account in what follows.

   \begin{figure}
   \centering
   \includegraphics[width=\columnwidth]{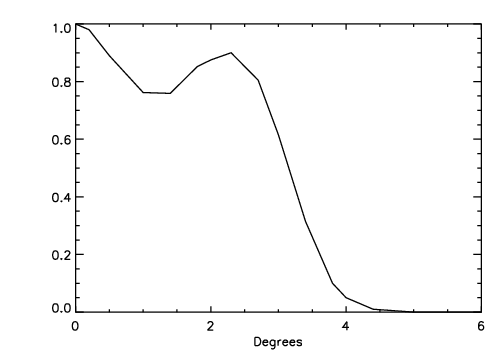}
   \caption{\label{beam_prof} FIRAS beam profile (from \url{http://lambda.gsfc.nasa.gov/product/cobe/firas_prod_table.cfm}).}
   \end{figure}

\paragraph{Convolution}
We carry out the beam convolution in the \healpix\ scheme. To simulate the
movement during an integration of an interferogram, the 
DIRBE-format data were further convolved by a 2\pdeg6  top-hat in the
direction perpendicular to the ecliptic plane (which is roughly the
FIRAS scanning direction).  

\subsection{FIRAS data: towards the \healpix\ projection}
The \Planck~maps are presented in the \healpix\ format~\citep{Gorski05} with a resolution $\Nside=2048$. For comparison with FIRAS, the convolved HFI maps are downgraded to $\Nside=32$. 
\COBE\ data are presented in the 
\COBE\ Quadrilateralized Spherical Cube projection (CSC), an approximately equal-area projection. 
For comparison with \Planck, the FIRAS  maps are regridded into the \healpix\ format using a drizzling re-projection code.

\subsection{Deriving the calibration gains and zero points}

We fit for the calibration coefficients $K$ and $O$ following:
\begin{equation}
\label{eq_firas_HFI}
F(\nu_0) / \cal{C} =  K \times H(\nu_0) + O  \, ,
\end{equation}
where $F(\nu_0)$ is the FIRAS brightness at frequency $\nu_0$, $H(\nu_0)$ is the HFI signal in pW at $\nu_0$,
$K$ is the gain calibration factor (for a source spectrum with $\nu I_{\nu}$ = constant), and $\cal{C}$ is the colour correction given in Eq.~\ref{cc_2}.
The calibration coefficients $K$ and $O$ are derived from a linear fit
of the FIRAS and HFI cosecant variations, restricted to intermediate
Galactic latitudes ($10\deg<|b|<10\deg$).  We avoid using the inner
part of the Galactic plane, as in these areas the spectral characteristics, averaged out in the FIRAS measurements, may present  angular scale variations that are not accurately accounted for in our processing of the HFI data (i.e., we evaluate $K$ and $\cal{C}$ at 7\deg\, resolution,  not  5\arcm). More importantly, we avoid the inner part to minimize the effect of the FIRAS beam uncertainties. The Galactic polar caps are also  not used, since the S/N ratio of the FIRAS data extrapolation is very low there.  We also mask regions where CO emission lines (removed from FIRAS measured spectra) are bright in the \cite{dame1987} map, and add a template of  CMB anisotropies to the FIRAS data.

\begin{figure}
\centering
\includegraphics[width=0.5\textwidth]{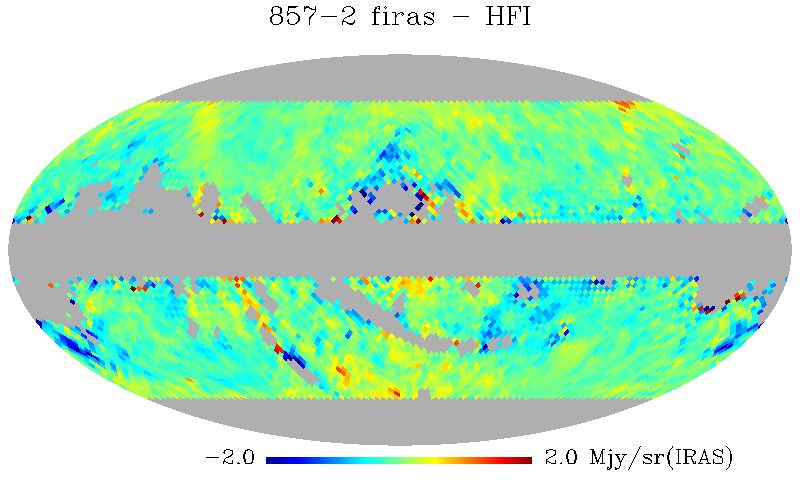}
\caption{FIRAS minus convolved-HFI data for the 857-2 detector (in MJy sr$^{-1}$). Only pixels used for the ``nominal'' calibration are shown (i.e., $10\deg<|b|<60\deg$, outside the CO mask, and for pixels with enough FIRAS coverage)}
\label{fig:diff_FIRAS_HFI_nominalregion}
\end{figure}

We use Eq.~\ref{eq_firas_HFI} to fit for both the gain and the offset
for each HFI detector, and use the measured offset to compute the zero
level. Statistical errors are dominated by the FIRAS errors (the HFI
errors are negligible). The error on $O$ is dominated by the systematic effect observed on the gain $K$ (see next section).  We use the dispersion of the fitted values in different parts of the sky to estimate the systematic error. This error is about 3\,\% at 857, 5\,\% at 545, 5\,\% at 353, and 10\,\% at 143\,\GHz.  At 217 and 100\,\GHz, since we have no way to address the real variations without removing the CO contamination, we also take 10\,\% as the systematic error. For the total error on $O$, we sum the statistical and systematic errors linearly. Errors for some individual bolometers are given in Table~\ref{tab:PZ}.

\begin{table}
\caption{Uncertainties on the zero points (in MJy\,sr$^{-1}$[$\nu I_{\nu}={\mathrm{constant}}$]). }
\begin{center}
\begin{tabular}{@{} c c c c @{}}
\hline
\hline
\noalign{\vskip 2pt}
Detector  & Statistical & Systematic & Sum \\
\hline
\noalign{\vskip 2pt}
857-1 &	0.23 & 	2.40	& 2.63 \\
545-1 &	0.67	& 0.65	& 1.32 \\
353-1 &	0.57	& 1.78	& 2.25 \\
217-1 &	0.57	& 1.08	& 1.65 \\ 
143-5 &	0.61	& 0.66	& 1.27 \\
100-1a &	0.85	& 0.42 &	1.67 \\
\hline
\end{tabular}
\end{center}
\label{tab:PZ}
\end{table}

Using the frequency maps and the dedicated component separation in the CIB fields \citep{planck2011-6.6}, we can compute the CIB mean value in those fields and compare them with the expected values (see Table~\ref{tab:PZ_CIB}). Although the error bars on the measured CIB at low frequencies are quite large, we see a systematic trend: the measured CIB is systematically lower than the expected CIB by factors of 2.4, 2, 1.4, and 1.5 at 857, 545, 353, and 217\,\GHz, respectively. 

\begin{table*}
\caption{Values computed in the CIB fields (averaged on N1, SP, AG, LH, and Bootes, see \citealt{planck2011-6.6} for more details) and expected CIB from \cite{lagache99} FIRAS measurements, and from the \cite{bethermin2012} model, as given in Table~\ref{tab:cib}.}
\begin{center}
\begin{tabular}{@{} c c c c @{}}
\hline
\hline
\noalign{\vskip 2pt}
Frequency & CIB from HFI & CIB measured from FIRAS & CIB from \cite{bethermin2012}\\
$[$GHz$]$ & $[$MJy\,sr$^{-1}]$ ($\nu I_{\nu}={\mathrm{constant}}$) &
$[$MJy\,sr$^{-1}]$ ($\nu I_{\nu}={\mathrm{constant}}$) &$[$MJy\,sr$^{-1}]$
($\nu I_{\nu}={\mathrm{constant}}$)\\
\noalign{\vskip 2pt}
\hline
\noalign{\vskip 2pt}
857 & 	0.29	& 0.71$\pm$0.23	& 0.64\\
545 &	0.18	& 0.37$\pm$0.12	& 0.35\\
353	& 0.09 &	0.13$\pm$0.04	& 0.13\\
217	& 2.2$\times$10$^{-2}$	& (3.4$\pm$1.1)$\times$10$^{-2}$
& 3.3$\times$10$^{-2}$\\
\hline
\end{tabular}
\end{center}
\label{tab:PZ_CIB}
\end{table*}

An example of a residual map is shown in Fig.~\ref{fig:diff_FIRAS_HFI_nominalregion}. We see that in the sky area used to compute the calibration coefficients the residual is close to zero, except for nearby bright regions (e.g., the Taurus cloud), where the FIRAS brightness is underestimated compared to HFI.

\begin{figure}
\centering
\includegraphics[width=0.45\textwidth]{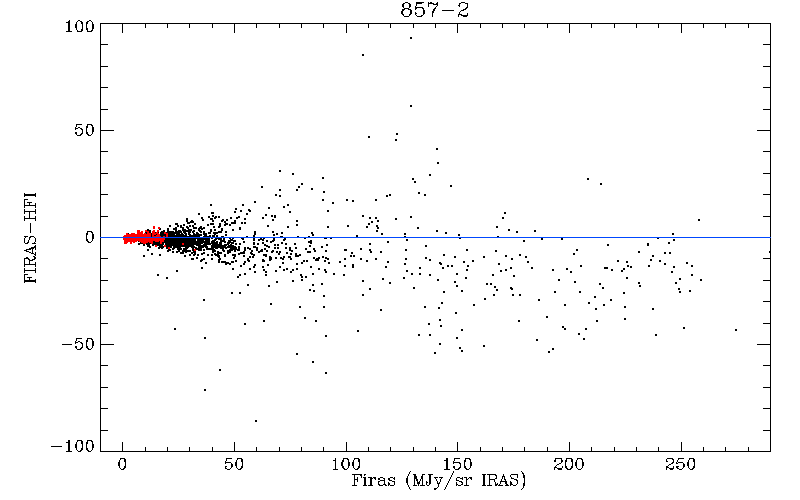}
\caption{Scatter plot of FIRAS minus convolved-HFI versus FIRAS (in MJy sr$^{-1}$). On average, the difference becomes more negative as the brightness increases. The red points are those used to compute the nominal calibration coefficients. At high brightness the HFI is overestimated relative to FIRAS.}
\label{fig:scatter_plot_FIRAS_HFI}
\end{figure}

\subsection{Systematic effects in the calibration coefficients \label{FIRAS_Syst_Effect}}
{  As already mentioned, we observe spatial variations of the FIRAS calibration gain (i.e., a  variation of $K$, and thus of $O$).
Indeed, the calibration coefficients recovered from the narrow part of the Galactic plane and the intermediate Galactic latitude regions (with lower gradients) differ by 10$-$15\,\% (see Fig.~\ref{fig:scatter_plot_FIRAS_HFI}).} 
To understand this unresolved discrepancy, we consulted the FIRAS team on: (1) the FIRAS beam knowledge and its potential changes with frequency, with the FIRAS beam to first order being independent of frequency (given by a geometrical optics dominated horn); and (2) non-linearity effects. None of these explain the discrepancy. We also investigated several other possibilities:
\begin{itemize}
\item Pixelization: FIRAS data and errors are given in the quadrilateralized spherical cube projection\footnote{FIRAS data in \healpix\ format are available in NASA's Legacy Archive for Microwave Background Data (\url{http://lambda.gsfc.nasa.gov}) but the covariance matrix is not provided}. In the nominal pipeline, we computed the calibration gain $K$ after reprojecting FIRAS data onto the \healpix\ grid. We have also tried
 reprojecting the HFI data onto the cube (using several schemes for the pixel decimation) to compare the HFI and FIRAS data. We find no difference in the photometric calibration.
\item FIRAS beam: beam uncertainty could result in some variations of $K$ where the signal is rapidly varying on the sky (Galactic plane, molecular clouds, bright cirrus regions). 
In the HFI calibration, we do not account for FIRAS beam variations
with frequency. But we tested several ``beam configurations'' to investigate
their impact on the calibration. First, we measured the beam window
function $B_{\ell}$ using full-sky FIRAS and HFI power spectra. A
good fit is obtained for a Gaussian with a ${\mathrm{FWHM}}=4\pdeg94$. We used this beam in the convolution rather than the ```nominal beam'' to cross-calibrate the two data sets. We also used other FWHMs (4\deg, 8\deg, and 10\deg). We could not find any beam that reconciles the FIRAS and HFI data.
\item Colour corrections: To compute $K$ we need to correct the data for
  the variation of the spectral energy distribution of dust emission
  across the sky. Working at $7\deg$ and having the FIRAS dust spectrum for each pixel, it was easier to compute the colour correction at the FIRAS resolution ($\cal{C}[7\deg]$). However,  
\begin{equation}
\cal{C}[7\deg] \int_{7\deg} S_\nu[5\arcm] d\Omega \ne \int_{7\deg} \cal{C}[5\arcm] S_\nu[5\arcm] d\Omega\,,
\end{equation}
so we checked (for several bolometers) whether the colour corrections could produce the observed spatial variations of $K$. We used the DR2 all-sky temperature map (from \citealt{planck2011-7.0}) obtained by fitting HFI+IRIS data with a spectral index equal to 1.8, to compute a colour-correction map for each HFI pixel, and modified the calibration pipeline to use this $\cal{C}[5\arcm]$. A strong variation of $K$ was still observed, comparable to the variation observed using $\cal{C}[7\deg]$. These refined colour corrections cannot explain the variation of $K$ across the sky. Note however, that using $\cal{C}[5\arcm]$ rather than $\cal{C}[7\deg]$ significantly changes the calibration coefficient for the 545\,\GHz\ channels, by ~6\,\% (although it does not change the 857\,\GHz\ coefficient).  
\item Zodiacal light: The FIRAS data set we are using has the zodiacal
    emission removed using the \COBE\ model. When comparing HFI and
    FIRAS data, zodiacal residuals are clearly visible in the
    difference map. We therefore redid the photometric calibration
    using the HFI data with the zodiacal light removed. The difference
    [${\mathrm{FIRAS}} - K \times {\mathrm{HFI}}$] does not show any
    zodiacal residuals. Removing the zodiacal emission decreases the
    calibration coefficient by less than 2\,\% at 857 GHz. However, it
    does not decrease the observed spatial variations of $K$.
\item Far sidelobes: we tested whether far sidelobes could have any impact on the photometric calibration, by looking at the detectors that have low far-sidelobe contamination (e.g., 857-2). We noticed that the spatial variation of $K$ is at the same order whatever the FSL contamination.
\item Time gain variations: we have searched, unsuccessfully, for any temporal gain variations using the individual calibration of the HFI all-sky survey maps.
\end{itemize}
We thus have no explanation for this effect other than a possible systematic bias in the FIRAS {\tt pass4} interstellar dust spectra.\\

If the  HFI brightness is calibrated using FIRAS, several results suggest that the HFI brightness is overestimated at high frequencies:
\begin{itemize}
\item The SEDs of sources and diffuse dust show an excess at 545\,\GHz\ over a smooth interpolation between higher and lower frequencies. A simple interpolation between 857 and 353\,\GHz\ shows the excess is about 11\,\%. Using a very simple dust model, a residual dipole is also present in the 545\,\GHz\ maps.
\item The CMB anisotropy power spectrum is detectable at 545\,\GHz\ and the {\tt SMICA} component separation method shows a (20.3$\pm$4.7)\,\% calibration discrepancy. The analysis of the FFP6 simulated data sets shows that the same method gives reliable results at all HFI frequencies.
\item The dipole calibration at 545\,\GHz, although quite uncertain, is also discrepant by about 20\,\% with the FIRAS calibration.
\item The measurements on planets (Mars, Jupiter, Saturn, Uranus, and Neptune) used for beam determination and pointing are higher than the models for the two sub-millimetre channels, at least for the deetctors that are not affected by non-linearity effects.
\end{itemize}
We therefore abandoned the FIRAS calibration and used the planet calibration instead.

\raggedright

\end{document}